\documentclass[useAMS,usenatbib]{mn2e}

\usepackage{graphicx}
\usepackage{aas_macros}
\newcommand{\ltsim}{\raisebox{-0.5ex}{$\;\stackrel{<}{\scriptstyle \sim}\;$}}
\newcommand{\gtsim}{\raisebox{-0.5ex}{$\;\stackrel{>}{\scriptstyle \sim}\;$}}
\newcommand{\unit}[1]{\ifmmode \:\mbox{\rm #1}\else \mbox{#1}\fi}

\title{The red halos of SDSS low surface brightness disk galaxies}

\author[N. Bergvall, E. Zackrisson and B. Caldwell]{{Nils 
Bergvall$^{1}$\thanks{E-mail:nils.bergvall@fysast.uu.se;ez@astro.su.se; brady.caldwell@fysast.uu.se}, Erik 
Zackrisson$^{2}$ and Brady Caldwell$^{1}$ }\\
$^{1}$Department of Physics and Astronomy, Uppsala University,
  Box 515, SE-751~20 Uppsala, Sweden\\
$^{2}$Oscar Klein Centre for Cosmoparticle Physics, Department of Astronomy, Stockholm University, 10691, Stockholm, Sweden\\
}

\begin{document}

\date{Received ; accepted }

\pagerange{\pageref{firstpage}--\pageref{lastpage}} \pubyear{2007}

\maketitle

\label{firstpage}

\begin{abstract}
%context heading (optional)
{The faint stellar halos of galaxies contain key information about the oldest stars and the process of galaxy formation. A previous study of stacked SDSS images of disk galaxies has revealed a halo with an abnormally red $r-i$ colour, seemingly inconsistent with our current understanding of the stellar populations inhabiting stellar halos. Measurements of this type are however plagued by large uncertainties which calls for follow-up studies.}
%  aims heading (mandatory)
{Here, we investigate the statistical properties of the faint envelopes of low surface brightness disk galaxies to look for further support for a red excess.}
%  methods heading (mandatory)
{A total of 1510 nearly edge-on, bulgeless low surface brightness galaxies were selected from the SDSS Data Release 5, rescaled to the same apparent size, aligned and stacked. This procedure allows us to reach a surface brightness of $\mu_r\sim$ 31 mag arcsec$^{-2}$. 
After a careful assessment of instrumental light scattering effects in the stacked images, we derive median and average radial surface brightness and colour profiles in $g,r$ and $i$.}
%  results heading (mandatory)
{The sample is then divided into 3 subsamples according to $g-r$ colour. All three samples exhibit a red colour excess in $r-i$ in the thick disk/halo region. The halo colours of the full sample, $g-r = 0.60\pm0.15$ and $r-i = 0.80\pm0.15$, are found to be incompatible with the colours of any normal type of stellar population. The fact that no similar colour anomaly is seen at comparable surface brightness levels along the disk rules out a sky subtraction residual as the source of the extreme colours. A number of possible explanations for these abnormally red halos are discussed. We find that two different scenarios -- dust extinction of extragalactic background light and a stellar population with a very bottom-heavy initial mass function -- appear to be broadly consistent with our observations and with similar red excesses reported in the halos of other types of galaxies.}
\end{abstract}


\section{Introduction}
\label{sec:intro}

%-------------------------------------------------------------
%\begin{figure*}

%\resizebox{\hsize}{15cm}{\includegraphics*{x.ps}}

\begin{figure*}
\centering
\begin{minipage}[c]{0.34\linewidth}
   \centering{\includegraphics[width=6.08cm]{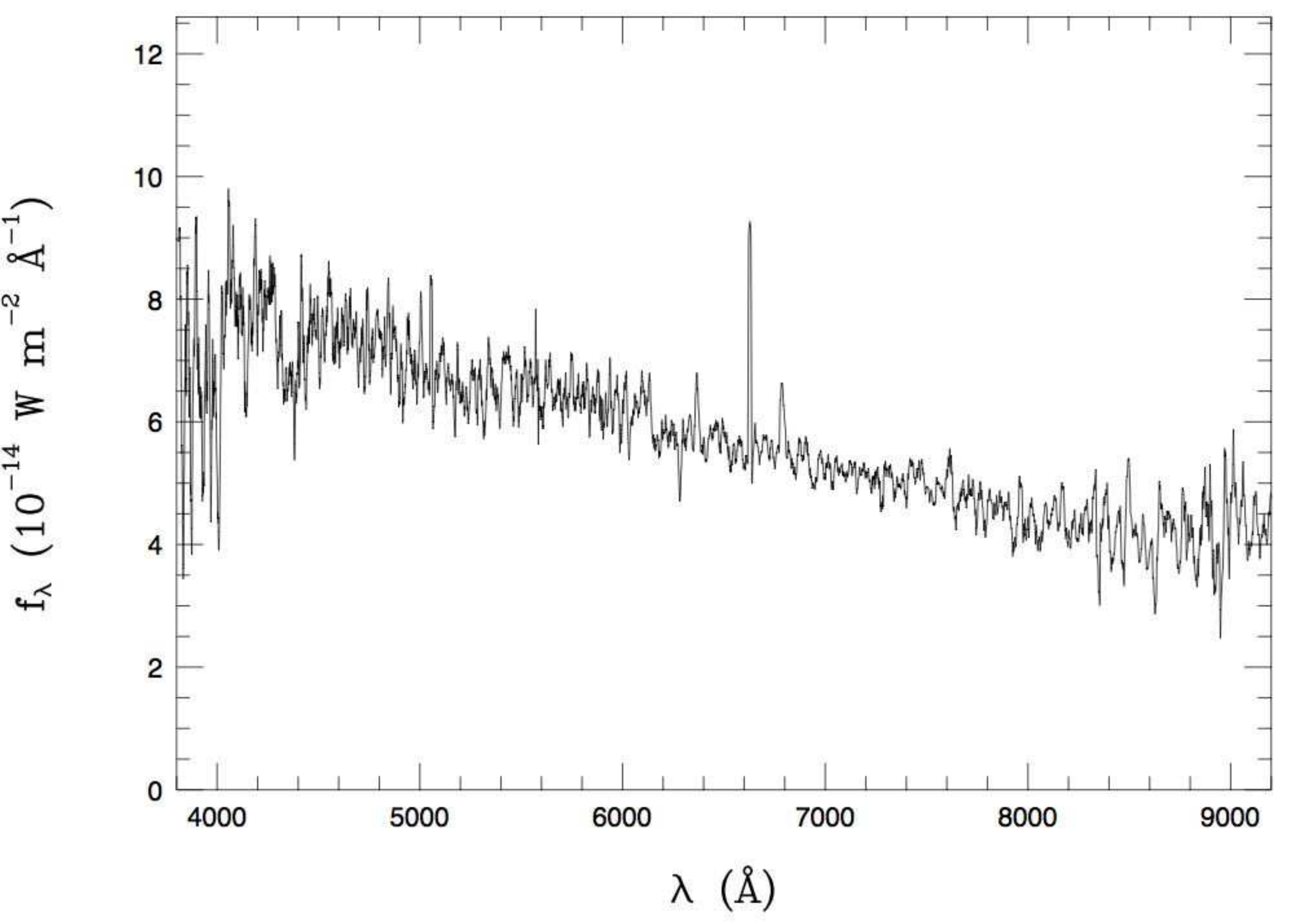}}
\end{minipage}%
\begin{minipage}[c]{0.33\linewidth}
   \centering{\includegraphics[width=5.6cm]{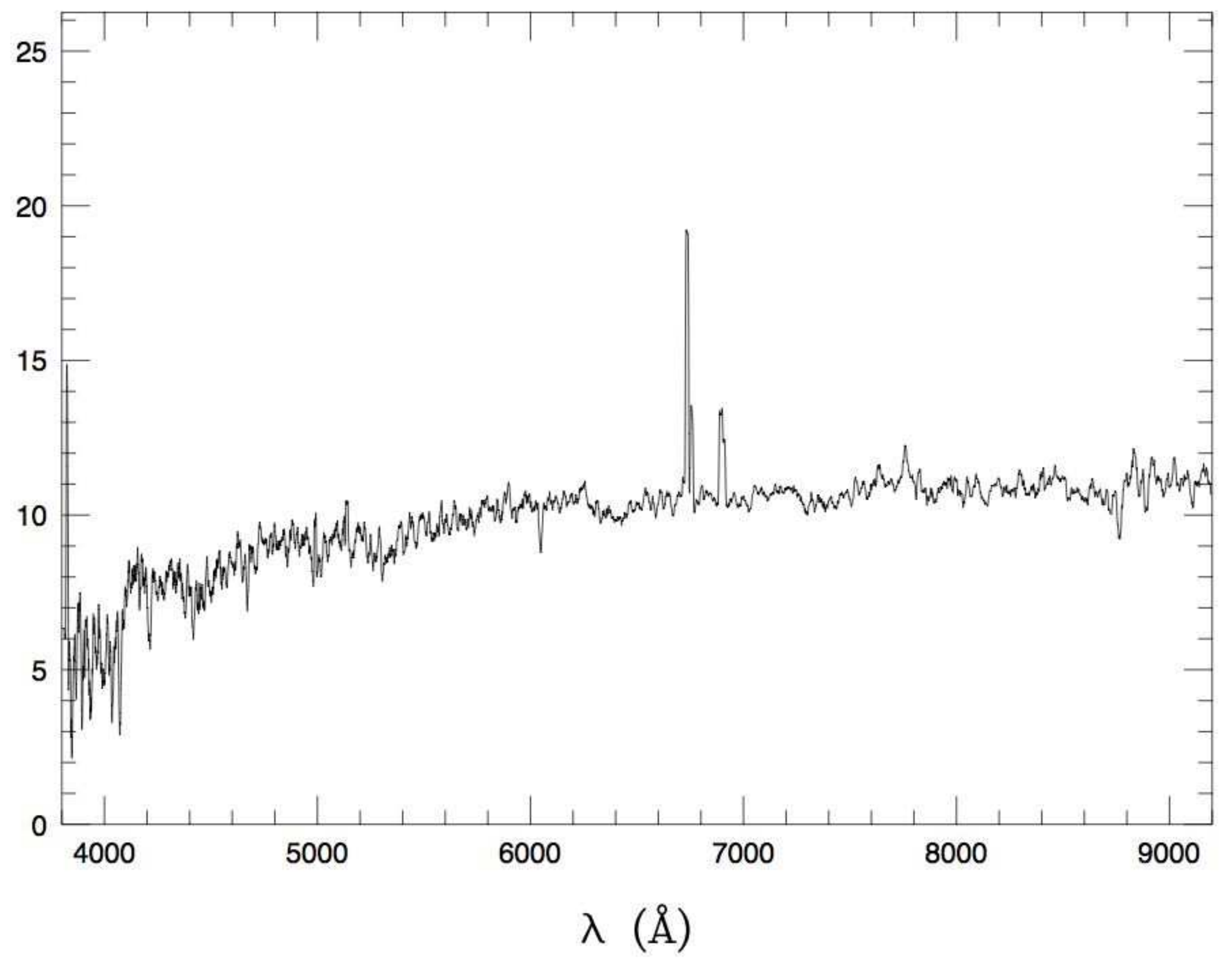}}
\end{minipage}%
\begin{minipage}[c]{0.33\linewidth}
   \centering{\includegraphics[width=5.6cm]{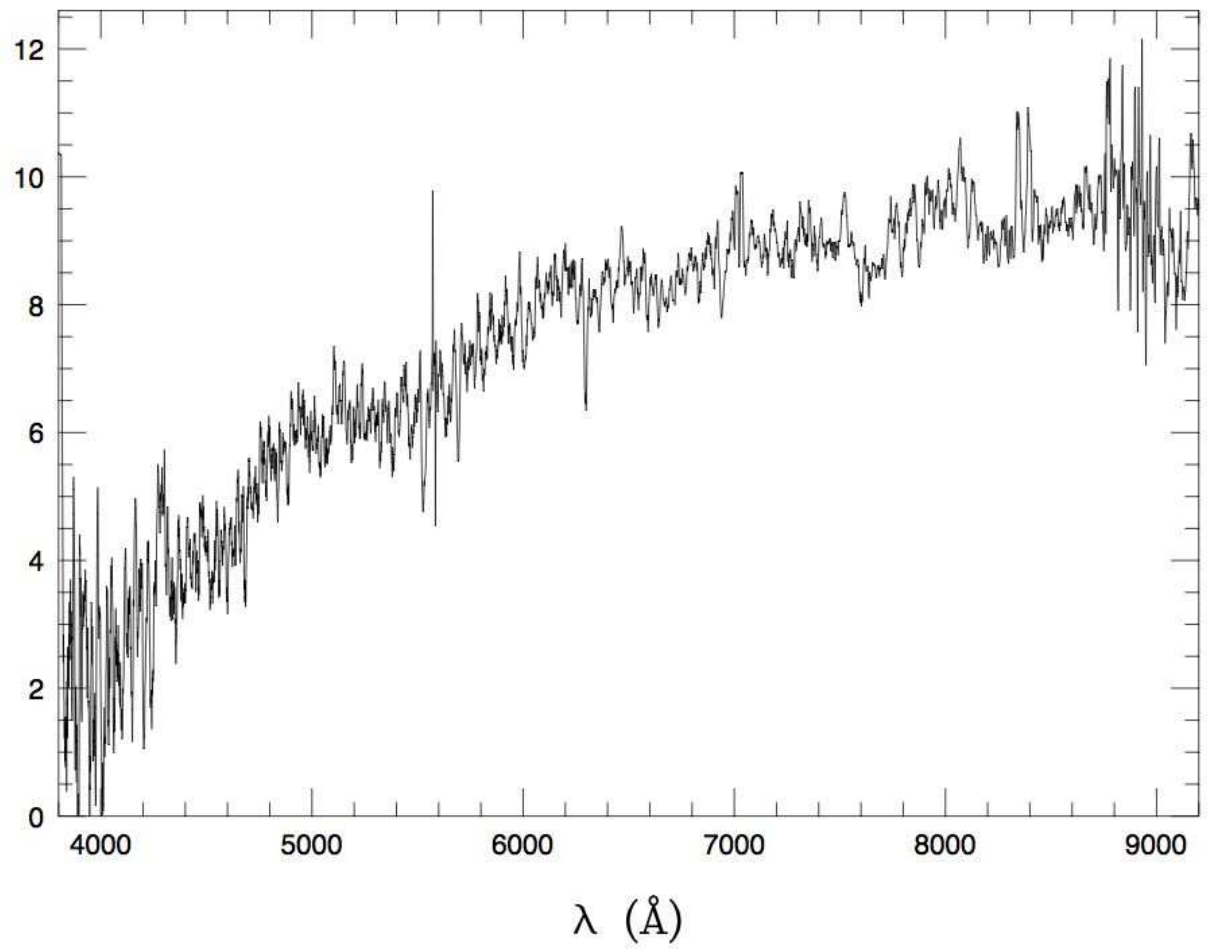}}
\end{minipage}   
   \caption{Characteristic spectra of the galaxies in the three subsamples 
A,B and C obtained from the SDSS. }
      \label{spectra}
\end{figure*}

The stellar halos of galaxies are nearly spherical, old and metal-poor structures with little net rotation. Due to the long orbital time scales of halo stars, these halos contain the fossil record of galaxy assembly, and observations of resolved halo stars in the Milky Way, Andromeda and other nearby galaxies  \citep[e.g.][]{2005ApJ...633..828M,2007ApJ...671.1591I,2008A&ARv..15..145H,2009MNRAS.396.1231R, 2009IAUS..254P..19D} have provided a wealth of information on such structures. Current simulations of stellar halos \citep[e.g.][]{2005ApJ...635..931B,2006MNRAS.365.1309H,2006ApJ...653.1180G, 2007ApJ...666...20P, 2008MNRAS.391...14D, 2008ApJ...673..215F, 2009ApJ...702.1058Z, 2009MNRAS.395.1455S} indicate that they form through a combination of disruption of accreted dwarf galaxies, and by migration of old stars outwards from the inner regions of galaxies. Models of this type are able to explain many of the features observed -- including the numerous streams from disrupted satellite galaxies, the inner-outer halo dichotomy, and the existence of stellar halos around dwarf galaxies. However, there is a lingering uncertainty concerning the integrated colours of halos.

Observationally, there are two complementary techniques for studying the halos of galaxies: direct star counts and surface photometry. Star counts can trace the structure of stellar halos out to much larger distances (or, equivalently, to fainter surface brightness limits) than surface photometry can, but is currently only applicable for galaxies at distances out to $\sim 10$ Mpc. Surface photometry (i.e. the study of the integrated light from large numbers of unresolved sources within the each system) can be used for far more distant galaxies and in principle gives a more direct measurement of surface brightness, since it is sensitive both to stars at masses below the detection threshold of current star counts and to diffuse light from the interstellar medium. This is, on the other hand, a far more challenging technique, which requires that the sky flux can be subtracted with excruciating accuracy and that various instrumental scattering effects can be kept under control. Curiously, a number of attempts to study the halos of galaxies through surface photometry have revealed halo colours that are too red to be straightforwardly reconciled with the halo populations captured through the study of resolved stars (see \citealt{2006ApJ...650..812Z} and \citealt{2007arXiv0708.0762Z} for a general discussion of this phenomenon). 

The first indication of a red excess in the halos of disk galaxies was found through BVRIJK photometry of the edge-on system NGC 5907.  The interest was initiated by the discovery of a faint extended $R$-band halo around this galaxy \citep{1994Natur.370..441S}, which triggered follow-up studies in the optical and near-IR \citep{1996A&A...312L...1L,1997Natur.387..159R,1998MNRAS.301..280J}. A common interpretation of the red excess was that the galaxy halo hosted an unusually large population of low-mass stars. Later, deep images in the optical \citep{1998ApJ...504L..23S,1999AJ....117.2757Z,2008ApJ...689..184M} revealed complex arc structures in the halo of the galaxy that, when mixed with scattered light from stars in the field, may have caused confusion in the analysis of previous data of lower S/N. Therefore the reality of a true, diffuse halo in NGC 5907 at the surface brightness levels where the red excess had been measured was questioned by  \citet{1999AJ....117.2757Z}. Still, an HST study of resolved stars in the halo \citep{2000AJ....119.1701Z} indicated a relative paucity of massive stars which could be interpreted either as a very low metallicity of the stellar population or as a bottom heavy initial mass function (IMF), inconsistent with the halo colours as observed from the ground. 

Surface photometry at wavelengths outside the optical at low brightness levels are valuable as a complement to the optical. If the red excess is of thermal origin, it is likely to show up also when comparing the optical to the near-IR fluxes. A few years ago \citep[see][]{2002A&A...390..891B}, we obtained deep optical/near-IR images of a handful of Blue Compact Galaxies (BCGs). When 
inspecting the faint halo regions we found a strong red excess in the near-IR, again
inconsistent with normal stellar populations. Dust emission or 
extinction could be ruled out as a major cause of the excess in our sample. The
infrared excess was later confirmed for a larger sample of BCGs \citep{2005mmgf.conf..355B}. The results from these studies relied on contemporary models of galaxy spectral evolution. An uncertain ingredient in these models is the late stages in the evolution of late type stars. Recent models predict redder colours of stars of intermediate masses which tends to reduce the discrepancy between observations and models \citep{2009arXiv0902.4695Z}. Still, we are left with an uncomfortable fine tuning of age of the halo stars and either a metallicity or dust reddening that must be higher than in the starburst region.

One problem with all these investigations is the difficulties of dealing with data 
at very low surface brightness levels, much below  1\% of the background sky. 
Under such conditions it becomes critical how the subtraction of the emission from the sky and scattered light is 
performed. This is a problem that we can never overcome as long as we use data 
from groundbased instruments. A second problem is due to the low photon flux rate in the studied regions, causing an uncomfortably large photometric inaccuracy. If we want to study the halos of individual galaxies, the only way we can avoid this problem is to use very large telescopes or extend the integration times to the extreme. 

An alternative approach was tested by \citet{2004MNRAS.347..556Z}, hereafter Z04. It is based on the statistical analysis of a large sample of galaxies with common fundamental properties (essentially morphologies, luminosities and stellar content). In such a fairly homogeneous sample one may assume (since one cannot measure each individual galaxy with sufficient S/N) that the physical properties of the faint halos of the galaxies are also similar. An increase in S/N can therefore be obtained by stacking all images after rescaling and aligning the individual images of the galaxies.
Based on data from the Sloan Digital Sky Survey (SDSS), Z04 tested this approach and digitally stacked 
images of 1047 nearly edge-on disk galaxies. As a result 
they managed to reach extremely low surface brightness levels, $\mu _r\sim$ 31 
mag arcsec$^{-2}$. When measuring the colours in the 
polar direction, they found that the colours ($g-r$= 0.65$\pm$0.1, $r-i$= 
0.60$\pm$0.1) of the faintest regions where reliable data could be obtained, were deviating in $r-i$ with 2$\sigma$ or about 0.2$^m$ 
from the reddest population observed in areas of higher surface brightness. Although acknowledging that a small amount of the red excess could be credited to light scattering effects, they still 
concluded that "$r-i$ [...] is difficult to reconcile with any theoretical models, 
even allowing for ad hoc modified IMFs dominated by low-mass stars and high 
metallicity". One might suspect that the excess they found in the optical ($i$ band) 
was a confirmation of the red excess we found in the near-IR. This remains to be 
confirmed, however \citep[see discussion in e.g.][]{2009arXiv0902.4695Z}.  While being confident with the existence of an abnormal 
red excess, Z04 point out that their "colour measurements provide 
inconclusive and troublesome results". Recently, there was a claim by  \citet{2008MNRAS.388.1521D} that the abnormal colours could have been caused by an improper treatment of the point spread function (PSF) when analysing the effects of light scattering. An independent confirmation or falsification of the results by Z04 
is therefore obviously of highest concern.

In the present investigation we carry out a follow-up of the work of Z04. Preliminary
results were presented by \citet{2007IAUS..235...82C}. The 
sample we select for our study is different from that of Z04  and includes more galaxies, thus increasing the S/N after stacking. It focuses on another type of galaxy, Low Surface 
Brightness Galaxies (LSBGs), and thus allows us to make more general conclusions about the phenomenon, if a red halo is found. We chose to work with LSBGs mainly for two reasons. Firstly, they   have a low 
dust content 
\citep[e.g][]{1999A&A...341..697B,2005AJ....129.1396H,2007ApJ...663..908R}, 
thereby essentially removing the potential problem of dust reddening.
Secondly, the influence of the light scatter from the bulge is less severe than in normal disks. This 
second argument is weaker however, since we do not know {\it a priori} if the red halo 
perhaps is physically correlated with the bulge component. An additional advantage is 
that we could learn about the properties of the spheroidal component of LSBGs, 
hitherto almost unknown.

The method we will use is similar to that used by Z04 with a few 
modifications as described below. We select images of LSBGs from SDSS in the $g, 
r$ and $i$ bands, scale and align and finally stack the images. In the following 
we will describe the selection criteria and reduction procedures in more detail.

\section{Data analysis}
\subsection{Selection criteria}

We selected a sample of nearly edge-on LSBGs from the SDSS Data Release 5 (DR5) database \citep{2007ApJS..172..634A}.

In the selection of targets we used the following criteria:

\begin{enumerate}
\item 23$<\mu_{r,50}<$25 mag arcsec$^{-2}$, where $\mu_{r,50}$ is the mean 
surface brightness inside the radius encompassing 50\% of the light in the 
$r$-band
\item $i$-band isophotal diameter of $a\geq$ 10 arcsec
\item $i$-band axial ratio of $b/a<$ 0.25
\item redshift $z<$ 0.2
\end{enumerate}

The red $i$ band was chosen in order to minimize the 
effects on the size parameters due to patchiness caused by young blue star 
forming regions and dust extinction. The axial ratio was chosen in order to avoid too much contribution from the inclined disk to the halo light. We made the choice with the aim to optimize the S/N of the halo light. The halo region of interest that we will discuss below starts at about 60 pixels in the vertical direction. At that point the inclined disc contributes about 20-30\% to the light in the $i$ band. At larger distances the relative contribution is considerably lower. The useful region for determination of the halo colour with the present axial ratio limit is $\sim$ 80-110 pixels. Should we increase the  limit, we would obtain more galaxies but the usefiul region would shrink and the reverse would be the case if we should decrease the limit. In the end we are limited by the value of the intrinsic flattening.  

The final sample contained 1510 galaxies. We used Petrosian magnitudes and divided the sample by $g-r $ colour into three subsamples, a "blue" sample (A), a "green" sample (B) and a "red" sample (C). Set Table~1 for details.  The goal of this 
subdivision was to search for possible correlations between the strength of the 
red excess and the total colour/luminosity. The maximum redshift was chosen to avoid uncertain $k$-corrections, uncomfortably noisy data and to permit a direct comparison with 
the study by Z04. In fact very few galaxies fulfilling the remaining criteria are found at higher redshifts. 

All images were corrected for atmospheric extinction based on primary standards 
observed regularily at Apache Point (where the SDSS observations are carried out). Galactic extinction was corrected for 
according to \citet{1998ApJ...500..525S}. No $k$-corrections were applied. This 
will affect the red sample the most since those galaxies have the largest median 
redshift. The correction is however relatively small for a normal galaxy of intermediate/late type. At the median redshift 
($<z>\sim0.1$) of the reddest sample, the $k$-correction is $<$0.2$^m$ in both 
colours (see e.g. \citealt{1995PASP..107..945F}). A $k$-correction of the halo 
colours is anyhow useless because, when it comes to regions showing a red excess, 
we do not know {\it a priori} the origin of the radiation.  In section 4 we will discuss how the colours are affected at different redshifts assuming various types of radiation sources. 

\begin{table}
\centering
\caption{Properties of subsamples A, B, C and the total sample, T.}
\begin{tabular}{lll}
\hline
      Name & No. of galaxies & Colour constraint \\
\hline
A & 447 & $g-r< $ 0.55  \\
B & 655 & 0.55 $\leq g-r<$ 0.75 \\
C & 408 & 0.75 $\leq g-r$ \\
T & 1510 &  \\
\hline
\end{tabular}
\end{table}

\subsection{Sample properties}

The following figures show comparisons between the properties of each subsample
and the full sample. Typical examples of spectra of galaxies from the three 
subsamples are shown in Fig. \ref{spectra}.  They show that there is a large difference in age of the stellar population dominating the light within the SDSS aperture ($\phi$=3") from a few 10$^7$ years to several Gyr.  

%-------------------------------------------------------------
\begin{figure}
\centering
\includegraphics[width=8cm]{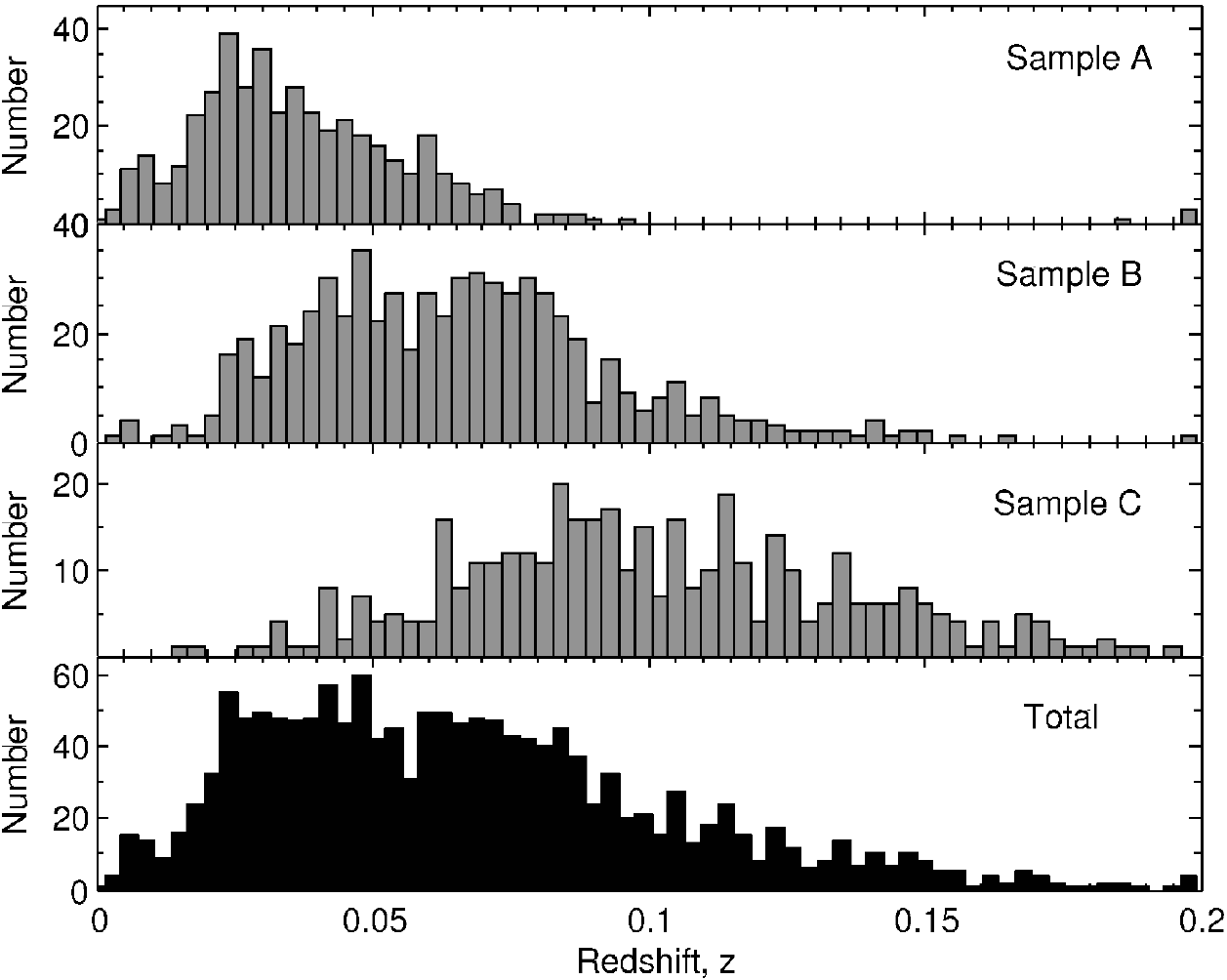}
   \caption{The redshift distribution of the subsamples 
   	      and the full sample.
           }
      \label{redshifts}
\end{figure}
%
%_____________________________________________________________
%-------------------------------------------------------------
\begin{figure}
\centering
\includegraphics[width=8cm]{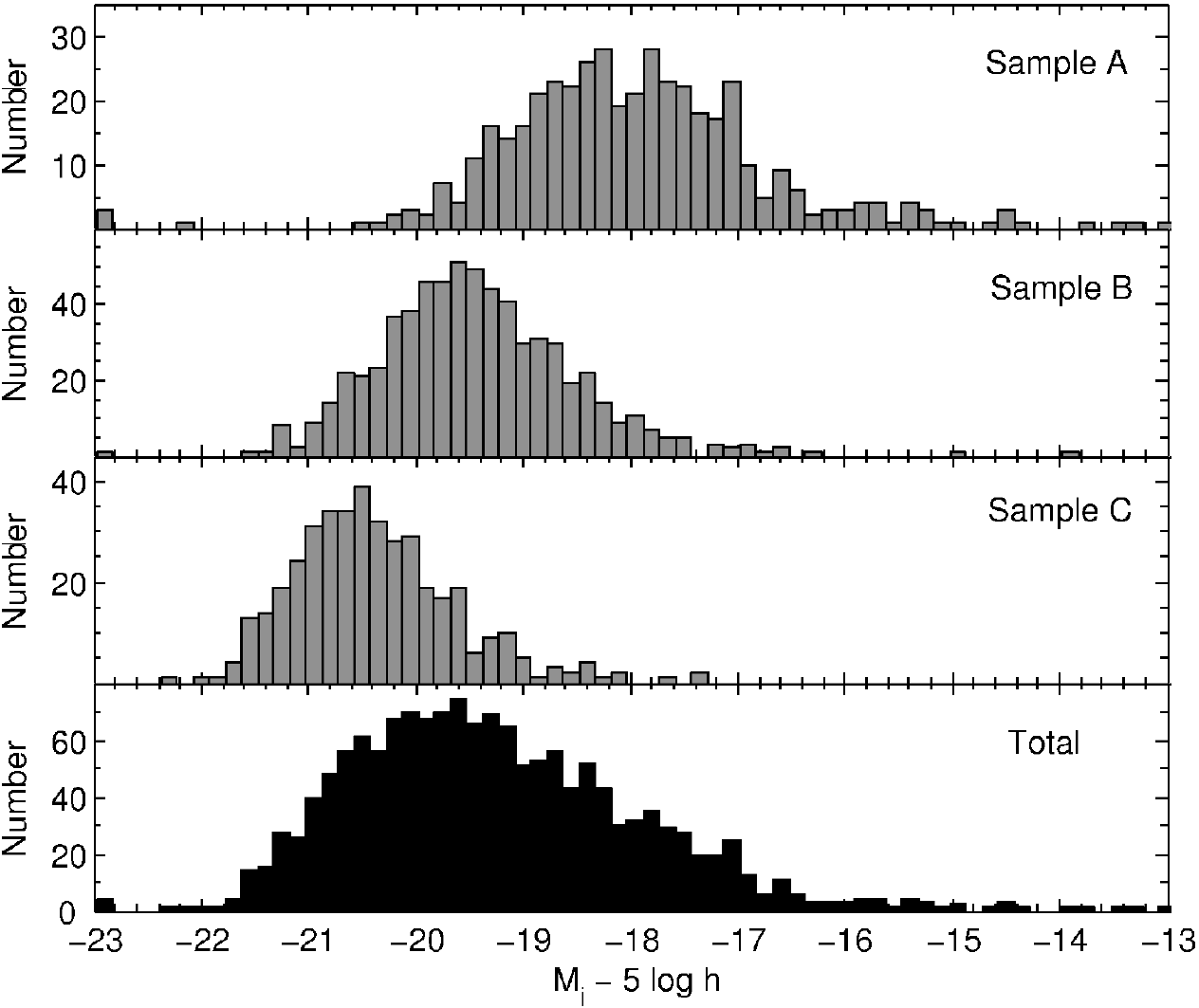}
   \caption{The distribution of absolute magnitudes in the $i$ band of the 
   	galaxies in the subsamples 
   	      and the full sample.
           }
      \label{magnitudes}
\end{figure}
%
%_____________________________________________________________
%-------------------------------------------------------------
\begin{figure}
\centering
\includegraphics[width=8cm]{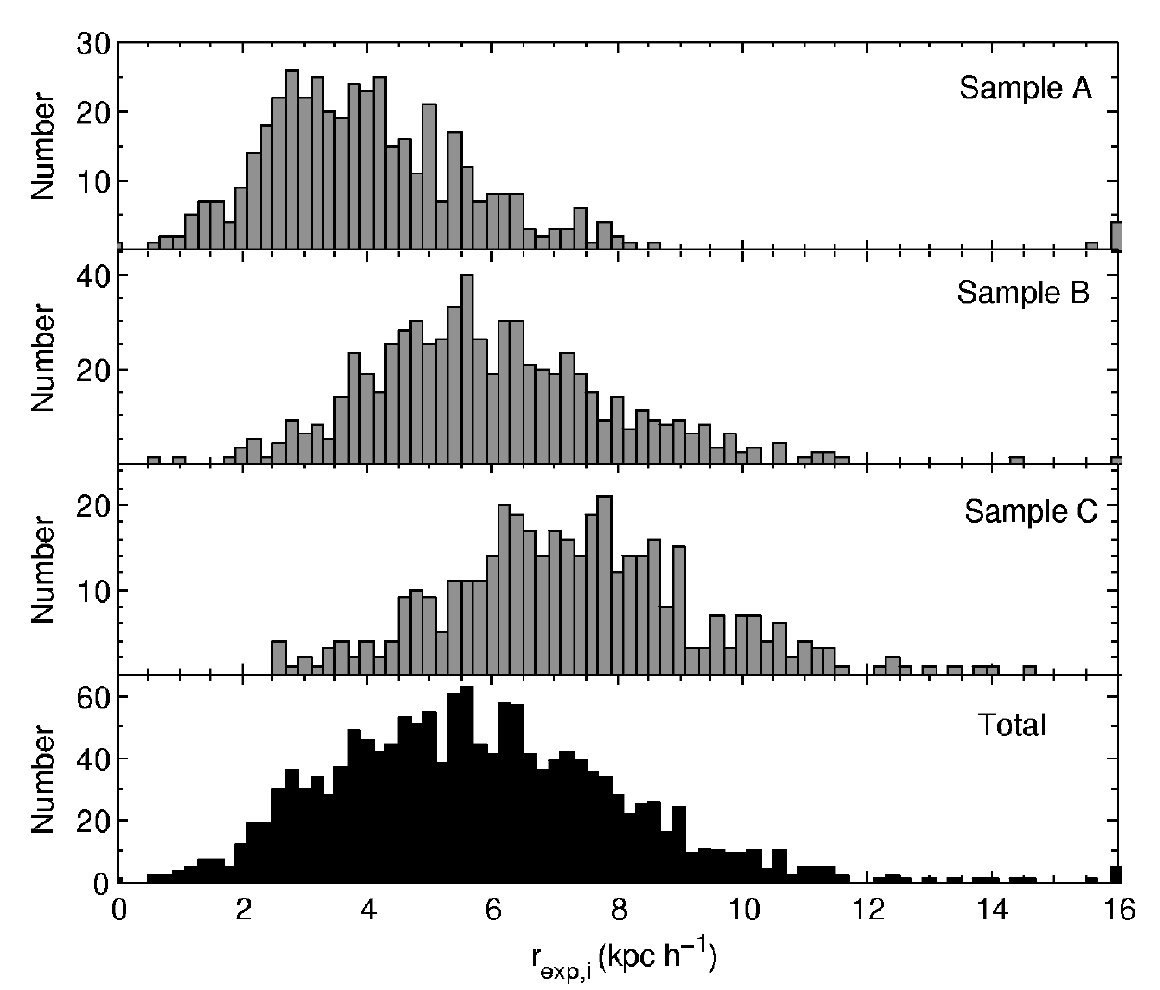}
   \caption{The distribution of exponential scalelengths of the galaxies in the 
subsamples and the full sample.
           }
      \label{scalelengths}
\end{figure}
%
%_____________________________________________________________
%-------------------------------------------------------------
\begin{figure}
\centering
\includegraphics[width=8cm]{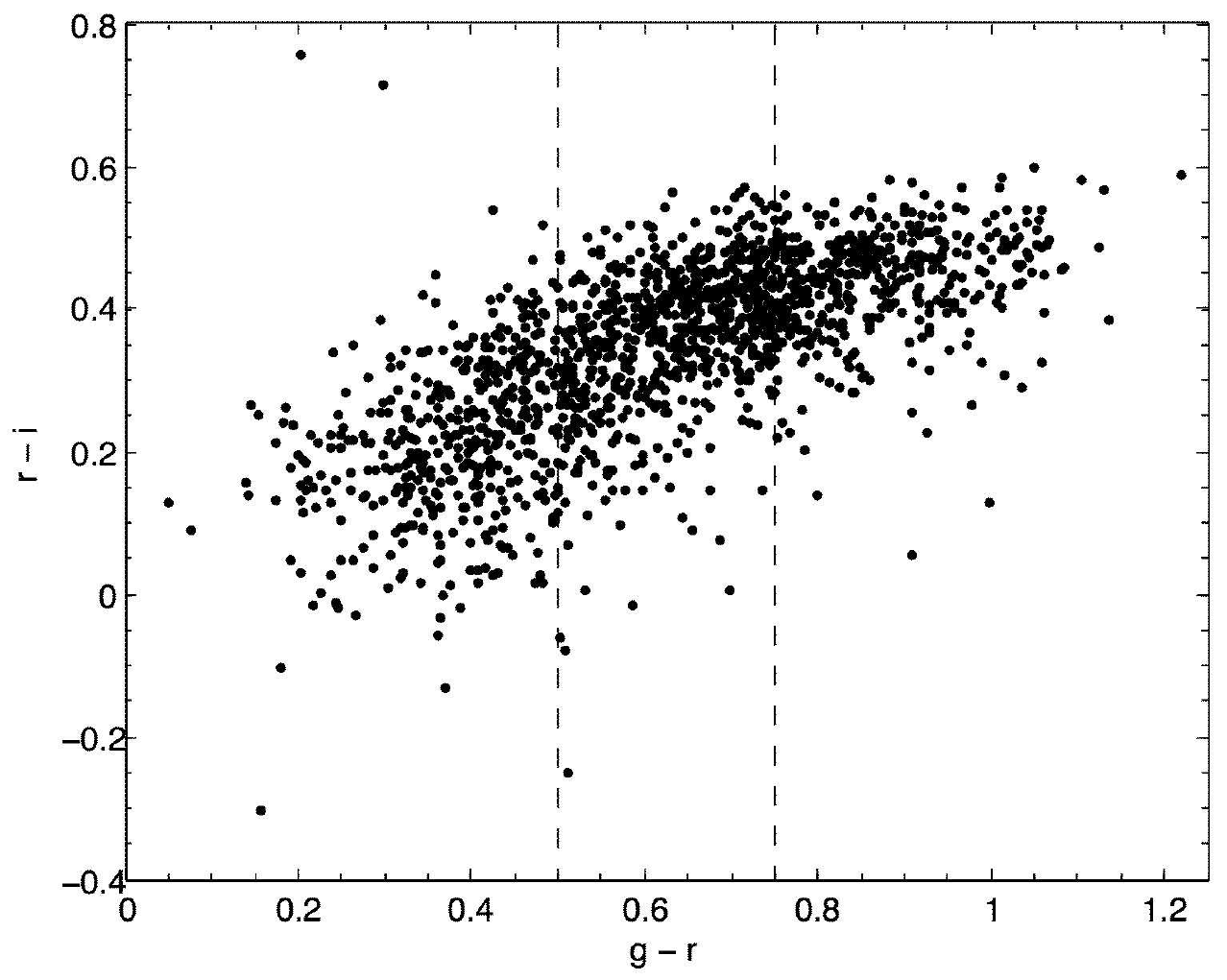}
   \caption{The two-colour distribution of the galaxies in the three 
subsamples, separated at colours $g-r$=0.55 and 0.75.
           }
      \label{colours}
\end{figure}
%
%_____________________________________________________________

The distribution of redshifts is shown in Fig. \ref{redshifts}. There is a clear 
trend towards higher redshifts as we go from the blue to the red sample, with a 
median redshift of z$\sim$0.06. The correlation between redshift and colour 
reflects the fact that the galaxies in the blue sample are smaller and therefore 
reach the limiting apparent diameter at smaller redshifts than the redder ones. 
There is also a related correlation between colour and absolute luminosity shown 
in Fig. \ref{magnitudes}, displaying the distribution of absolute magnitudes. 
The bluer Sample A falls within the classical range of typical LSB galaxy 
luminosities of -13 $\leq M_B\leq $-19 
\citep{1996ApJS..105..209I}.  These galaxies are normally gas rich and globally star forming. The peaks of the green and red samples are moving 
more toward normal spiral galaxy type absolute magnitudes. 

Fig. \ref{scalelengths} presents the exponential  scalelength distributions. The range of scalelengths is
from 1 to 16 kpc h$^{-1}$ with a median value of 5.5 kpc h$^{-1}$. This is somewhat larger than that of high surface brightness disk galaxies studied by \citet{1997ApJ...482..659D,2003ApJ...582..689M} and {\it twice} the size of the Z04 sample. It confirms and strengthens the results from a previous study of LSB galaxies by \citet{1995MNRAS.273L..35Z}. In this study, the difference in scale length was interpreted as consequence of a larger spin for a given mass in the LSB sample.

The colour-colour distribution is shown in Fig. \ref{colours}. The scatter of 
the whole
sample is most apparent in the blue subsample and most constrained in the
redder LSB galaxies. This may be explained as an effect
of the patchiness of the blue LSB galaxies due to a larger fraction of star 
forming regions.

\subsection{Stacking preparations}

Since our main result, the red excess, could be regarded as controversial, we want to make sure that our data retrieval and processing can be reproduced by anyone who reads this article. In the following we will therefore describe the procedure in detail.

The galaxy target field images were obtained from the SDSS archive. The 
resulting raw SDSS images are of size 2048x1489 pixel$^2$ or 
13.5x9.8 arcmin$^2$. Each of these frames were bias subtracted and flat field 
corrected. For each galaxy we obtained the galaxy centres, position angles, 
scalelengths and atmospheric extinction corrected magnitudes from the SDSS 
database. The so prepared images were then passed through a number of processes one by 
one where they were masked, shifted and rescaled before finally being stacked, 
averaged and median filtered, all with the combined use of the ESO MIDAS command language, version 08FEBpl1.1 on MAC/OSX and 
SExtractor v2.4.4 \citep[Source Extractor,][]{1996A&AS..117..393B} packages in the following way. 

First SExtractor was used to automatically remove all objects in the frame except for the target galaxy itself. A first fit to the sky background 
was initially made based on a course grid of resolution 512x512 pixels. Since the 
typical diameter of a target galaxy is about 10\% of the grid size we did not 
expect serious problems with fluctuations in the fitted sky on the size of the 
halo we were exploring. After the fit, the background was subtracted and all 
objects, defined as at least 5 connected pixles with fluxes $\geq$ 1.5 times the 
local pixel mean error in the sky level, were automatically analysed, identified 
as stars or galaxies and all objects except for the target galaxy were then flagged into a separate file.  The galaxy was now the only object (defined as 
above) present in the frame. The resulting 
dummy file had the galaxy moved to its center and rotated so that the major axis 
of the galaxy became horizontal.  The image was then scaled to a standard exponential 
scalelength of 20 arcseconds while conserving surface brightness. The central 
1024x1024 pixels were extracted into another file: the image subfile. Because the 
first sky fit was made with a coarse grid over a region many times larger than 
the galaxy, weak local gradients at scales below the size of the mesh could still 
be present. Therefore a second sky subtraction was performed in the smaller 
field, based on a grid with a mesh size of 256x256 pixels, excluding the flagged 
regions and the target. The mesh size was thus half the size used for the first fit to the sky background. New objects were now identified and 
flagged. To make sure that the wings of the smaller stars and galaxies were 
completely removed we flagged 5 extra pixles around the flagged regions around 
all detected objects.

All further pruning was done by eye to identify objects not suitable for 
stacking. Some false detections were obvious as problems with the SDSS PHOTO
reduction pipeline. Stars superposed on the stacked galaxy that were not identified by 
SExtractor were flagged. This process is somewhat subjective since at low 
relative brightnesses it is difficult to distinguish between a superposed star 
in the Milky Way and a luminous star cluster in the galaxy. Potential target 
galaxies were rejected if they had obvious warps, excessive dust lanes, bright 
bulges, were in interaction with nearby companions or showed other irregular 
morphologies. A further set was discarded due to the closeness of nearby bright 
stars. Fainter structures were identified after filtering the images with a 
gaussian filter of $\sigma$ = 5 pixels. The filtering makes it easier to detect 
wings from bright stars just outside the field borders and other low frequency 
variations. Unwanted regions were thus flagged or the whole image was deleted. 
The returned rough sample was culled to a total of 1510 galaxies: 447 in
sample A, 655 in sample B and 408 in sample C.  The relative fraction of galaxies fulfilling the selection criteria that were finally included in the stacking were 62\%, 71\% and 69\% for the A,B and C samples respectively.

%-------------------------------------------------------------
\begin{figure}
\centering
\includegraphics[width=7cm]{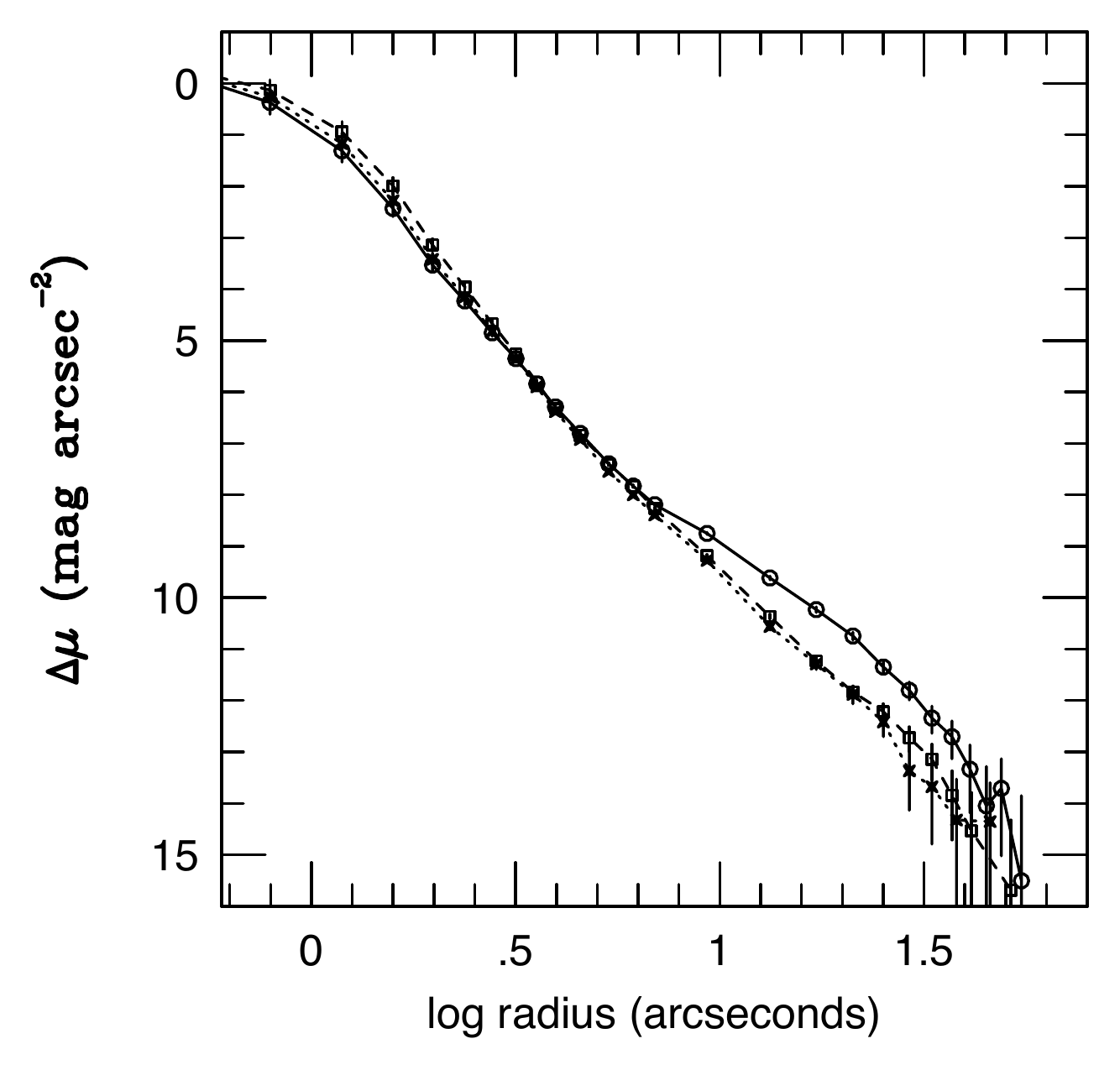}
   \caption{The PSF profile in $g$ (dashed), $r$ (dotted) and $i$ (solid)}
      \label{psf}
\end{figure}
%
%_____________________________________________________________

\subsection{The image stacking}

The technique we use for image stacking largely follows that used by Z04. The 
stacking process aims to increase the signal/noise (S/N) by median filtering a 
large number of rescaled and aligned images of assumingly conforming galaxies. The difference between our procedure and Z04, is that we use both averaging and median filtering while they use the mode, based on an empirical relation. If
there is a wide diversity in morphologies and luminosity gradients one may worry 
about what the result will tell us, at least if we use median filtering. 
Therefore it is important to make comparisons between the results from median 
filtering and averaging. We will show below that from this aspect, the result we 
obtain is quite robust. 

\begin{figure*}
\centering
\begin{minipage}[c]{0.3\linewidth}
   \centering{\includegraphics[width=5.25cm]{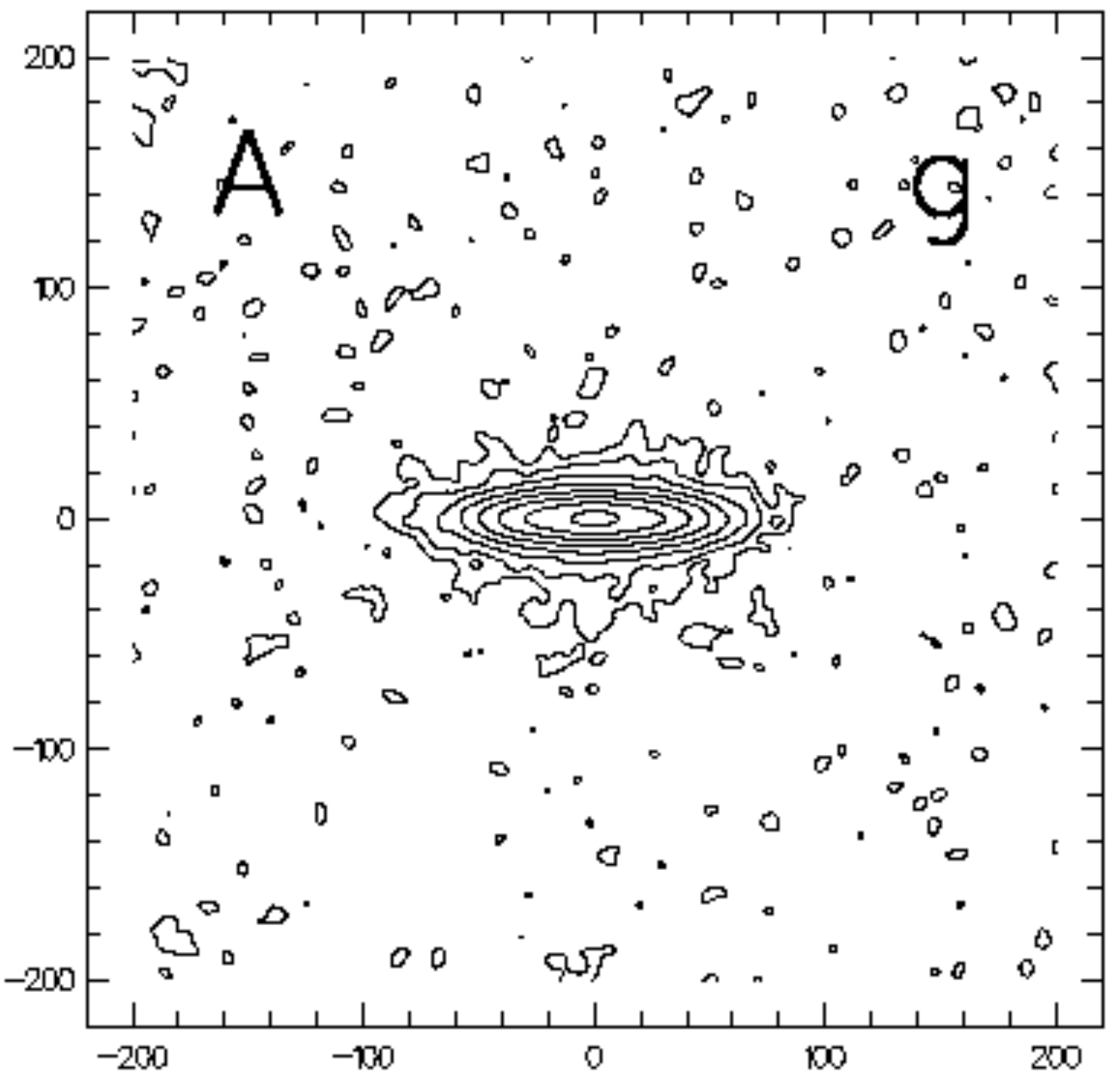}}
\end{minipage}%
\begin{minipage}[c]{0.3\linewidth}
   \centering{\includegraphics[width=5.25cm]{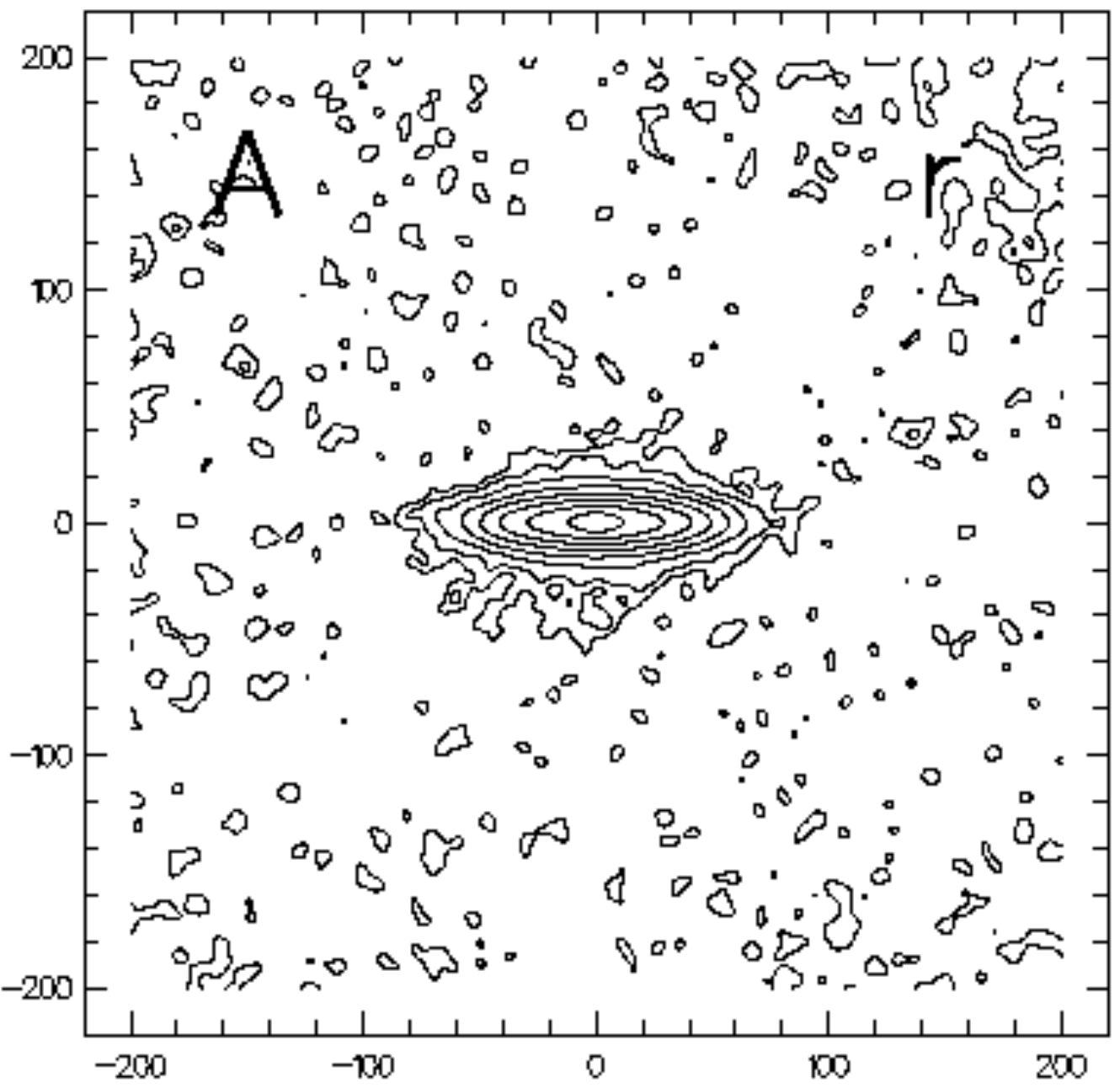}}
\end{minipage}%
\begin{minipage}[c]{0.3\linewidth}
   \centering{\includegraphics[width=5.25cm]{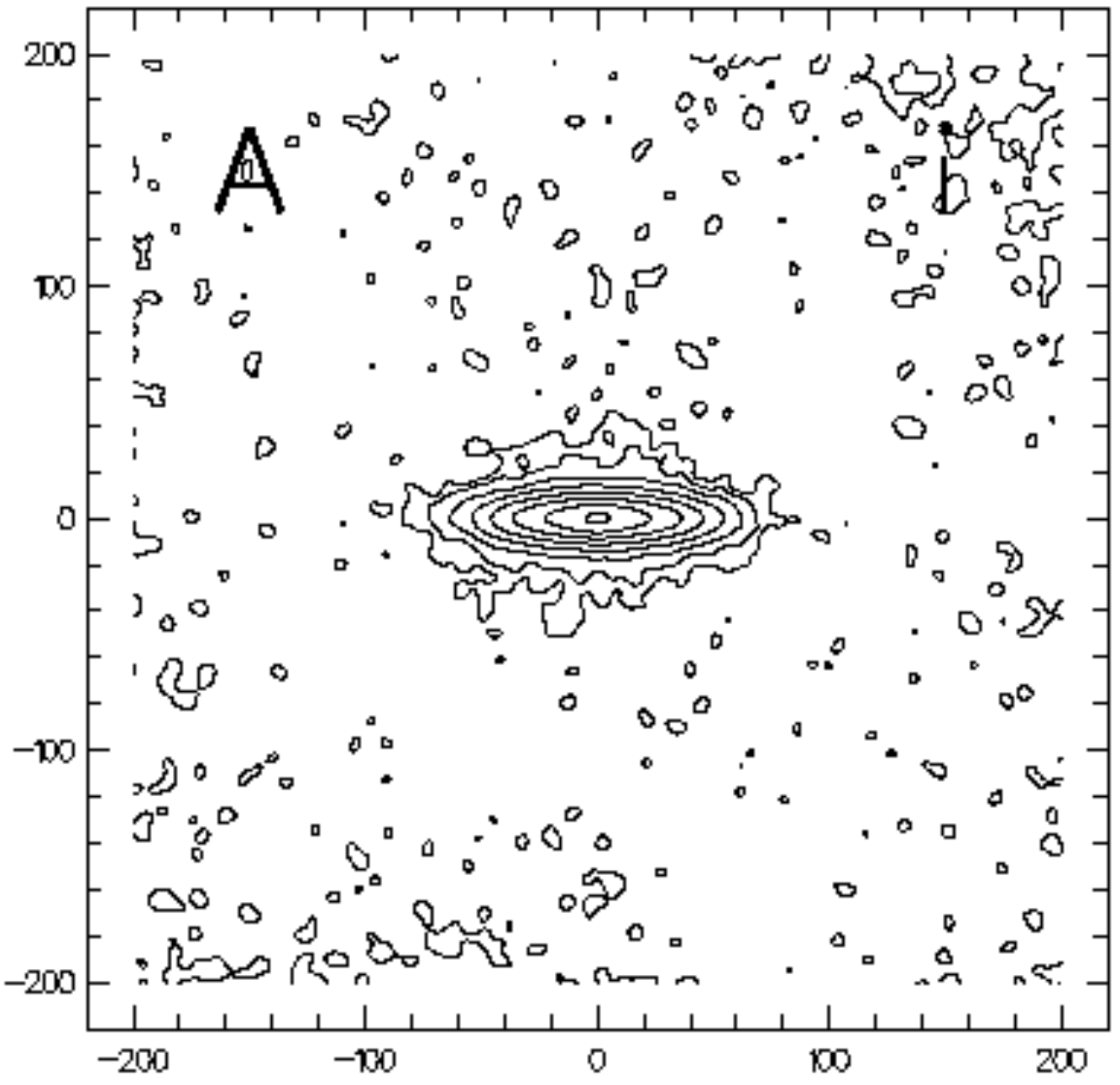}}
\end{minipage}   
\begin{minipage}[c]{0.3\linewidth}
   \centering{\includegraphics[width=5.25cm]{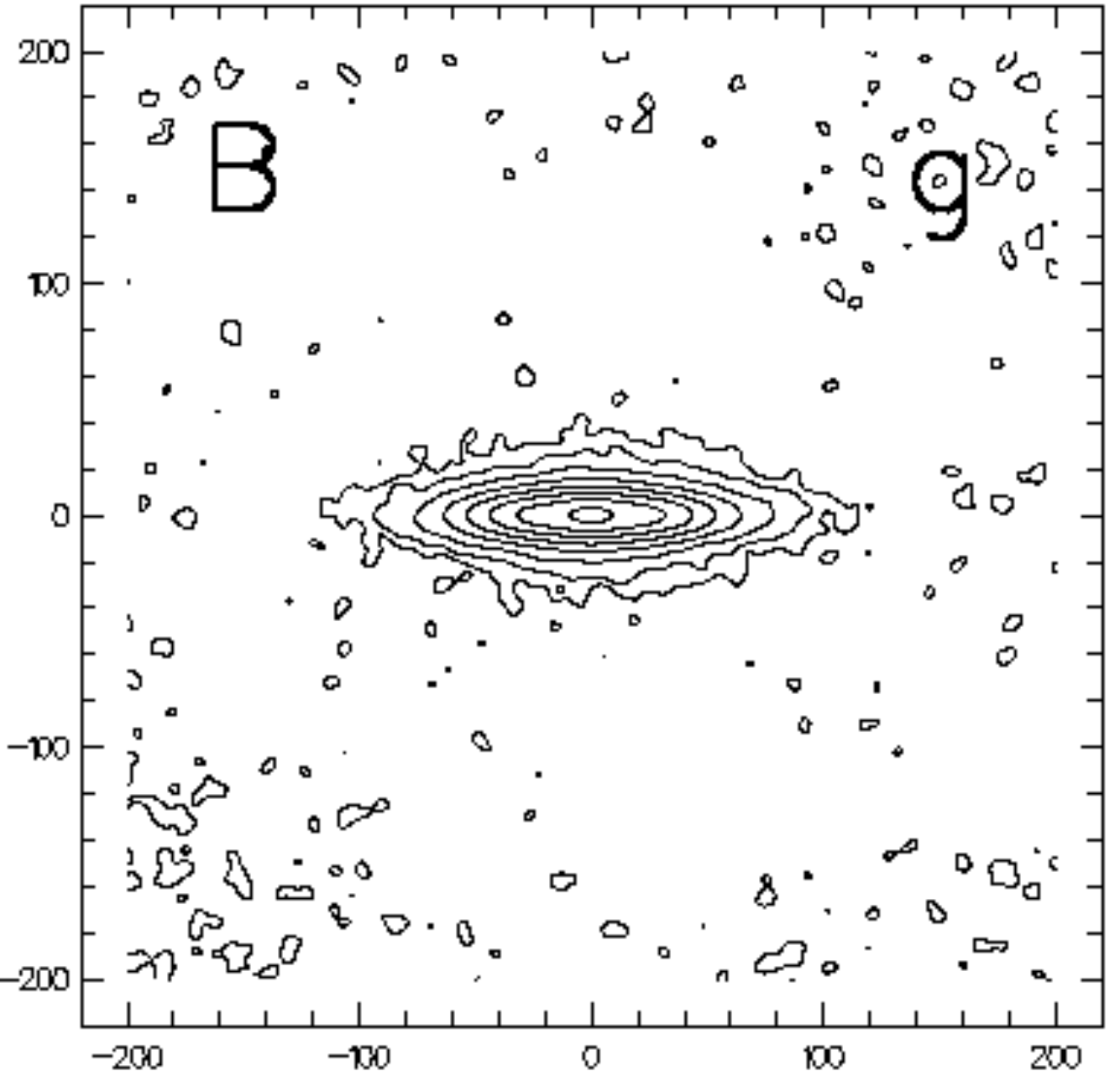}}
\end{minipage}%
\begin{minipage}[c]{0.3\linewidth}
   \centering{\includegraphics[width=5.25cm]{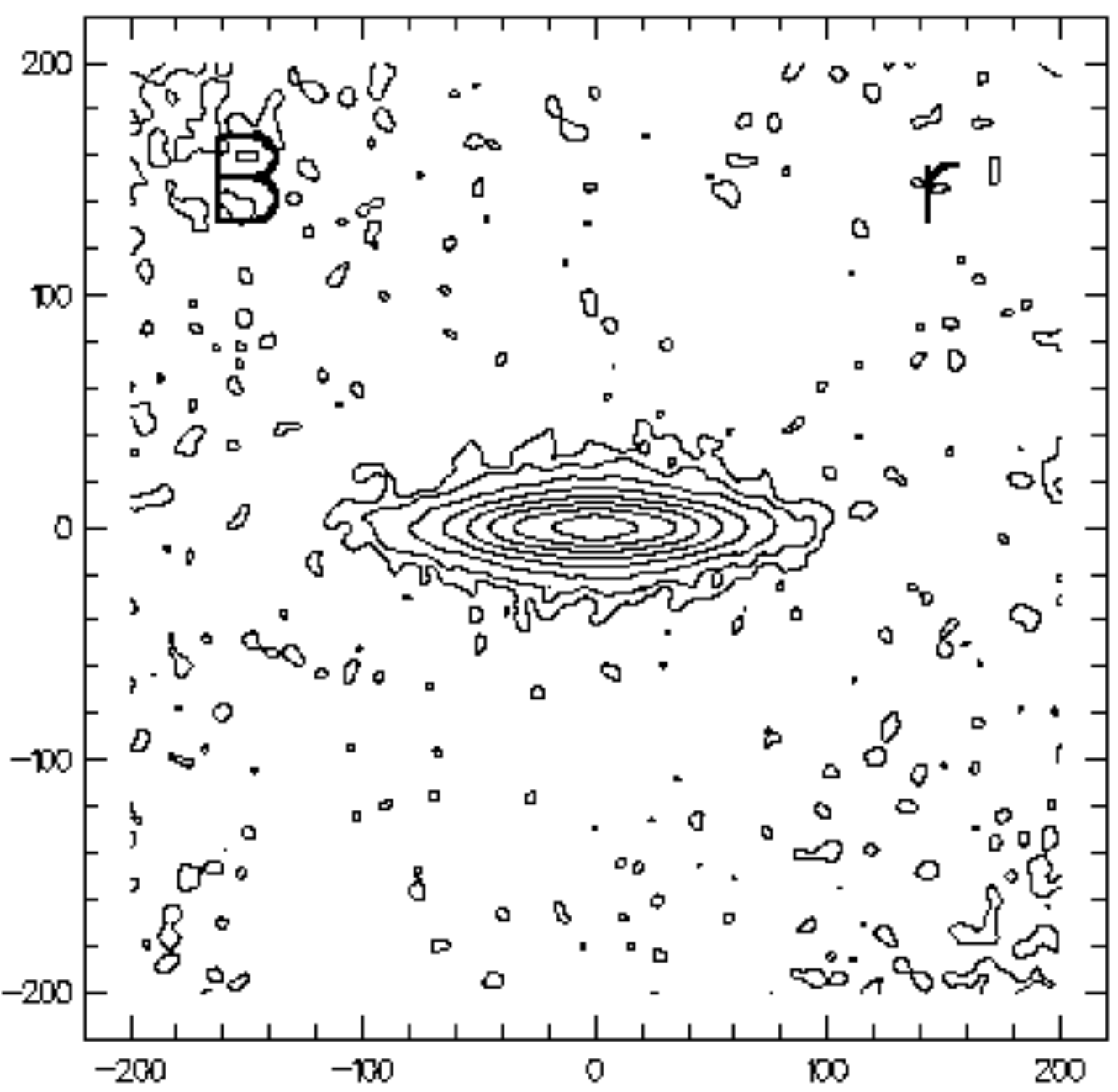}}
\end{minipage}%
\begin{minipage}[c]{0.3\linewidth}
   \centering{\includegraphics[width=5.25cm]{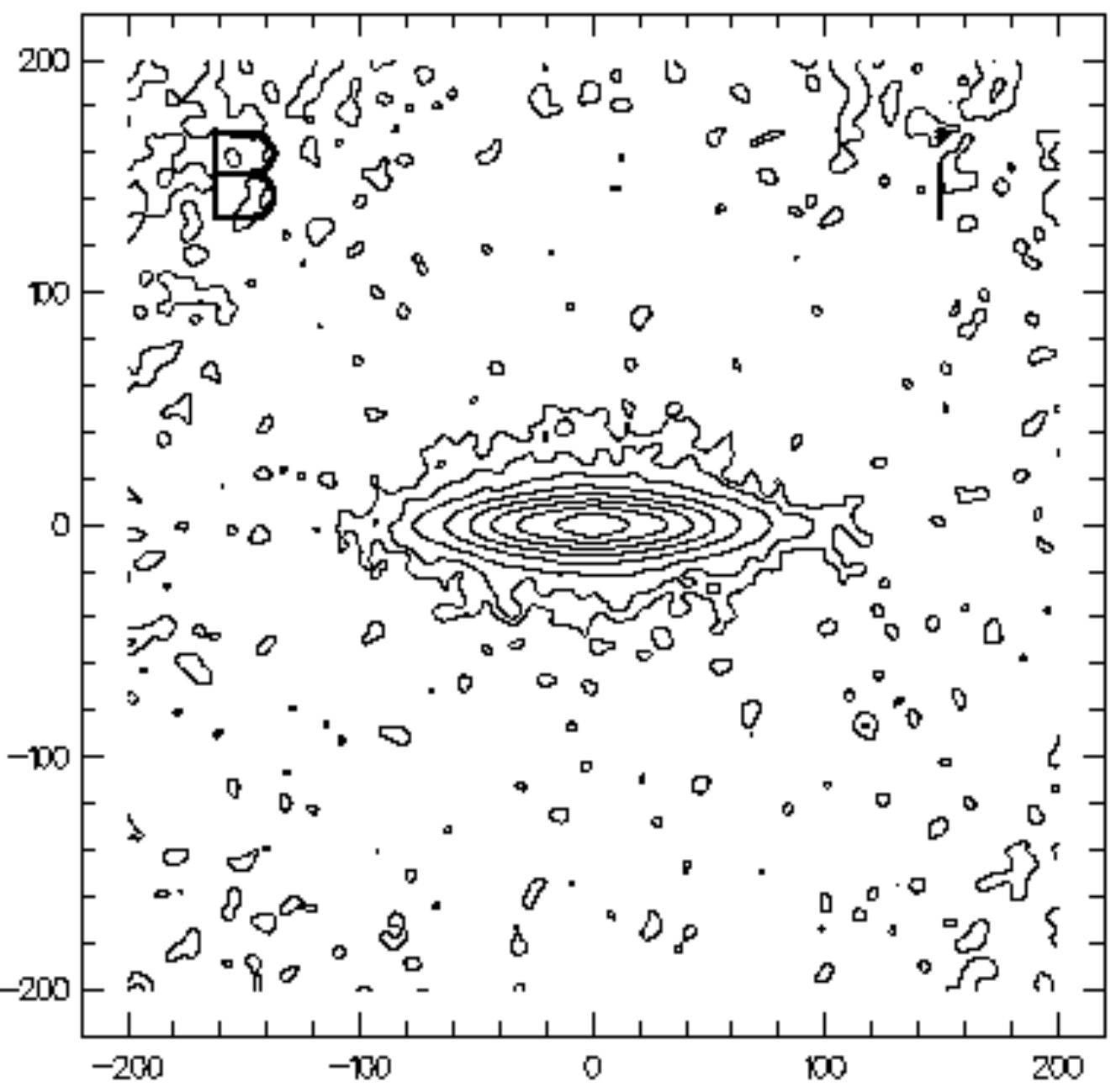}}
\end{minipage}
\begin{minipage}[c]{0.3\linewidth}
   \centering{\includegraphics[width=5.25cm]{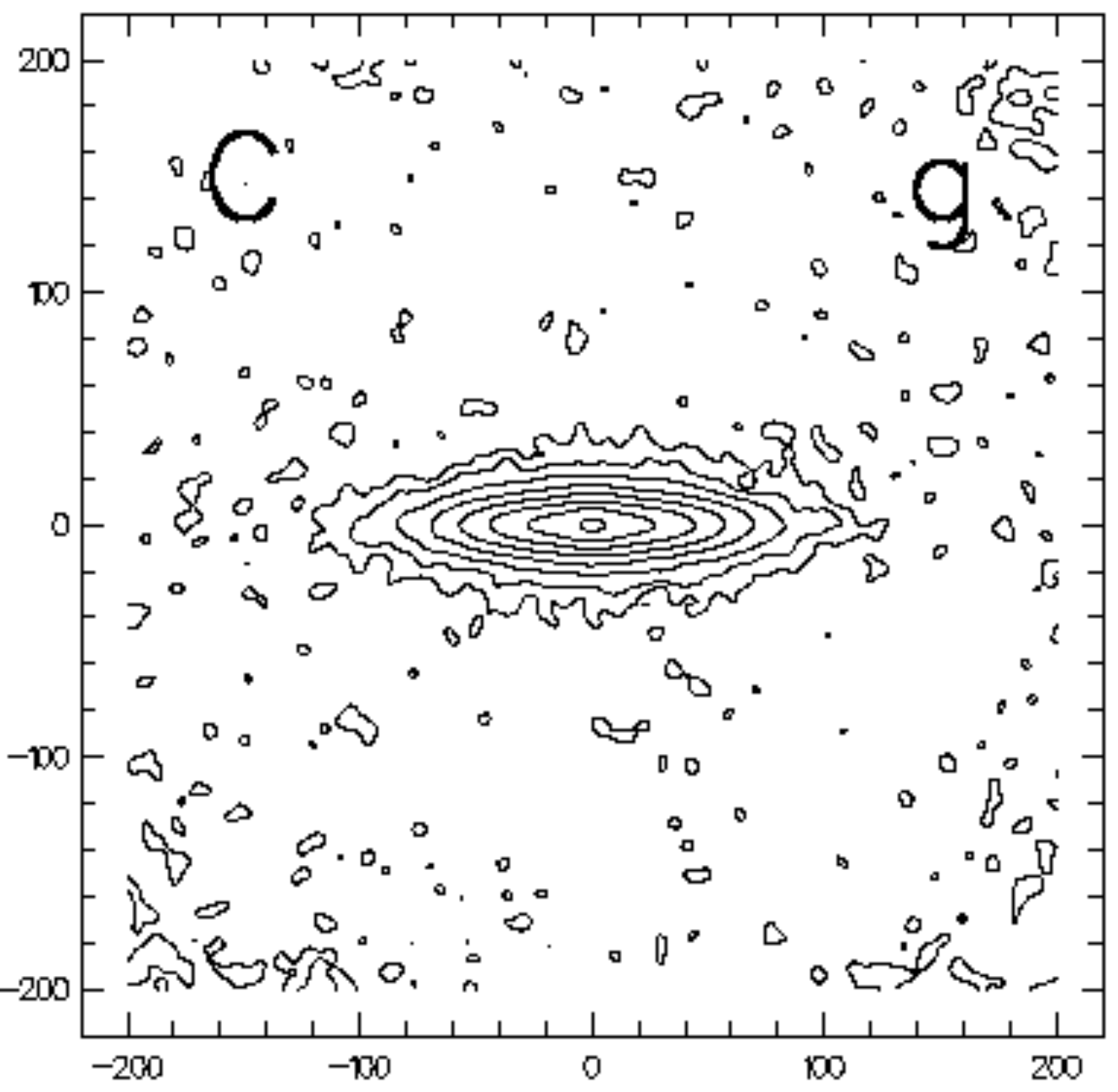}}
\end{minipage}%
\begin{minipage}[c]{0.3\linewidth}
   \centering{\includegraphics[width=5.25cm]{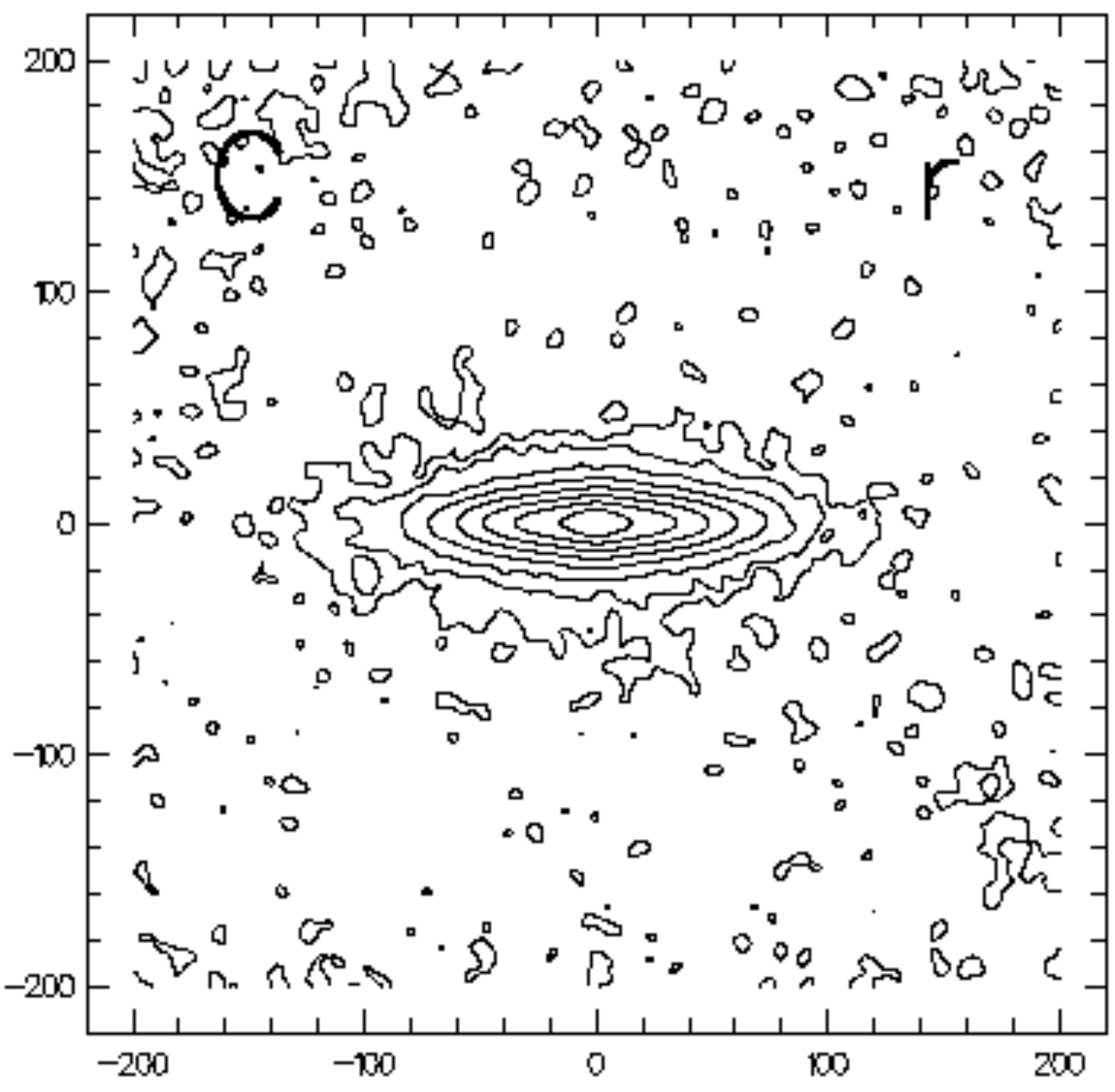}}
\end{minipage}%
\begin{minipage}[c]{0.3\linewidth}
   \centering{\includegraphics[width=5.25cm]{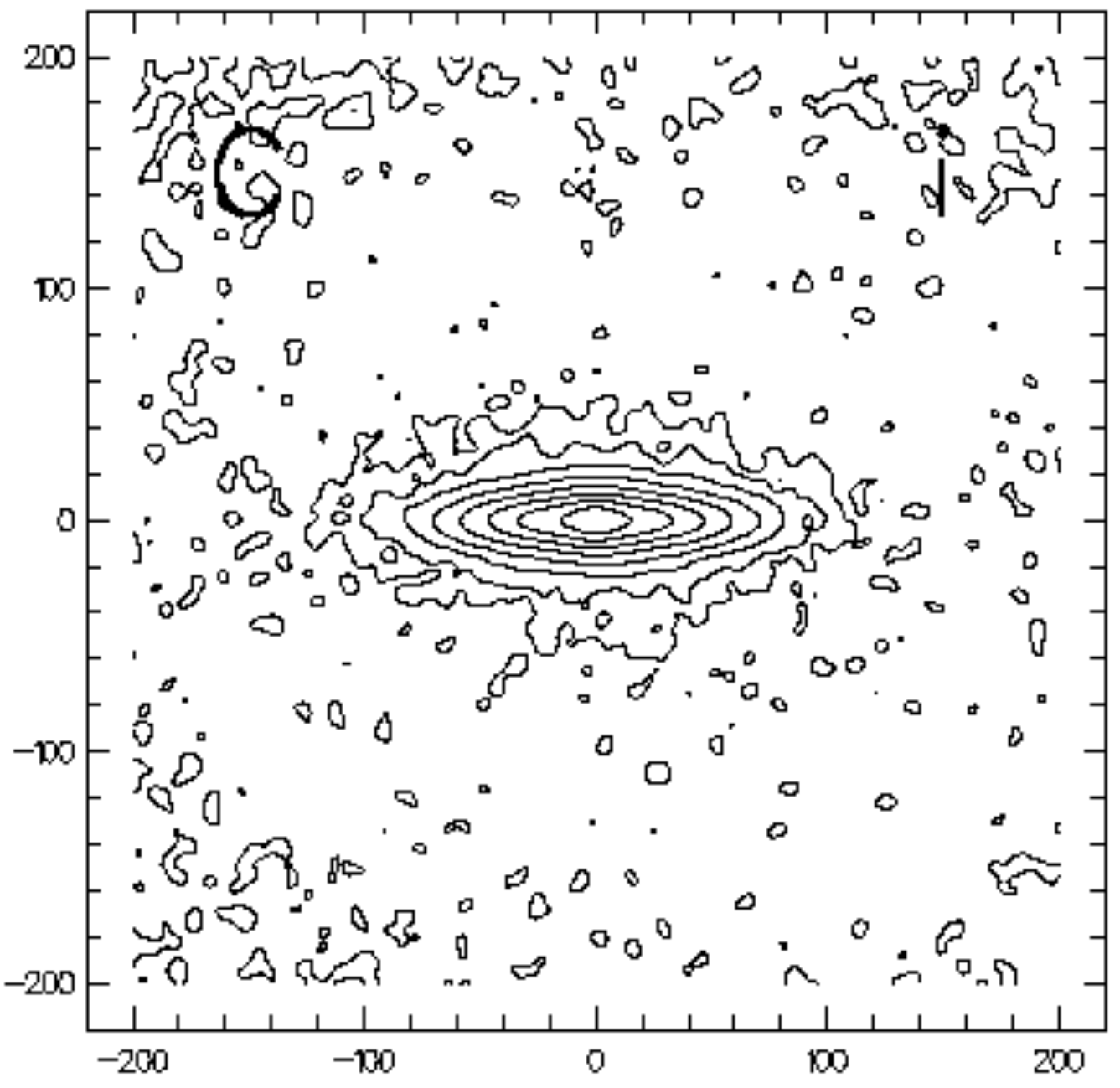}}
\end{minipage}   
\begin{minipage}[c]{0.3\linewidth}
   \centering{\includegraphics[width=5.25cm]{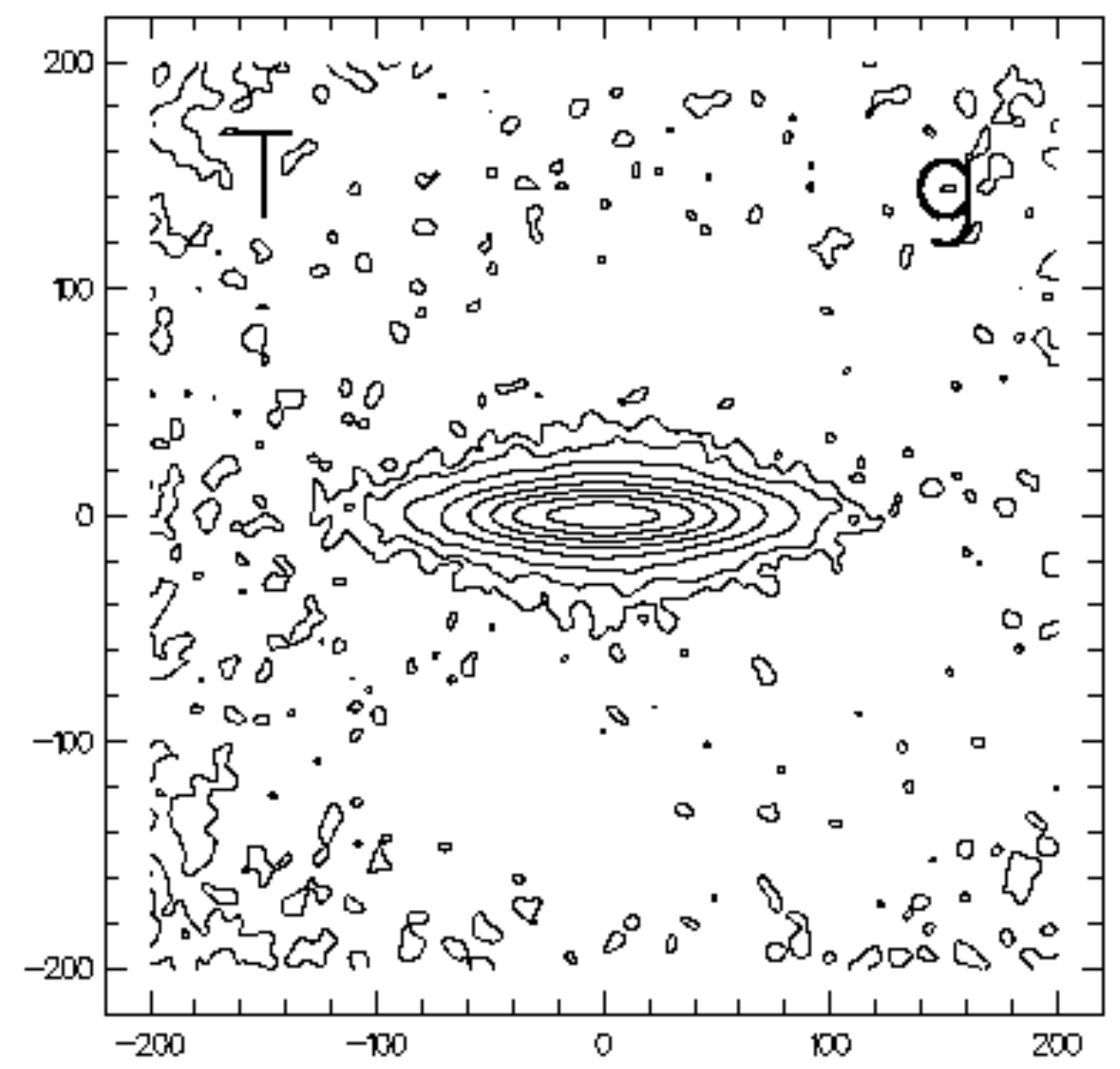}}
\end{minipage}%
\begin{minipage}[c]{0.3\linewidth}
   \centering{\includegraphics[width=5.25cm]{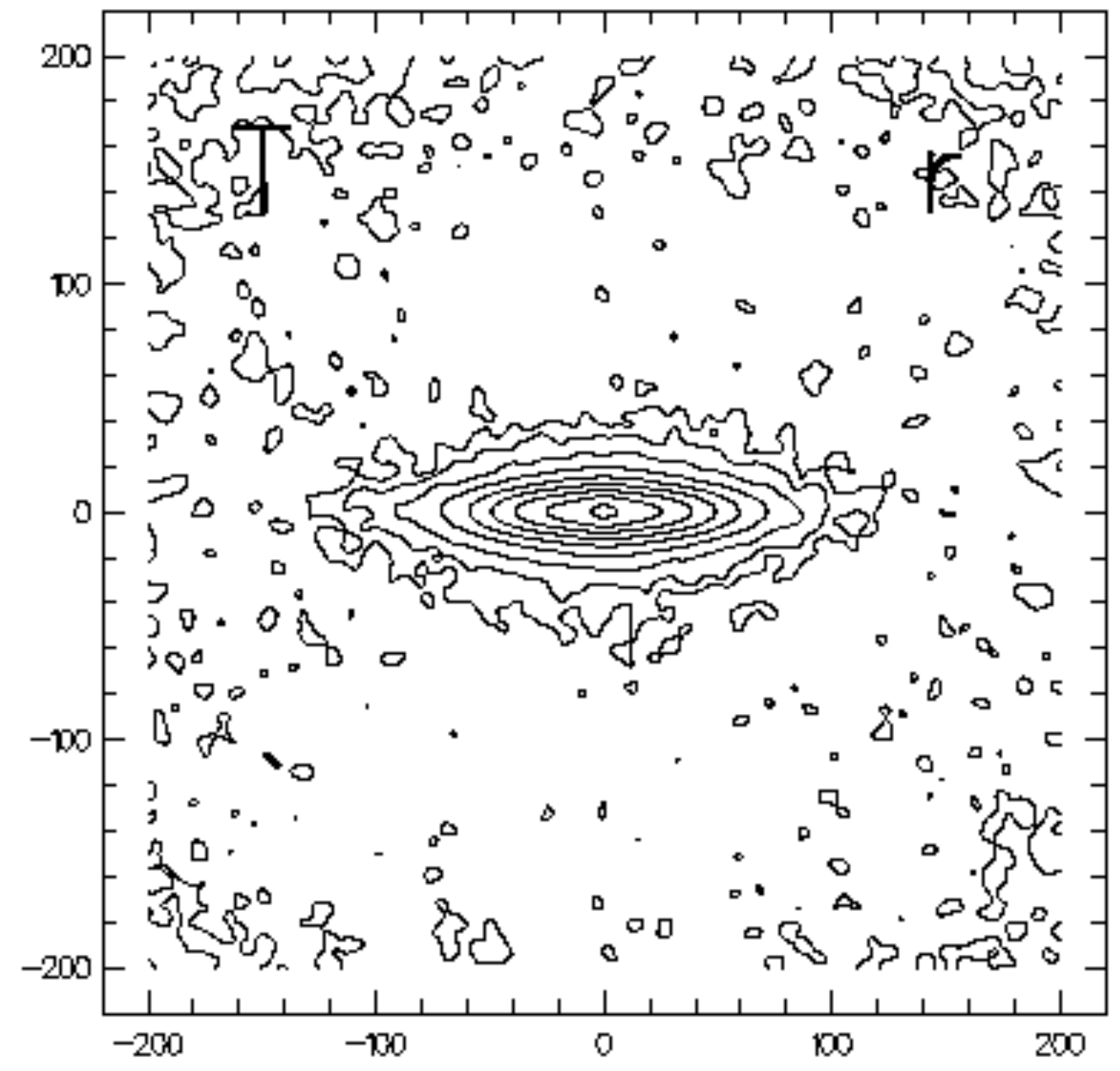}}
\end{minipage}%
\begin{minipage}[c]{0.3\linewidth}
   \centering{\includegraphics[width=5.25cm]{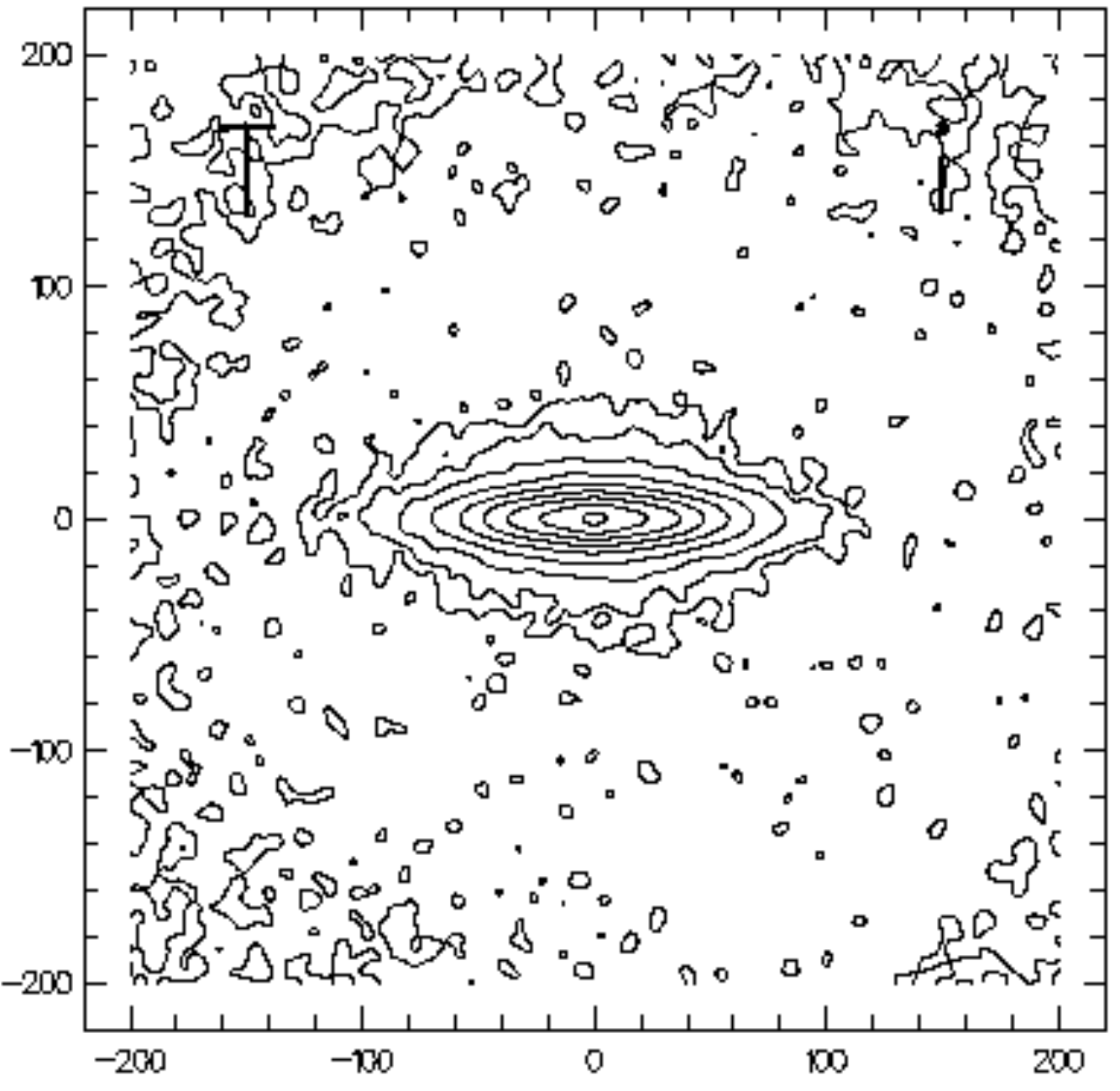}}
\end{minipage}

\caption{Isophotal contour plots of the stacked images of the subsamples (A, 
B and C) and the total sample (T). The images have been smoothed with a gaussian filter with $\sigma$=7 pixels. The contour levels are plotted in 
steps of 1 magnitude. The faintest level in the $g,r$ and $i$ images of each sample 
is A) 30.1, 29.8, 30.1 B) 30.0, 29.7, 29.2 C) 29.5, 29.2, 28.7 T) 30.6, 30.1, 
29.7 mag arcsec$^{-2}$ respectively.}
      \label{contours}
\end{figure*}

The image stacking technique consists of three basic steps: (1) masking out all 
unwanted surrounding objects (2) aligning, rotating and rescaling the LSBG 
images to make them superposable (3) combining the images to produce an average 
and median. All the above steps were combined in an in-house developed programme 
written in MIDAS. This programme processes the SDSS images 
through the use of the MIDAS and SExtractor software packages and local 
FORTRAN routines. The median and average image of the resulting individual 
masked, centered and rescaled
images were then computed using MIDAS routines. The S/N within the areas we study of each individual galaxy is proportional to the square root of the number of photons received, i.e. the surface area. In the stacking procedure, the images were therefore given weights 
proportional to the original scalesize squared. This has the consequence that the integrated weight given galaxies at a certain distance is independent of distance. Thus, nearby galaxies are strongly overrepresented in the stacked image. The ratio between the largest and smallest galaxies in the subsamples is typically $\sim$ 10. However, no single large galaxy dominates any of the three subsamples. After the stacking, a flat surface fit to the 
background sky in 15 boxes around the galaxy was made and subtracted. A final background correction was carried out in connection to the integration of the luminosity profiles (see below).

\subsection{The error in the chosen sky level}

As a check to find out whether the sky noise is purely Poissonian or structured on some specific scale we carried out a simple test. We compared the pixel-to-pixel standard deviation $\sigma_{sky}$ calculated from data obtained in boxes of various sizes, up to approximately the scale of the image, in our case avoiding the corners that contain more stellar residuals than the rest of the image. In a completely structureless (white noise) sky, $\sigma_{sky}$ should be independent of the size of the box. We found this to be close to fulfilled so apparently the median filtering and the simple sky subtraction was sufficient to remove measurable correlated structures outside the image of the galaxies. The error in the background zeropoint level (after subtracting the sky values from the data) was therefore set equal to the mean error of the mean of the median values of the pixels in the 15 boxes used for the sky subtraction.

%_____________________________________________________________

\subsection{The Point Spread Function}

A serious problem when dealing with weak signals in images is the influence of 
scattered light from the target and objects in the environment. The importance 
of this potential problem can be estimated if the shape of the PSF is known. This problem was discussed at length in the investigation by Z04. To tackle the problem, the authors modelled the PSF in the different 
filters by a combination of a gaussian core plus an exponential wing. They 
concluded that the influence of the PSF on the data in the red 
halo region was insignificant.

Here we have chosen a more direct approach. To obtain the PSF, we stacked bright 
stars, one from each galaxy field after cleaning and rejecting bad images in the 
same manner as before. The result in shown in Fig. \ref{psf}. Obviously there is 
a clear difference in the PSF between the $i$ window and the $g$ and $r$ 
windows. The PSFs of SDSS images have independently been measured by 
\citet{2005ApJ...622..244W} and \citet{2008MNRAS.388.1521D}.  Our results agree very well with theirs down to the levels that might be 
relevant in this investigation, i.e. about 15 magnitudes
below peak surface brightness. The PSF profiles have been derived with the same 
procedure as the galaxy profiles, and the tight agreement between our results and 
those of Wu et al. and de Jong supports the robustness of our method. The shape of the PSF is markedly different from that derived by Z04. The one we derive is significantly more extended and clearly differs between $i$ and the other filters at distance from the centre that in the mean is close to the distance at which Z04 find the red colour excess. This potential problem is discussed in detail in the paper by de Jong.

\subsection{Correcting for the PSF}

\subsubsection{Deconvolution}

We used two independent methods to correct for the degrading of the images due to light scattering, as quantified by the PSF. The first one was simply to deconvolve each individual image before stacking. We carried out a deconvolution of the full sample using the MIDAS software. The deconvolution was done before the rescaling and stacking of each individual image using the task DECONVOLVE in MIDAS, based on the iterative algorithm by L.B. \citet{1974AJ.....79..745L}. Before the deconvolution, all images were rebinned to twice the original stepsize in order to avoid long processing times. The size of the PSF image after rebinning was 95x95 pixels. The deconvolution increases the high frequency noise considerably due to the influence of the background objects and may cause unwanted ringing effects in the data. The problems are most severe in the deconvolution of galaxies of small apparent sizes.  When we inspected the results from the deconvolution we found structures at low fluxes that appeared artificial. We will therefore not discuss these results other than when it concerns data with weak radial gradients at  flux levels safely above the background noise.

%The result is shown in Fig. \ref{deconvolve}. As seen in the Figure, the PSF has a strong influence on the colour profile in $r-i$. Still, as we will see, there is still a red colour excess in the corrected data.

\begin{table*}
\centering
\caption{Model galaxy photometric parameters.}
\begin{tabular}{lllllllllll}
\hline
Filter & \multicolumn{3}{c}{Inner disk} & \multicolumn{3}{c}{Outer disk}& Truncation radius & \multicolumn{3}{c}{Flattened spheroidal}   \\
& $\mu_0$ & $h$ & $b/a$ & $\mu_0$  & $h$ & $b/a$ &  $r_{trunc}$ & $\mu_0$ & $r_{e}$ & $c/a$ \\
& mag arcsec$^{-2}$ & pixels & &  mag arcsec$^{-2}$ & pixels & & pixels & mag arcsec$^{-2}$ & pixels & \\
\hline
g &  23.81 & 41 & 0.12 & 20.55 & 18 & 0.12 & 95 & 21.92 & 132 & 0.34 \\
r & 23.42 & 48 & 0.16 & 19.80 & 17 & 0.16 & 86 & 21.13 & 146 & 0.34  \\
i & 23.37 & 51 & 0.16 & 22.06 & 17 & 0.16 & 86 &  20.39 & 100 & 0.50 \\
\hline
\end{tabular}
\\
\begin{flushleft}
$\mu_0$: Central surface brightness\\
h: Exponential scalelength\\
b/a: Observed (projected) minor/major isophotal axes ratio\\
r$_e$: Effective radius
\end{flushleft}
\end{table*}

\subsubsection{Using a model of the luminosity distribution as a template}

The other method we used to correct for the PSF is based on a model of a typical sample galaxy. A similar procedure was recommended (although not practised) by de Jong as a proper way to account for the PSF problems in the Z04 sample. Our model was constructed in the following way. The 35 apparently largest galaxies were selected, deconvolved, scaled to the same apparent scale sizes and stacked.  The deconvolution seemed to work well for these galaxies, having large apparent sizes. The resulting stacked image was used as a template to a model that was derived iteratively from a mixture of disks and flattened spheroidals with different flattenings, radial surface brightness dependencies and relative peak fluxes. A good fit to the template galaxy was obtained from a simple mixture of a spheroidal with a flattening of b/a=1/3-1/2 and an r$^{1/4}$ surface brightness dependence superposed on an exponential disk. The exponential disk had two sections with different slopes, i.e. a "truncated disk"-like morphology. No bulge component was necessary to obtain a satisfactory agreement between the model and the template. The structural parameters of the model were adjusted to optimize the fit to the template. In the end the deviations in surface flux in regions of the size of the seeing disk were less than about 15\%. The parameters of the final model are shown in Table~2. For this subsample one pixel corresponds to $\sim$ 0.32 arcsecs. This model was scaled to the same size as each galaxy in the sample and then treated equally in the stacking procedure. In a parallel sequence we convolved the model with the PSF after scaling and then stacked the images. In the end we thus obtained one image produced by the pure template images and the other by the PSF downgraded images. The ratio between these images could then directly be used to correct the images of the sample galaxies. The procedure seems to be quite independent of the model, except for the central region where the luminosity gradients are large. As an additional test we simply used the stacked image as a model, i.e. based on already degraded images. The result of the halo colours was very similar to the more appropriate method described above. The PSF
corrections mainly affect the fainter parts of the $r-i$ colour profile, making it 
bluer after the correction (see also Sect. 3.3). In the discussion below we will mainly refer to data corrected for the PSF using this method.

\section{Results}
\label{sec:derived}

\subsection{Luminosity distribution}

The stacked images are shown as contour plots in Fig. \ref{contours}. A slight 
increase in the noise and the mean level of the sky is seen towards the edges. 
This is due to influence of objects outside the frame limits where the flagging 
procedure was not completely successful. The typical enhancement of the sky 
region due to this is $\sim$10\% of the signal at the faintest reliable part of 
the luminosity profile. As is seen in the plots, the background surface in the 
region close to the galaxy is essentially flat. The zero-point
error of the sky background estimated from the statistics of the boxes used 
for the sky subtraction is typically about 1\% of the surface brightness in 
the faintest regions where we consider the photometry to be reliable. 

We find no significant difference in the morphologies of the three subsamples from the contour maps. Small differences between the samples can be found if we inspect the result we obtain by dividing the images in one sample with the corresponding images in another sample. We then find that the C sample has a more prominent contribution from the central region compared to samples A and B.  

\begin{figure*}
\centering
   \begin{minipage}[c]{0.5\linewidth}
   \centering{\includegraphics[width=8.5cm]{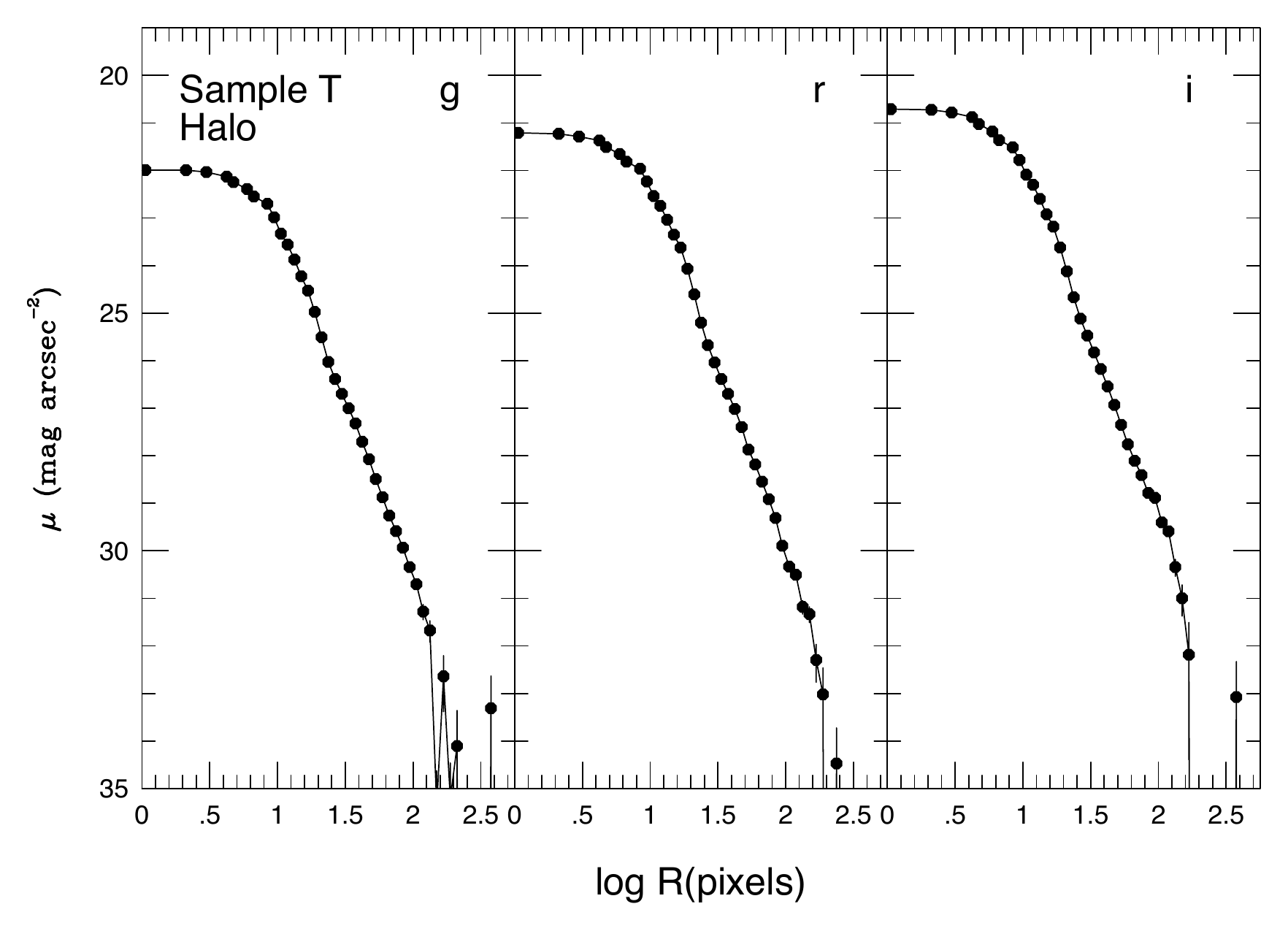}}
   \end{minipage}%
   \begin{minipage}[c]{0.5\linewidth}
   \centering{\includegraphics[width=8.5cm]{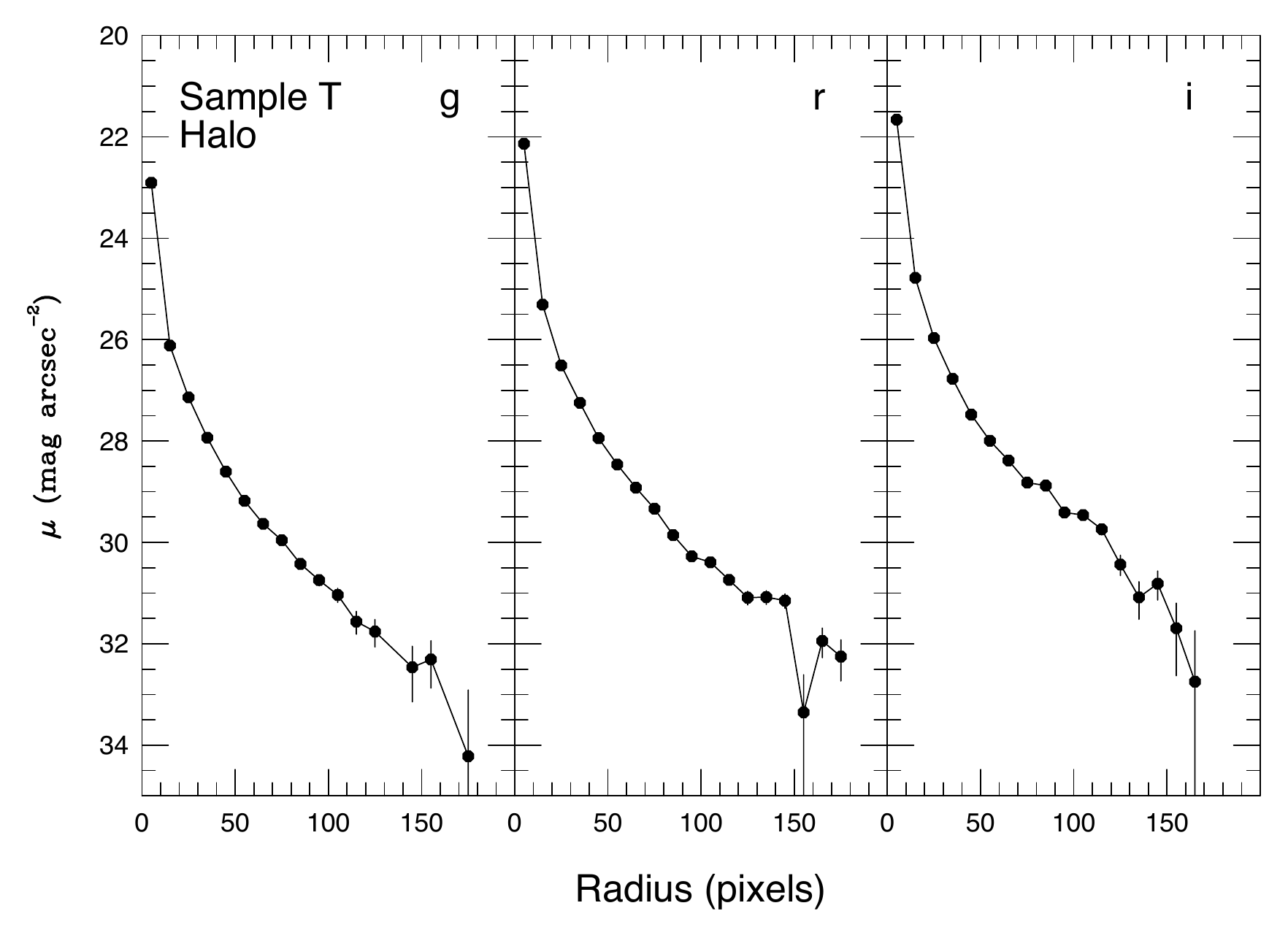}}
   \end{minipage}
    \caption{Luminosity profiles along wedges of 60$\degr$ opening angles 
in the polar direction of the stacked images of the 
total sample (T). The bars indicate mean errors as described in section 3.2. Profiles are presented in logarithmic and linear radial scale.}
      \label{lp_polar}
\end{figure*}

\subsection{Luminosity profiles}

We will now discuss the luminosity and colour profiles of the stacked images as derived from three regions of the images. We will call them {\it halo, thick disk and thin disk}. The halo is the faintest and most extended region of the stacked images. It is so faint that its structure is difficult to determine. Its presence is mainly revealed in the 1-D luminosity profiles. We will assume that it is spherical. The thick disk is defined by the faintest regular isophotes of the stacked images. It is close to ellipsoidal in shape but must be significantly flatter than the halo. There may be a smooth transition between the thick disk and the halo. The thin disk is the dominating morphological structure, defined by the faint isophotes of the individual images. It is clearly separated from the thick disk/halo in the $r-i$ colour profiles. More details are given below.

The halo plots were obtained from data in 
two opposite wedges along the polar axis of opening angle of 60$\degr$. The halo profile is measured in circular arc strips. These measurements correspond to those of  Z04. The equatorial data along the thin disk were measured in elliptical strips in horisontally oriented wedges with an opening angle of 30$\degr$. The median inclination of a sample of infinitely thin disks with random orientations 
restricted to an observed minor/major axis ratio of $b/a <$0.25 (one of the selection criteria) is  $\sim$ 81$\degr$ or ${b/a}$=0.16. 
The measured axis ratio of the $i$ image of the full sample, based on the isophote encompassing the faintest regions of the thin disk, is $b/a$ = 0.24, indicating a finite thickness of the disk. These data can be used to calculate  the {\sl intrinsic} minor/major axes ratio of the disk, p, from $$ cos(i)^2 = \frac{(b/a)^2-p^2}{1-p^2}$$ where $i$ is the inclination \citep{1926ApJ....64..321H,1950MeLu2.128....1H}. We then obtain  {\sl p} $\sim$ 0.18 . This is slightly flatter than what was obtained by \citet{2006AJ....131..226Y} for high surface brightness disk galaxies. We have used the observed axis ratio of 0.24 when we integrate the luminosity profile of the disk. 

As we discussed in the previous section, we obtained a satisfactory fit to the template galaxy using a model galaxy based on two components only - a truncated thin disk and a flattened spheroidal. At least at these isophotal levels (only 35 galaxies were used for the template), no bulge or halo component was needed. As we will show below, when we analyse the total sample (1510 galaxies) this conclusion largely persists. One must keep in mind however that our model of the structural properties is quite simple. In the faint region the stacked images seem to fit well to a flattened spheroidal but we cannot exclude that an equally good fit would be obtained with a mixture of a thick exponential disk and a more extended spheroidal halo component. {\ For the sake of simplicity, we call this dominating flattened structure the thick disk.} It has a flattening in the PSF corrected image in the $i$ band of b/a = 0.39$\pm$0.01, based on the three subsamples. The corresponding intrinsic flattening is 0.36. Thus the ratio of the flattenings between the thick and thin disk is 2.05, slightly lower than the corresponding value for normal disks of 2.35 \citep{2006AJ....131..226Y}. LSB galaxies normally have flatter disks than normal galaxies which generally is interpreted as an evolutionary effect during the slow collapse of the disk \citep{2005MNRAS.358..503K}. The observed flattening of the thick disk is significantly smaller in the $g$ and $r$ bands, 0.26 and 0.29 respectively. This condition makes the $i$ band flux become more prominent relative to the other bands along the polar direction. We call this phenomenon the {\it "red excess"} in the discussion below. The luminosity/colour profiles of the thick disk (see below) are based on data collected in vertically oriented wedges with an opening angle of 120$\degr$. We integrated in elliptical strips corresponding to a flattening of 0.39.

\begin{figure}
\centering
   \includegraphics[width=8.5cm]{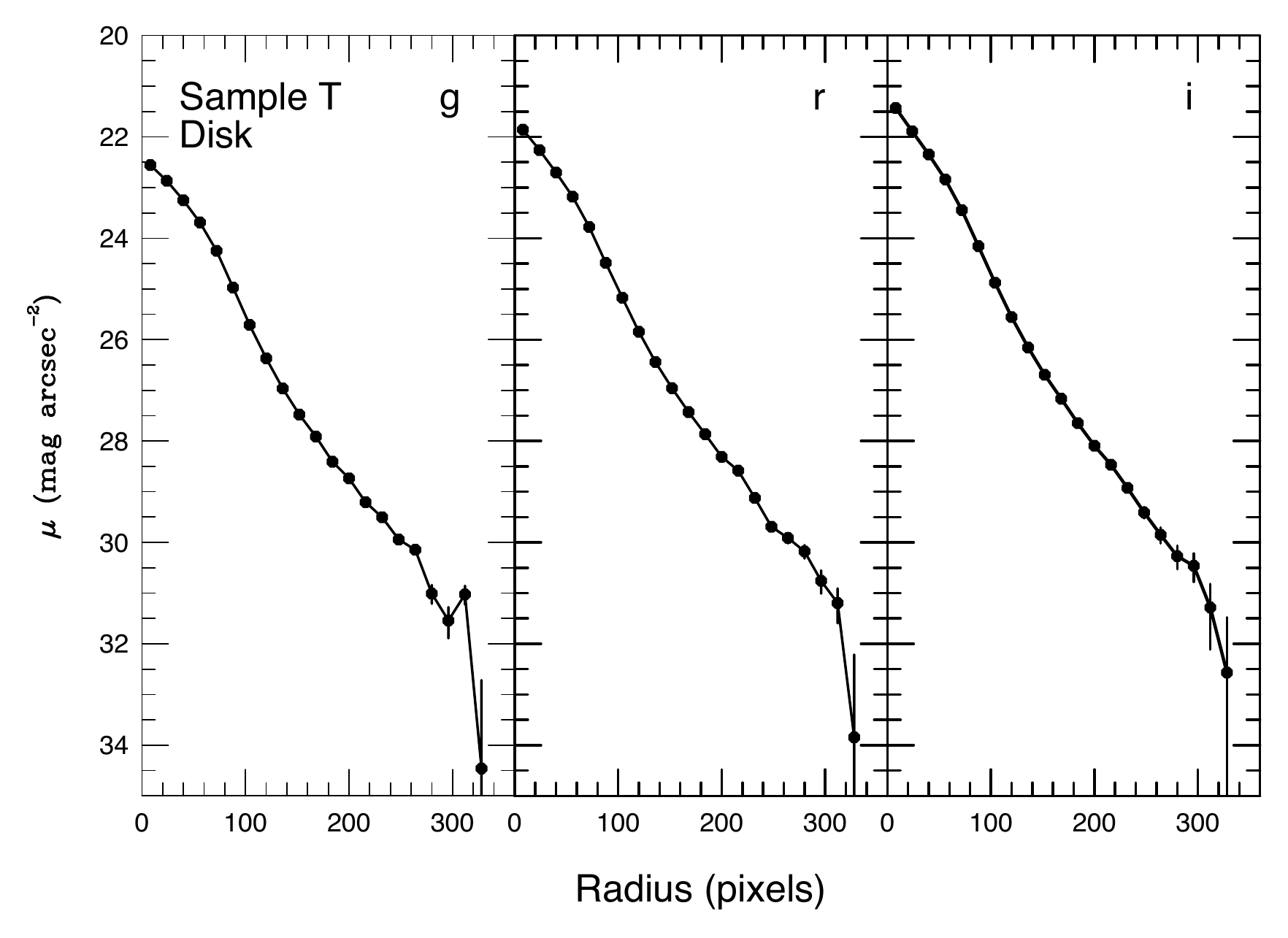}
\caption{Luminosity profiles along the major axis of the thin disk in sample T. The opening angle is  30$\degr$. A disk flattening of b/a=0.24 is assumed. The bars indicate mean errors.The physical scale in this and the following diagrams based on the full sample is approximately 500 pc pixel$^{-1}$. }
      \label{lp_disk}
\end{figure}

The luminosity profile of the total sample  along the polar direction is shown in log-log and log-lin form in Fig. \ref{lp_polar}. The corresponding equatorial profiles along the disk 
are shown in log-lin form in Fig. \ref{lp_disk}. Errors are based on statistics of the sky 
counts in the strip used to derive the luminosity profile at that specific radius. Thus the mean error in each bin was calculated from the mean error in the mean of the pixel values to which was added the estimated mean error in the sky level in quadrature.
All data in the diagrams have been corrected for the effects of the PSF. 

The disk profile in $i$ follows an exponential distribution in surface brightness. Subtracting this component would leave a structure resembling an extended bulge component. However, as we show in our model, it can be explained by the dominance of the thick disk in the central region. In the polar direction the curves display a behavior typical of a Sersic profile with n$>$1 except in the outermost parts where they tend to be more exponential-like.

\subsection{Colour profiles}

We will now discuss to what extent we can find any support for a red colour excess at the so faint surface brightness levels. We will start by discussing the thin disk colours and then go to the thick disk and halo data.

\subsubsection{Thin disk}

\begin{figure}
\centering
   \includegraphics[width=6cm]{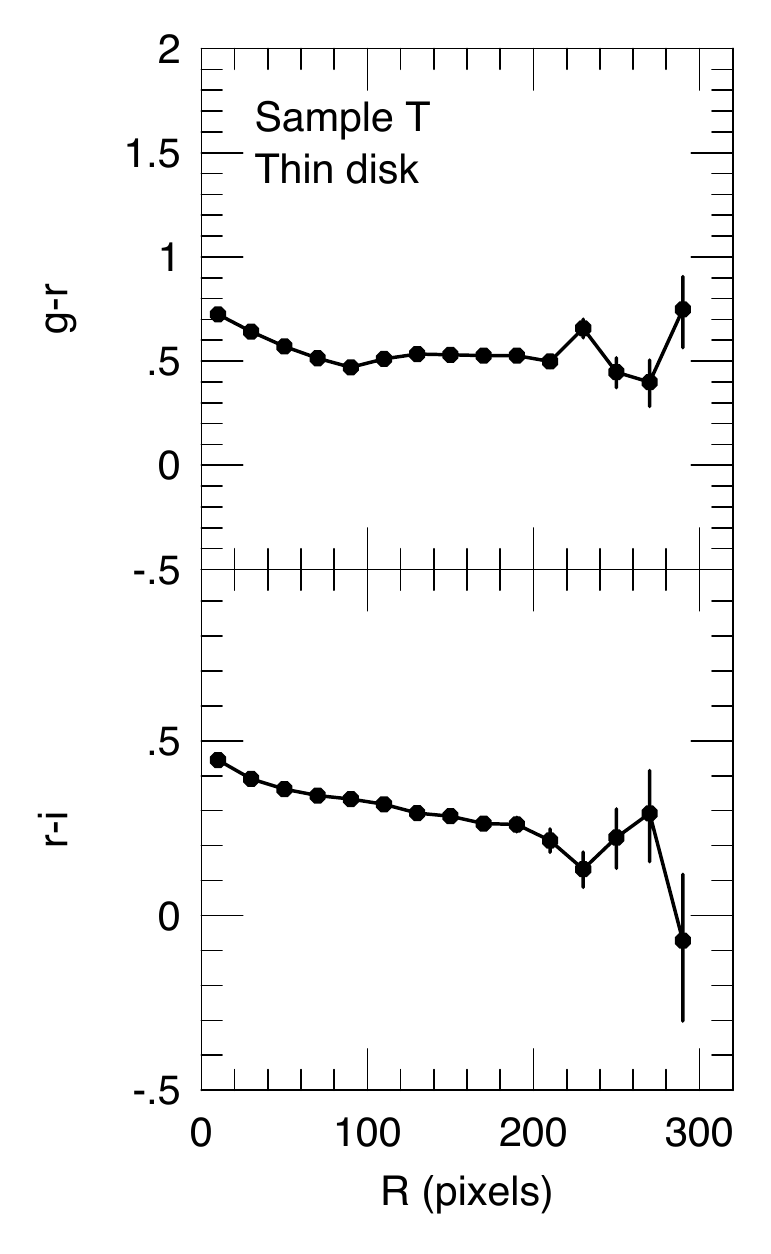}
\caption{Colour profiles along the major axis of the thin disk of sample T. The opening angle is 30$\degr$.  A flattening of b/a=0.24 is assumed. The bars indicate mean errors. }
      \label{col_thin}
\end{figure}

Fig. \ref{col_thin} shows the equatorial colour profiles of the thin disk of the full sample. After correction for a small amount of dust reddening, the central colours are in agreement with a middle-aged/old stellar population. The corresponding age and metallicity depend on the amount of reddening assumed. The colours then drop smoothly to bluer colours and end up at $g-r\sim$ 0.5, $r-i\sim$ 0.2, corresponding to a normal stellar population with a standard IMF (cf. Fig. \ref{Salpetercol}). An age-metallicity degeneracy prohibits us from being more specific about the age, but most likely the metallicity is low in the outer disk which would mean a high mean age of the population. We see no significant trends in the colours becoming redder at low surface brightness levels.

\subsubsection{Thick disk and halo}

\begin{figure*}
\centering
\begin{minipage}[c]{0.5\linewidth}
   \centering{\includegraphics[width=6cm]{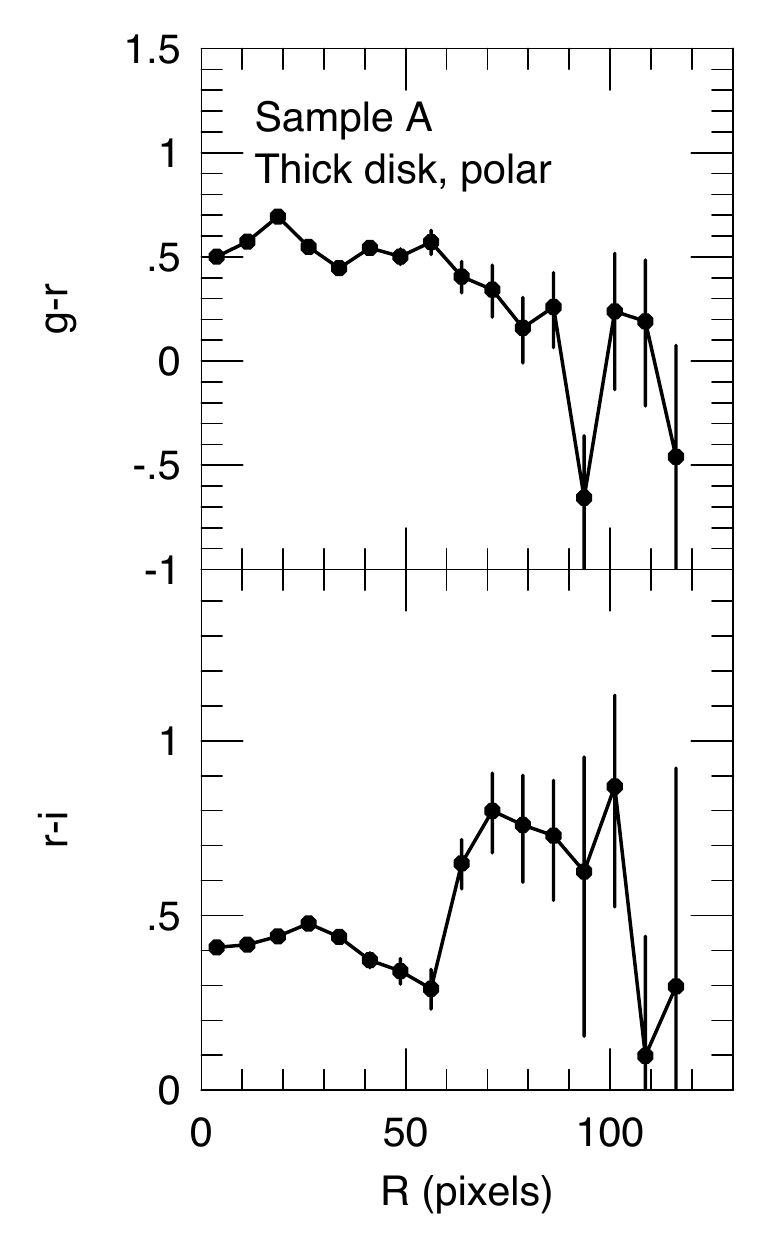}}
\end{minipage}%
\begin{minipage}[c]{0.5\linewidth}
   \centering{\includegraphics[width=6cm]{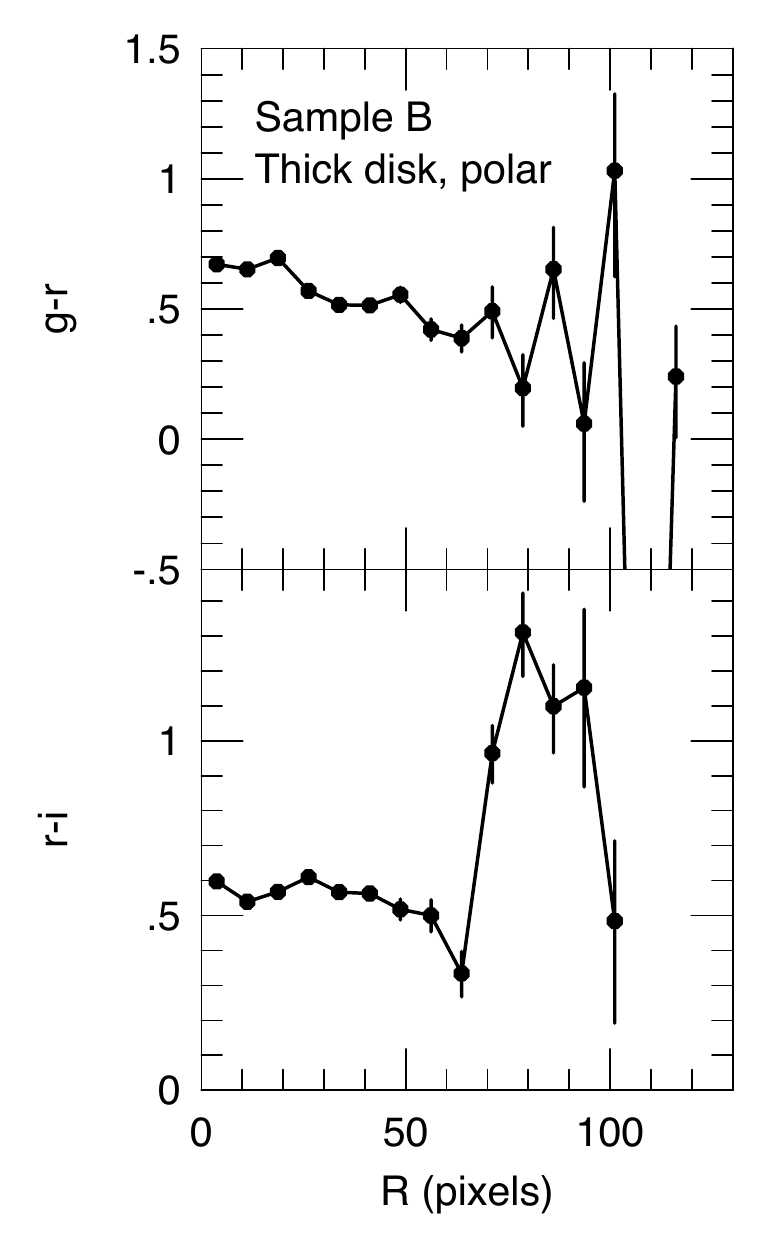}}
\end{minipage}\\[20pt]
\begin{minipage}[c]{0.5\linewidth}
   \centering{\includegraphics[width=6cm]{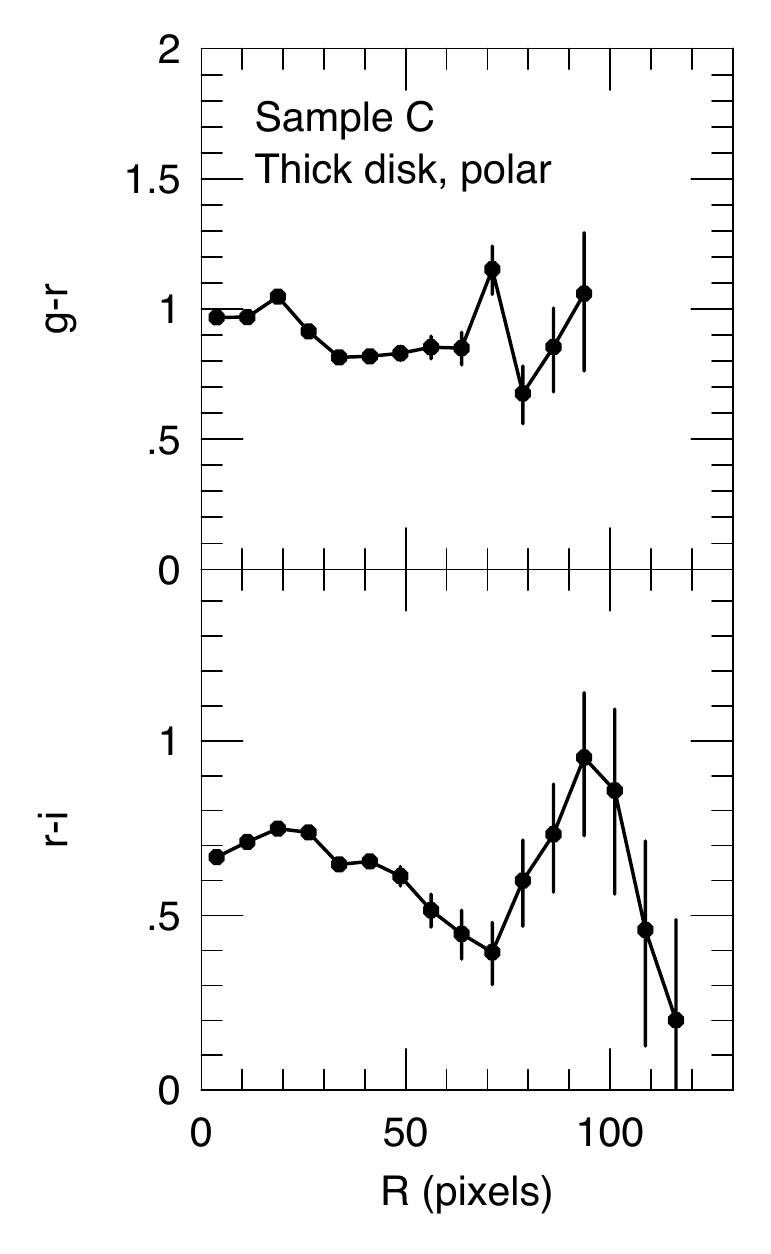}}
\end{minipage}%  
\begin{minipage}[c]{0.5\linewidth}
   \centering{\includegraphics[width=6cm]{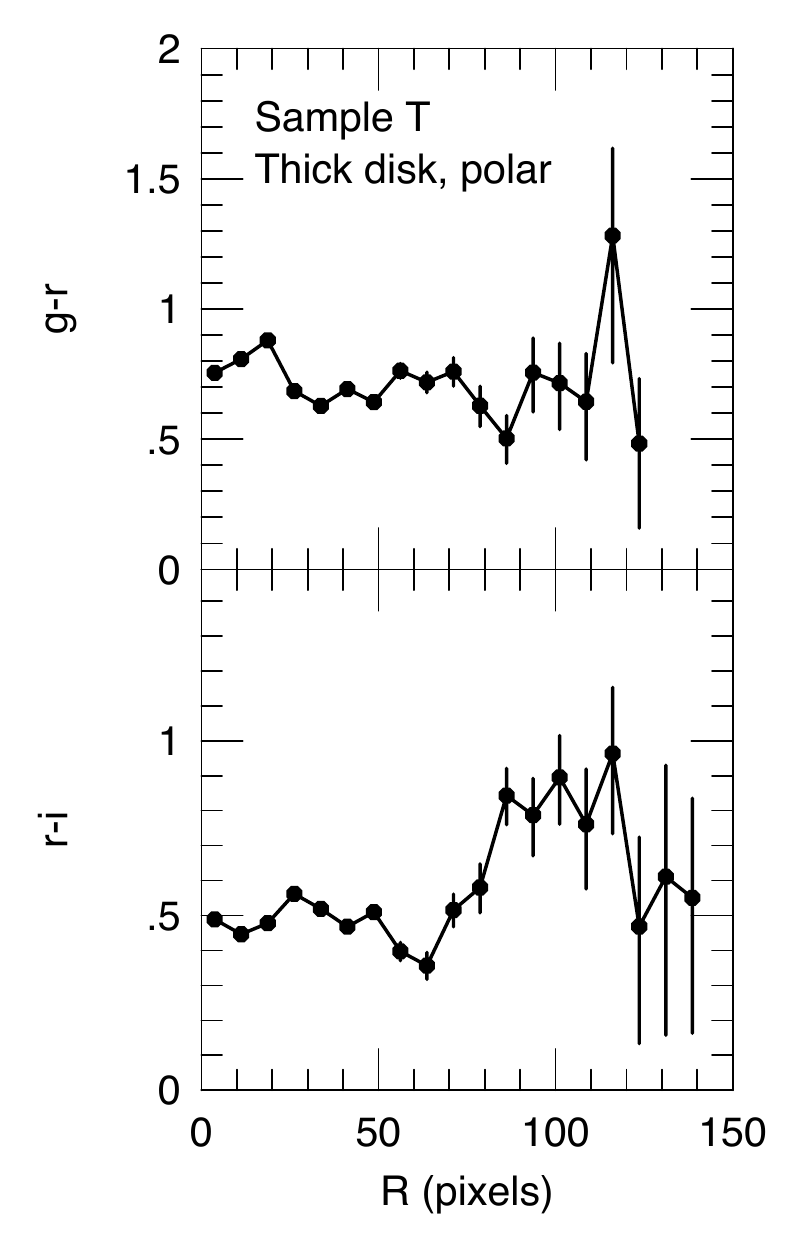}}
\end{minipage}
\caption{Colour profiles in the polar direction of the thick disk in the subsamples and the total sample. The integration is performed along two opposite wedges with opening angle 120$\degr$. A disk flattening of b/a=0.39 is assumed.  The bars indicate mean errors. Only data with errors below 1 mag have been plotted. The mean surface brightnesses in the region 100-110 pixels of sample T are $\mu_g$=30.9, $\mu_r$=30.2 and $\mu_i$=29.3 mag arcsec$^{-2}$. Please note the difference in radial extension between the subsamples (A, B and C) and the full sample (T).}
      \label{col_thick}
\end{figure*}

We begin our discussions with the subsamples A, B and C. Fig \ref{col_thick} shows the colour profiles of the thick disk. Here the colours become somewhat bluer as we go from the centre outwards. At about 100 pixels they suddenly rise to a very red colour in $r-i$ while they do not change significantly in $g-r$. The change occurs at a distance corresponding to the edge of the thin disk and is strongest in the B sample. There the colours become extreme, $g-r$ = 0.4$\pm$0.10, $r-i$ = 1.2$\pm$0.15. It is exceedingly difficult to reconcile these colours with a normal stellar population, even if we allow the slope of the IMF to take extreme values. Compared to normal galaxies in the SDSS, the disagreement is extreme. For instance, galaxies of the reddest morphological type in the SDSS (E galaxies) have $g-r$=0.83$\pm$0.14 and $r-i$=0.41$\pm$0.05 \citep{2001AJ....122.1238S}. The fact that the red excess occurs at a distance from the centre that increases with colour (largest in sample C) may indicate that the disk is becoming more dominant in the redder galaxies.

\begin{figure}
\centering
   \includegraphics[width=6cm]{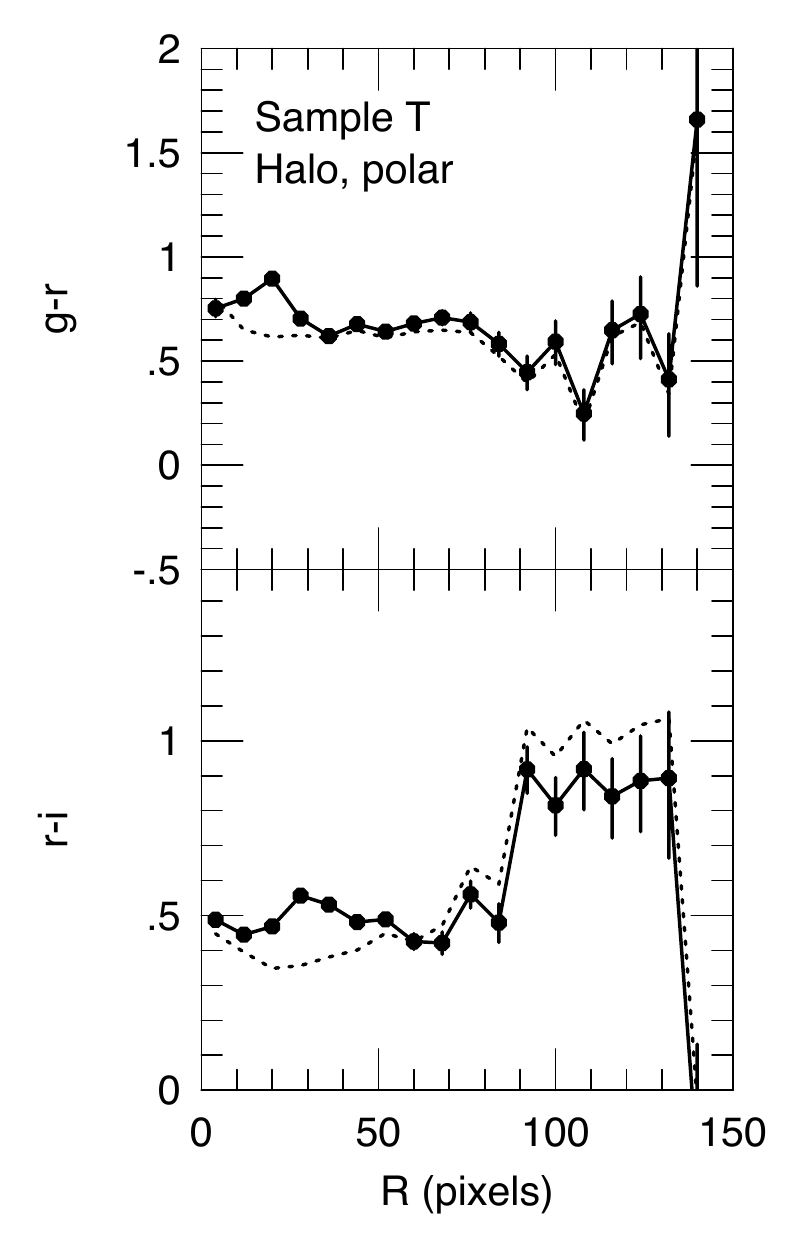}
\caption{Colour profiles along the polar direction of sample T. The opening angle is 60$\degr$. A spherical shape is assumed. The bars indicate mean errors. The dashed line shows the colour profile before correction for PSF effects. The mean surface brightnesses in the region 100-120 pixels are $\mu_g$=30.8, $\mu_r$=30.2 and $\mu_i$=29.4 mag arcsec$^{-2}$.}
      \label{col_halo}
\end{figure}

\begin{figure}
\centering
\includegraphics[width=6cm]{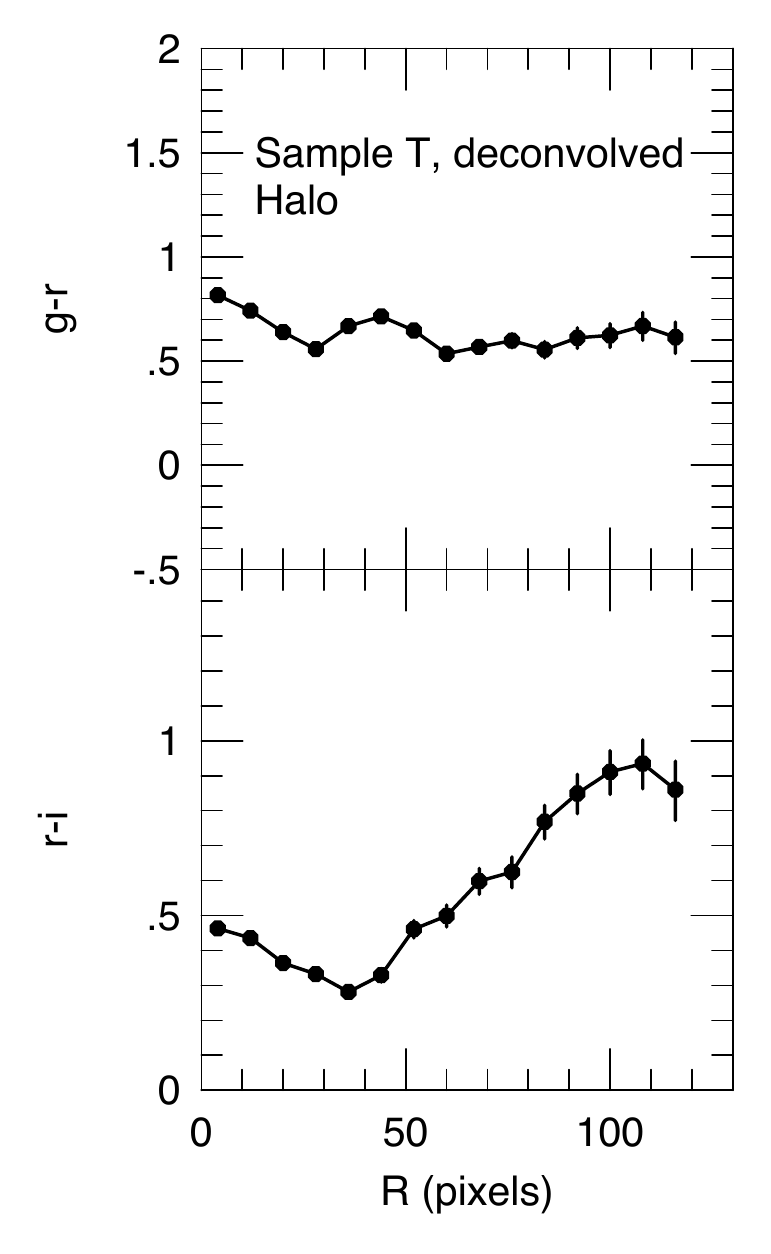}
   \caption{The halo colour profile after PSF deconvolution. Bars are mean errors.
           }
      \label{Tcol_deco}
\end{figure}

Fig. \ref{col_thick} also shows the corresponding data for the full sample. The mean $r-i$ is somewhat lower but the red excess is even more apparent. In the separate A-C samples the halo colours are too noisy to use for a rigid analysis but in the full sample the red excess emerges as a prominent structure (Fig \ref{col_halo}). In the diagram we have also plotted the colour profile {\it before} we correct for the effects of the PSF. We can see that although the correction for the PSF is highly significant, the red excess persists after the correction. As an additional support of the existence of the red excess we also look at the profile based on the deconvolved images. As we noted above, we do not trust the profiles at faint surface brightness levels but as is seen from Fig. \ref{Tcol_deco}, the colour profile at intermediate regions agree quite well with the data just discussed. The  colour in the halo, $r-i\sim$0.9, is close to the colours derived from the model corrected data. We finally note that all the thick disk subsamples, as well as the full samples show a weakly significant drop in $r-i$ at very large distances from the centre. This may indicate that there is a transition from the red halo to a bluer component or that reduction problems become serious at these extremely faint levels. The uncertainties are however too large to allow for further analysis.

\subsection{Further tests for bias in the thick disk/halo colours}

We have shown that the PSF will affect the colours such that the colour of the 
faint halo regions will become redder. Corrections for 
this reddening apparently are not sufficient to bring down the colours to values 
of a normal old stellar population. We have also argued that the reduction and 
stacking procedure is not likely to have caused the observed excess. The fact 
that the excess occurs at a radius where the galaxy is about to drown in the 
noise of the background however, made us concerned about other potential problems.

\subsubsection{Contamination by background objects}

In the process of cleaning the images free from stars and background galaxies to 
finally derive a flat background with a low noise level, we ran into the problem of
how to deal with faint objects superposed on the galaxy, contaminating the 
results. It is extremely difficult to distinguish between star clusters and/or H II 
regions in the target galaxy and superposed background galaxies of luminosities close 
to the detection limit. The faintest background objects were therefore not 
removed from the galaxy images after rotation, alignment and the second sky 
subtraction. Thus we treated the regions defined by the target galaxies differently from the regions free of bright and large objects. In the first case we did not remove the faint small objects while in the second case we did, both in the automatic identification procedure and in the median filtering. It would have been more consistent if we would have chosen not to remove the field objects, but that would have
increased the uncertainty in the zeropoint of the background flux leaving us
with unwanted structures caused by distant galaxy clustering. 

Although most of the 
influence of the superposed background galaxies should be removed by the median 
stacking one may ask if it is possible that the red excess is due to 
contamination of high redshift galaxies. Indeed, for example an elliptical 
galaxy, the colours rapidly become very red in both $g-r$ and $r-i$ at 
intermediate redshifts \citep[see e.g.][]{2003AJ....125..580C}. Since $i$ is the 
brightest part of the spectrum of the redshifted galaxies we would see the excess in $r-i$ first while the 
light from the LSB target galaxy would still dominate the $g$ band, so the $g-r$ 
would not be much affected. Moreover, if most of the light comes from galaxies 
at redshifts slightly above z=1, $g-r$ would become bluer while $r-i$ would 
still be very red. From this perspective, it would seem plausible that this is the 
explanation of the excess. However, there are counter arguments. 

\begin{figure*}
\centering
   \begin{minipage}[c]{0.5\linewidth}
   \centering{\includegraphics[width=6cm]{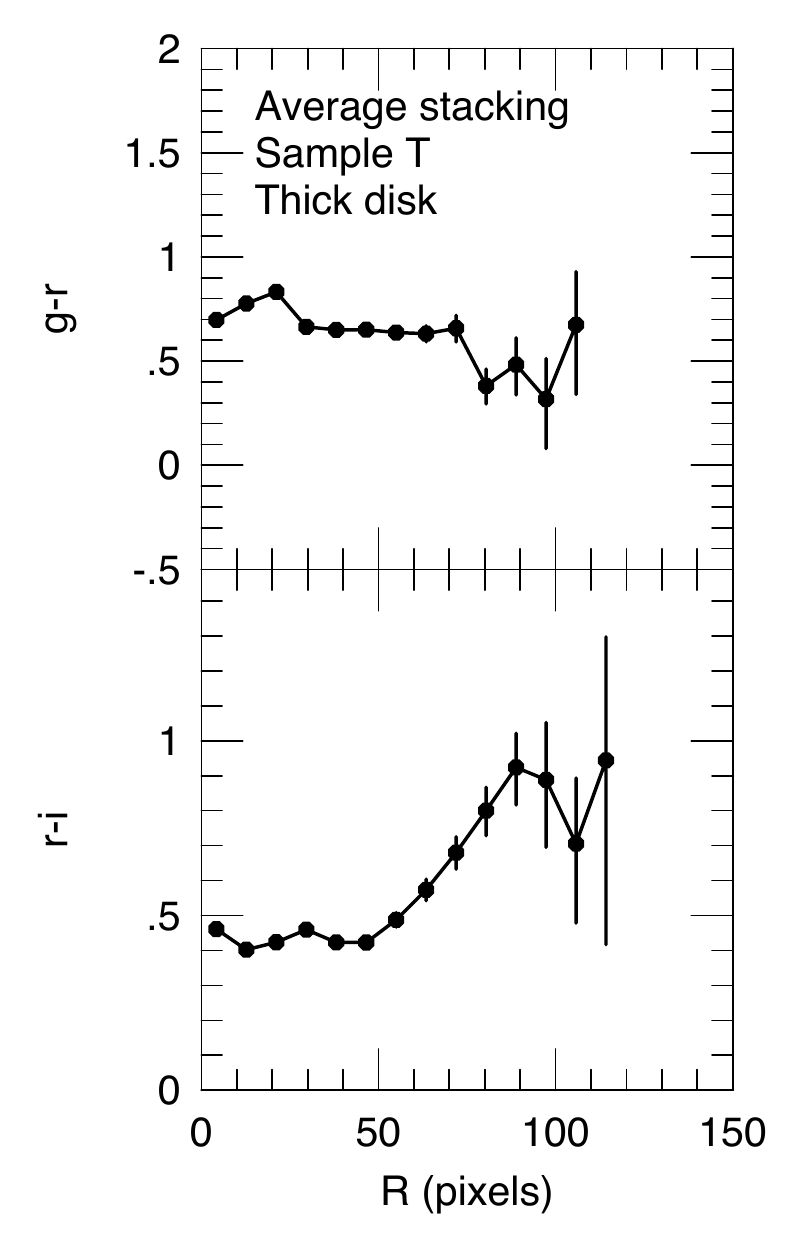}}
   \end{minipage}%
   \begin{minipage}[c]{0.5\linewidth}
   \centering{\includegraphics[width=6cm]{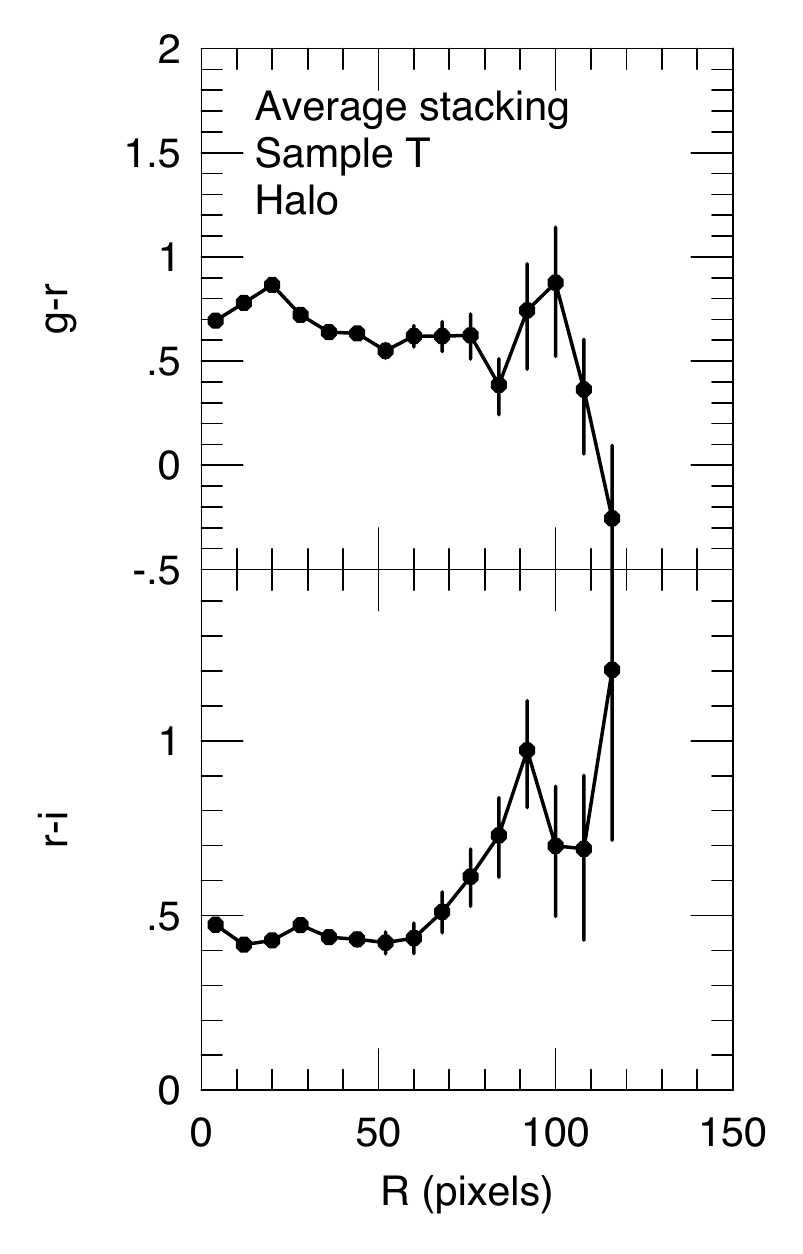}}
   \end{minipage}
    \caption{Colour profiles based on the average of the stacked 
images of the full sample. Thick disk and halo distributions. Bars are mean errors.}
      \label{col_aver}
\end{figure*}

In the stacking procedure we used median filtering to obtain the final result. If 
we instead had done an averaging of the images, the possible contribution from 
background galaxies would have increased and the colours would have become even redder. In 
Fig. \ref{col_aver} we show the $\it average$ colours of the full sample. 
Although the scatter is somewhat  larger, the general trend is the same as in the median filtered 
image. What might seem worrying is that the red excess starts at smaller radii in the 
averaged image, as we would expect if the background objects have more 
influence. Since the colour further out is the same in the averaged and median 
filtered image within the errors, we would have to conclude that most of the light in the outermost region 
is due to background objects. This is in conflict with the precondition that 
SExtractor would tend to miss out objects that are superposed on a halo with a 
surface brightness significantly above the sky level but remove them at regions of lower surface brightness. The result would be that the red excess would show up at regions of surface brightness slightly above the background but then fall back to normal at even lower surface brightnesses. There is also a second argument that could explain why the average colours become redder at a smaller radius than in the median filtered image. Since the galaxies are not in fact entirely similar in structure and stellar content,  we would expect that, if a red excess exists at faint levels, it would appear at different radii in different galaxies. The average would immediately be affected when we reach the radius where the "first" galaxy with red excess enters, but the median would not. Therefore we would expect to see the upturn in $r-i$ at a smaller radius in the average filtered image. 

%
%_____________________________________________________________

%-------------------------------------------------------------

Perhaps the strongest argument against a background bias emerges when we compare 
the halo colour profiles Fig. \ref{col_halo} and the disk colours, shown in Fig. 
\ref{col_thin}. We can clearly see the red excess in the halo profile but not in 
the disk profile at the same surface brightness. This cannot be explained by 
problems with background contamination. Therefore we conclude that 
it is far more likely that the red excess is intrinsic to the LSB galaxy sample.

\subsubsection{Inhomogeneities in the halo properties}

The small difference between the average and median stacks indicates that the distribution of halo-to-halo properties must be fairly symmetric. While halo properties may well vary substantially from galaxy to galaxy, the red excess cannot be due to a small fraction of bright and abnormal halos within the sample of galaxies. Moreover, the light sources responsible for the red colours must be relatively smoothly distributed throughout the halos. If the halo light were concentrated into a small number of randomly positioned pixels within each halo, the light from such sources would be taken into account in the average stack, but ignored by the median stack. The quantitative difference between the average and median properties could in principle also be used to constrain scenarios in which the red halo colours are attributed to a limited number of massive star clusters with abnormal stellar initial mass functions \citep{2008ApJ...687..242Z}, but such an analysis is outside the scope of the current paper.

%-------------------------------------------------------------
%
%_____________________________________________________________

\subsubsection{Improper sky subtraction}

We finally looked at the potential problems with the sky subtraction.  Although all the objects found by 
SExtractor were flagged, there will still be a weak remnant border of the 
transition region between objects (e.g. a star and sky). It is therefore an obvious risk that 
this remnant will produce an artificial signal in the sky. Bright stars superposed on the galaxy are removed by hand and will therefore probably produce a different remnant from that which is produced by the automatic stellar removal. As a consequence, the sky around and 'under' the galaxy would look slightly different. After the sky removal there would therefore be a region between the galaxy and the sky that could show up either as a positive or negative remnant. But as can be seen from Fig. \ref{lp_polar} and \ref{lp_disk}, there is a smooth transition from the galaxy luminosity profiles to the sky noise. Thus, these biases must be too weak to significantly influence the results.

\section{What is the cause of the red excess?}
Here, we explore a number of possible explanations for the anomalous halo colours, involving both stellar and interstellar effects.
We will demonstrate that while a standard halo population is unable to account for the colours, there are no less than four different mechanisms (a bottom-heavy stellar IMF, nebular emission, extended red emission and extinction of extragalactic background light) that in principle can account for the data. However, only two of these (a bottom-heavy IMF and extinction of extragalactic background light) can also account for similar red excesses detected by others at longer wavelengths. 

\subsection{A stellar halo with a standard initial mass function}
\begin{figure*}
%  \vspace*{174pt}
	\includegraphics[width=8cm]{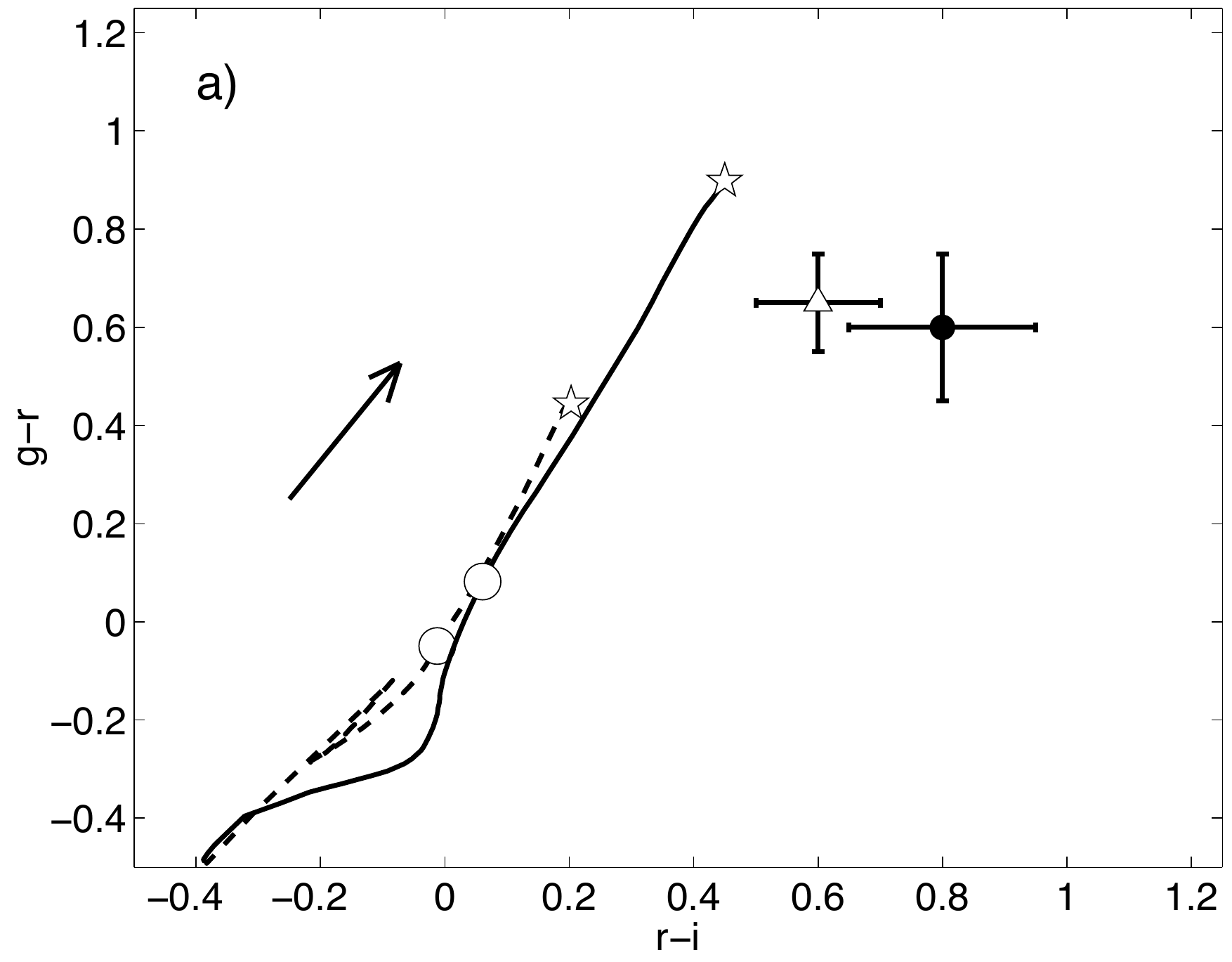}\includegraphics[width=8cm]{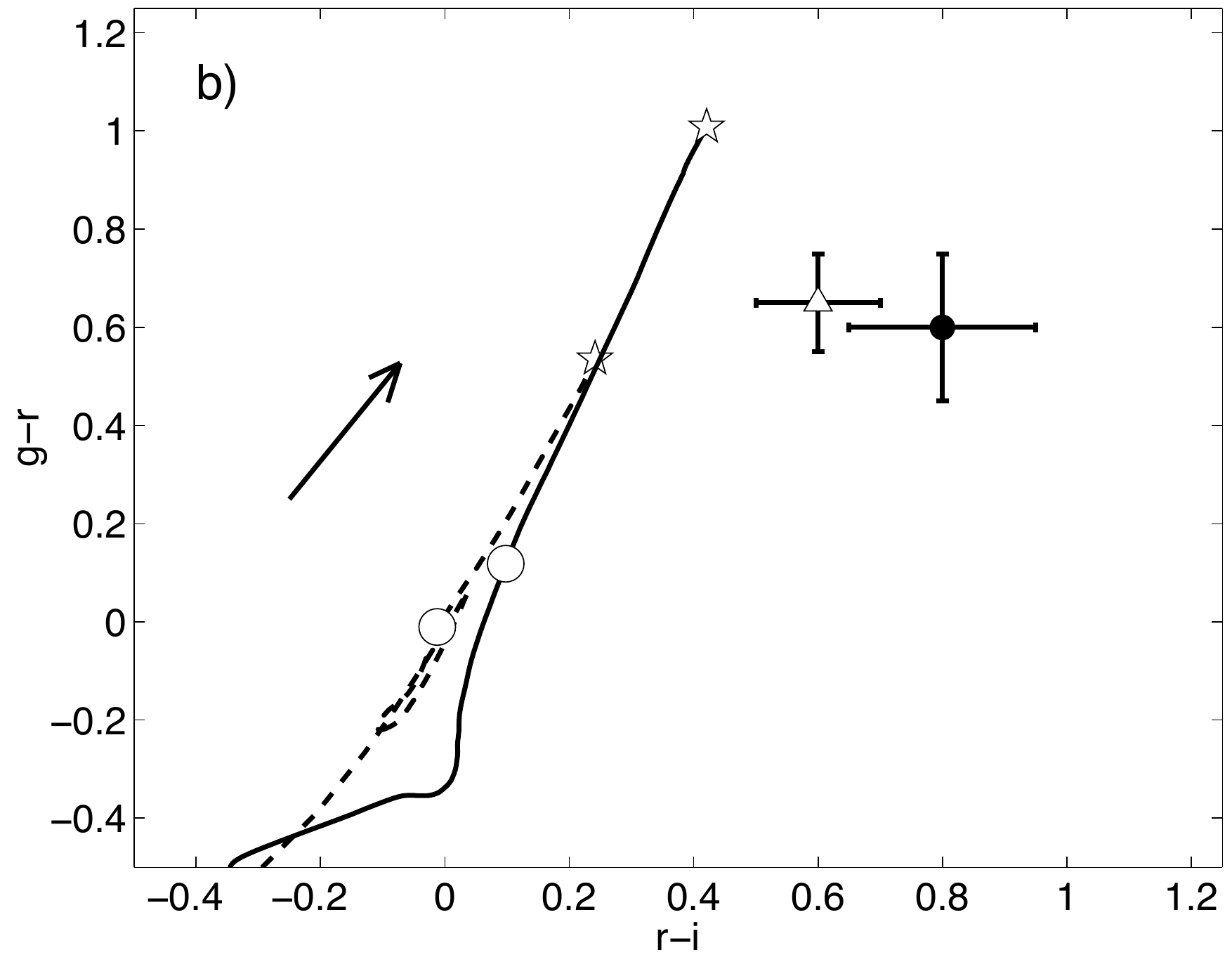}\\
	\includegraphics[width=8cm]{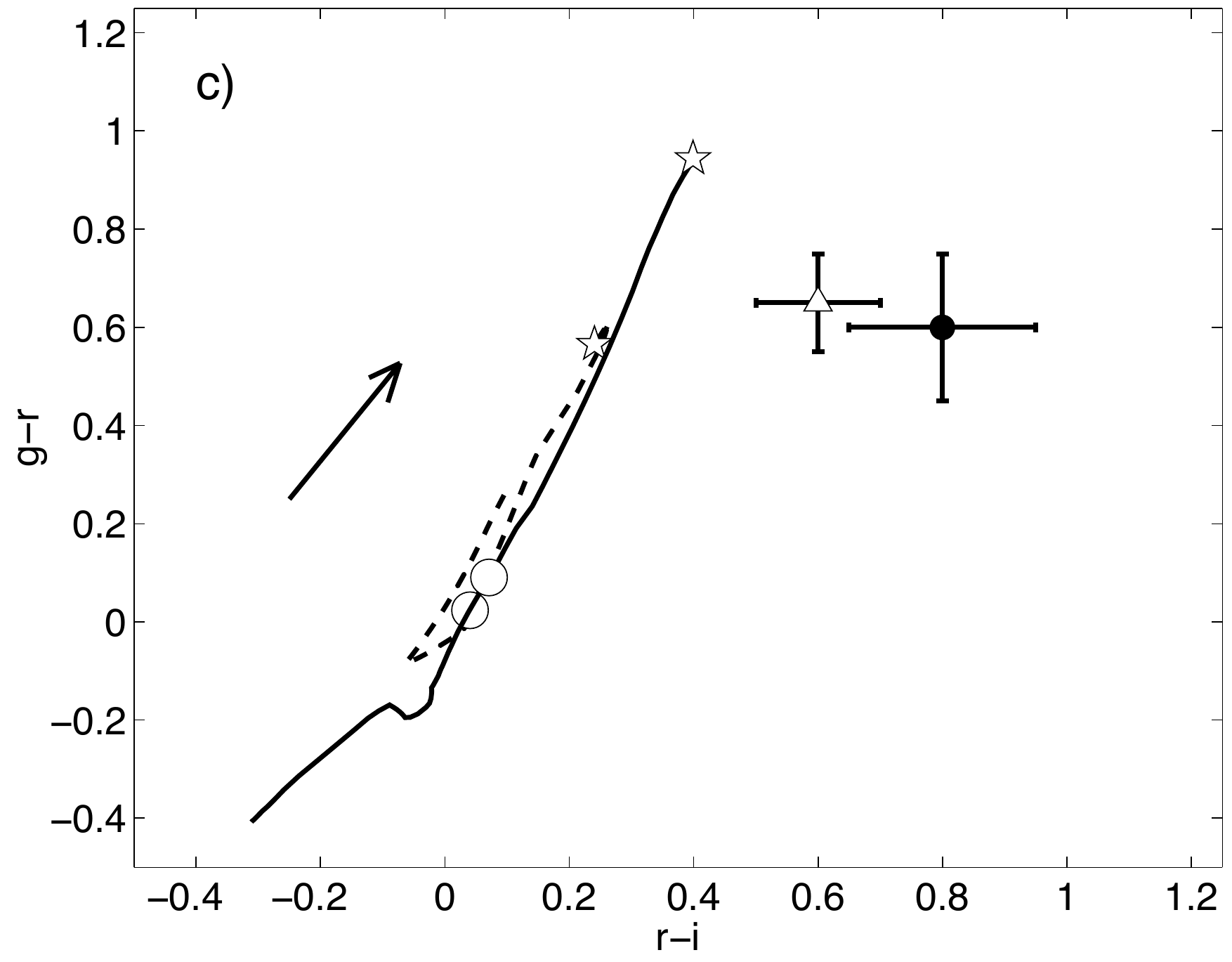}\includegraphics[width=8cm]{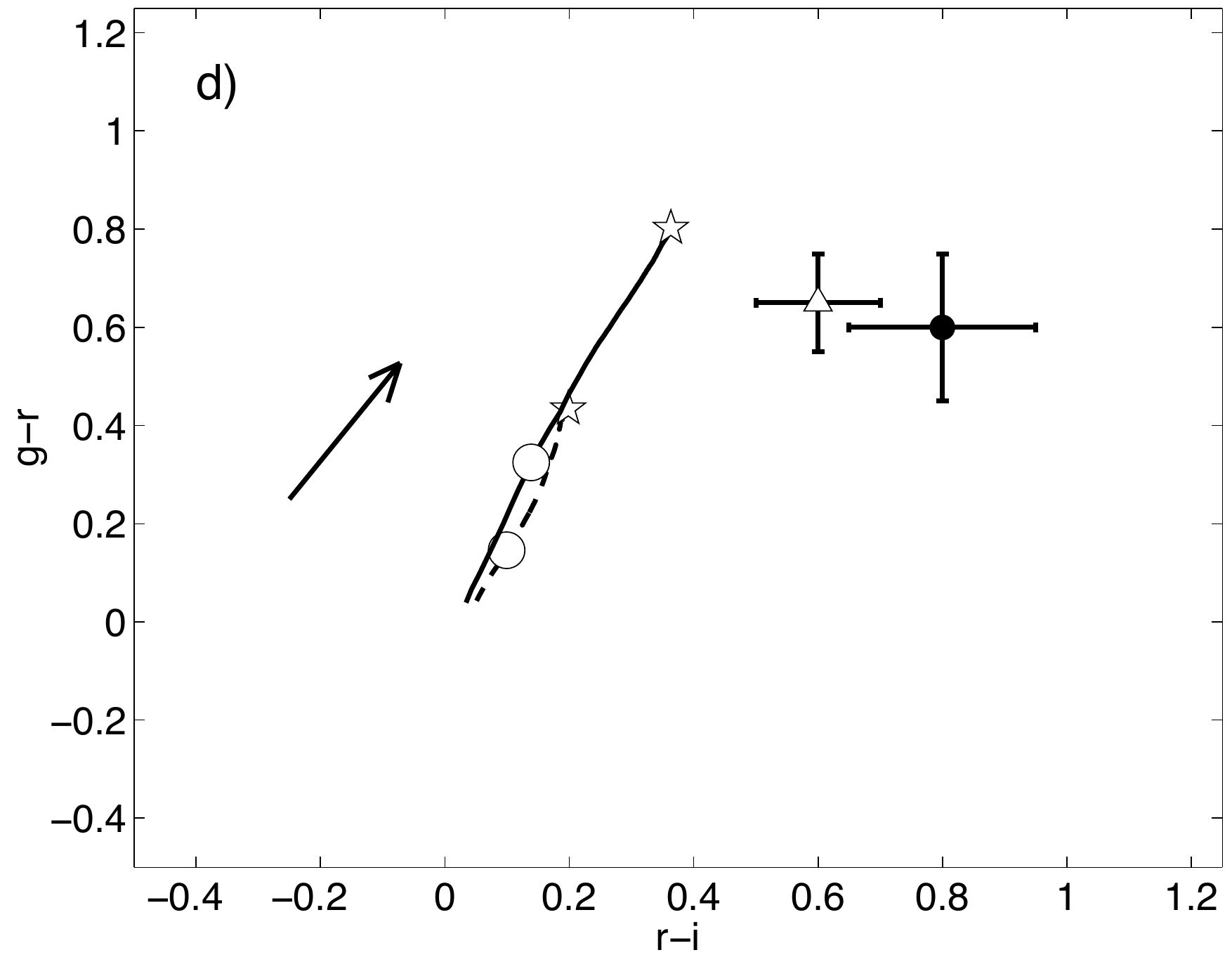}
\caption{The $r-i$ versus $g-r$ colours of the halo derived from sample T (filled circle with error bars), compared to the predicted spectral evolution of stellar populations with Salpeter IMFs and exponentially declining star formation rates ($\mathrm{SFR}(t)\propto \exp(-t/\tau)$, $\tau=1$ Gyr). Also included are the halo colours derived by Z04 for stacked high surface brightness galaxies (white triangle with error bars). For each model, the evolution of the lowest (dashed line) and highest (solid line) metallicities have been plotted. The markers along these evolutionary tracks indicate ages of 1 Gyr (circle) and 14 Gyr (star). The different panels correspond to different models: {\bf a)} P\'EGASE.2 \citep{1999astro.ph.12179F} {\bf b)} \citet{2003MNRAS.344.1000B} {\bf c)} \citet{2008A&A...482..883M} {\bf d)} the \citet{2008MNRAS.387..105L} model for binary star evolution. The Li \& Han model sequences are limited to ages $\geq$ 1 Gyr and therefore do not extend to as blue $r-i$ for low ages as the other models. None of these models for standard stellar populations are able to reproduce the observed halo colours. The arrow on the left side of each plot represents the dust reddening vector in the case of a Milky Way extinction curve \citep{1992ApJ...395..130P} and a $g$-band extinction of $A(g)=1.0$ mag. Since this arrow runs almost parallel to the age vector at ages above 1 Gyr, any reddening due to dust which may be present in the halo cannot reconcile the observed halo colours with a standard stellar population.}
\label{Salpetercol}
\end{figure*}

Current simulations of the formation of stellar halos around disk galaxies predict that the halo should start dominating over the disk and bulge at $\mu_V\approx 29$ mag arcsec$^{-2}$ \citep{2006ApJ...653.1180G}, which corresponds to $\mu_i\approx 28.6$ mag arcsec$^{-2}$ if one assumes $g-r=0.4$, $r-i=0.2$ as expected for a standard, low-metallicity halo population (see Fig.~\ref{Salpetercol}). This is in good agreement with the surface brightness at which our red halo colours turn up ($\mu_i\approx 28$--29 mag arcsec$^{-2}$; see Figs.~\ref{lp_polar} and \ref{col_halo}). However, we argue that the halo colours at that point are inconsistent with any kind of normal stellar population.

In Fig.~\ref{Salpetercol}, we compare the halo colours from the T sample (filled circle with error bars) to the predictions from four different spectral evolutionary models: P\'EGASE.2 \citep{1999astro.ph.12179F}, \citet{2003MNRAS.344.1000B}, \citet{2008A&A...482..883M} and \citet{2008MNRAS.387..105L}. Here, we have assumed a Salpeter IMF ($\mathrm{d}N/\mathrm{d}M\propto M^{-\alpha}$ with $\alpha=2.35$ throughout the mass range 0.08--120 $M_\odot$) and an exponentially decaying star formation rate $\mathrm{SFR}(t)\propto \exp(-t/\tau)$ with $\tau=1$ Gyr (suitable for an early-type system). While the Salpeter IMF is well-known to overpredict the number of stars with masses below $1\ M_\odot$ in many resolved stellar populations (which are better represented by e.g. the \citealt{2001MNRAS.322..231K} IMF), this mainly affects the mass-to-light ratio and has a negligible impact on the integrated colours. For models which do not allow exponentially decreasing star formation rates, we have implemented this star formation history by summing up the time steps (with appropriate weights) for an instantaneous burst (i.e. single-age) stellar population. While a pure single-age population would produce slightly redder colours at the highest ages, they are still nowhere near the observed colours. Moreover, the perfectly coeval onset and quenching of star formation associated with this scenario seems unrealistic given the huge spatial scales ($\sim 10$ kpc) involved in the halo. For each model, two different evolutionary sequences are plotted, corresponding to the lowest and highest metallicities allowed. In a plot of $r-i$ vs. $g-r$, the predicted colour sequences of intermediate metallicities typically lie inbetween those shown. 

P\'EGASE.2 and the \citet{2003MNRAS.344.1000B} model represent some of the most widely used spectral synthesis codes on the market, and were previously used by \citet{2006ApJ...650..812Z} to argue that a Salpeter-IMF stellar population could not account for the halo colours presented by Z04. For comparison purposes, we include them here as well. The \citet{2008A&A...482..883M} model represents the current state of the art in the treatment of thermally pulsating asymptotic giant branch stars, which are very important for the interpretation of near-infrared colours \citep[e.g.][]{2005MNRAS.362..799M,2009MNRAS.396L..36T}. The \citet{2008MNRAS.387..105L} model is one of the few models on the market that take the evolution of binary stars into account. While the prescriptions and assumptions going into these models differ, it is evident from Fig.~\ref{Salpetercol} that the differences between their predictions for the evolution of the $g-r$ and $r-i$  colours are relatively modest, at least at high ages. Moreover, all models -- regardless of the metallicity -- fail to explain the very red $r-i$ colours of the halo from sample T. Also shown in Fig.~\ref{Salpetercol} are the halo colours derived by Z04 for stacked high surface brightness galaxies (white triangle with error bars). These halo colours are marginally consistent with ours, but both sets of measurements are clearly anomalously red compared to the models. Is it then possible that this discrepancy is simply due to some shortcoming inherent to {\it all} models? This does not seem likely, since the red halos are considerably redder than other old stellar populations like globular clusters and elliptical galaxies (see Z04), which these models have no problem in explaining. 

The model predictions are formally valid for redshift $z=0$, whereas the median redshift of the T sample is $z\approx 0.06$. As long as the IMF is Salpeter-like, the $k$-corrections on the observed colours are, however, very small. Using the \citet{2001A&A...375..814Z} spectral evolutionary code, we have assessed the maximum $k$-corrections on the $g-r$ and $r-i$ colours to be $|\Delta(g-r)|<0.1$ mag and $|\Delta(r-i)|<0.03$ mag for a $Z=0.001$, $\tau=1$ Gyr stellar population at ages higher than 1 Gyr. Since these corrections are smaller than the observational error bars on the halo colours, redshift effects have no impact on our comparison of models and data.

The arrows in the the lefthand parts of Fig.~\ref{Salpetercol}a--d represent the dust reddening vectors expected from a Milky Way extinction curve \citep{1992ApJ...395..130P} with $g$-band extinction  $A(g)=1.0$ mag. Dust reddening makes the colours shift in a direction which is largely parallel to that of the age vector at ages higher than 1 Gyr in this diagram, and therefore fails to reconcile the observed halo colours with normal stellar populations. Estimates of dust extinction in low surface brightness galaxies \citep[e.g.][]{1999A&A...341..697B, 2007ApJ...663..908R} indicate far less extinction than this. Current estimates of the opacities of halos based on studies of background objects (which are likely to suffer more extinction than stars sitting inside the actual halo) also suggest much lower optical extinction values \citep{1994AJ....108.1619Z, 2009arXiv0902.4240M}. Hence, it seems highly unlikely that the extinction in the halo could be as high as $A(g)=1.0$ mag. We therefore conclude that the observed colours of halos around low surface brightness galaxies cannot possibly be reconciled with any {\it normal} type of stellar population.

\subsection{Insufficient mass sampling of a standard stellar population}
When comparing the observed colours of galaxies to the predictions of spectral evolutionary models, it is customary to assume that the IMF is well-sampled for all stars that contribute significantly to the integrated flux. This approximation is reasonably safe as long as the mass of the stellar population analyzed is $M\gtsim 10^4$--$10^5 \ M_\odot$ (assuming a standard IMF), but starts to break down as one moves to lower population masses, due to substantial scatter in the number of the high-mass stars present \citep{2003MNRAS.338..481C,2004A&A...413..145C}. 
\begin{figure}
\includegraphics[width=8cm]{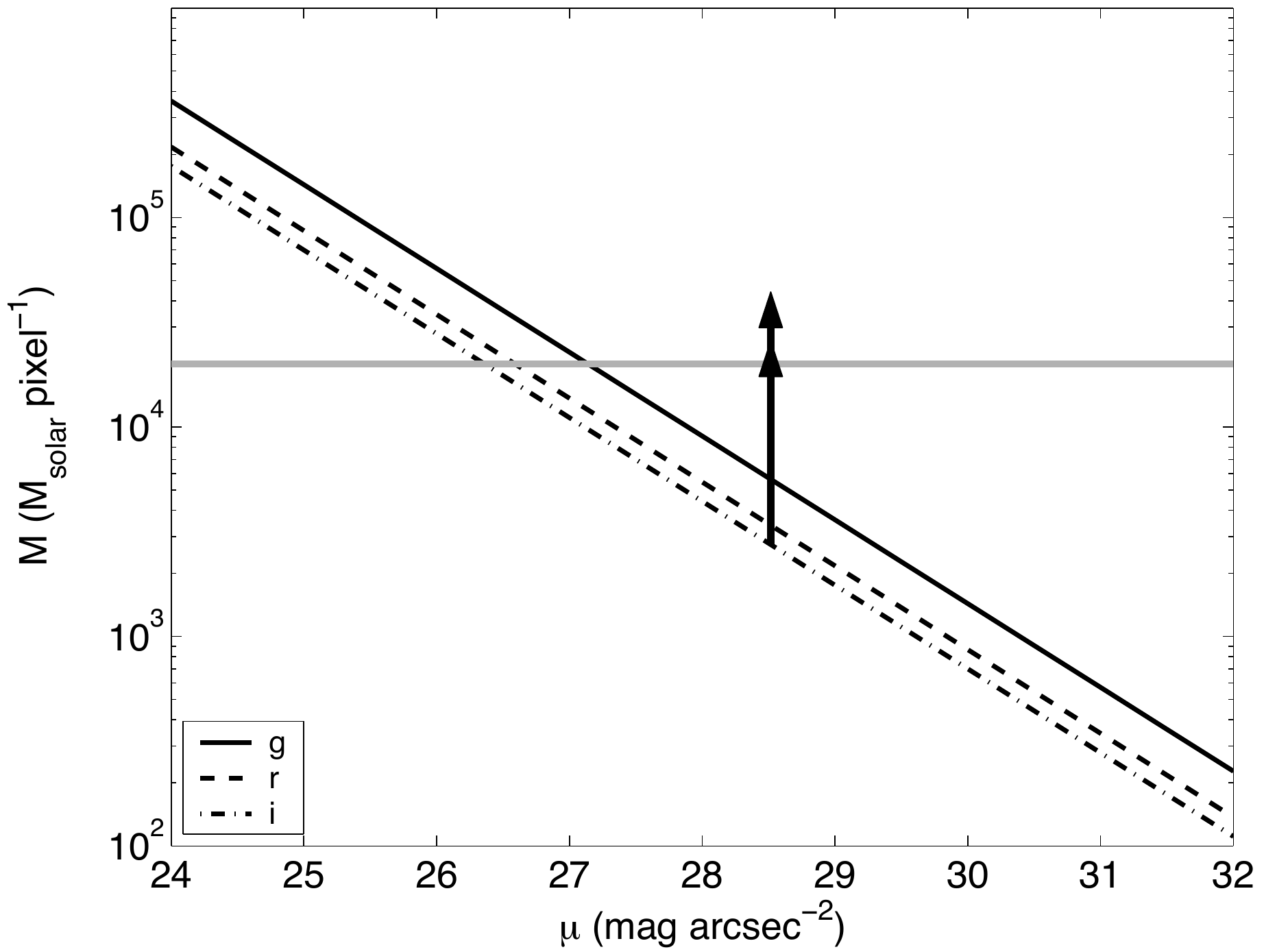}
\caption{The estimated population mass per resolution element in the T sample as a function of surface brightness. The different line types represent pixel masses derived in filters $g$ (solid), $r$ (dashed) and $i$ (dash-dotted). The gray horizontal line indicates the approximate initial stellar population mass below which IMF sampling effects become important when observing an old ($\geq 1$ Gyr) stellar population with a standard IMF at $gri$ wavelengths \citep{2004A&A...413..145C}. In the surface brightness range where the red halo colours are derived ($\mu_i\approx 28$--29 mag arcsec$^{-2}$), the mass per pixel is only a few times $10^3\ M_\odot$. However, surface brightness fluctuations due to an insufficiently sampled IMF can be dampened within the typical SDSS seeing disk, which may boost the mass relevant for the comparison by up to a factor of $\approx 10$ (large arrow). The conversion between current and initial population mass can boost the mass by another $\approx 40\%$ (small arrow). The initial mass per resolution element may therefore be a few times $10^4\ M_\odot$, which indicates that we are slightly above the surface brightness limit where IMF sampling effects are expected to become important.}
\label{Mstars}
\end{figure}

While the halo regions that we analyze are very large, the integrated luminosity of each pixel rapidly drops as a function of distance from the disk. This means that the stellar mass (as derived from the luminosity) contained in each resolution element will eventually fall below the safe limit as one moves outwards. At that point, one expects the observed colours to start deviating from the predictions of models which assume a well-sampled IMF, like the ones used in the previous section. The tell-tale signature of this would be a skewed flux distribution within the halo, with a few bright stars sprinkled across a more uniform background of faint low-mass ones. When the images of many such halos are combined, a median stack will reject the bright outliers and favour pixels containing the fainter and more common stars. Since high-mass and low-mass stars have very different spectra, a strong colour bias will be introduced. Could this then be the reason for the anomalous colours that we observe?  

There are a number of arguments that make this seem implausible. To begin with, the observed halo colours seem to be insensitive to whether the images have been stacked by the average or median pixel flux (see Figs.~\ref{col_thick}, \ref{col_halo} \& \ref{col_aver}). Since a stack based on the median will reject the extreme outliers in these distributions, and thereby ignore the few pixels which happen to contain the rare high-mass, high-luminosity stars, the mean and median stacks are expected to give rise to radically different results if the halo light stems from a stellar population with a standard IMF. Moreover, high-mass stars (red giant branch and asymptotic giant branch stars) tend to be significantly redder (in e.g. $r-i$) than stars on the upper main sequence. Hence, one would expect a bias of this type to render the halo anomalously blue, which is the opposite of what we observe. Therefore, the anomalously red halo colours cannot be attributed to insufficient mass sampling of a normal stellar population. 

While IMF sampling effects appear not to be an issue for the present study, one may ask at what surface brightness limit such effects are likely to become important? To address this, we in Fig.~\ref{Mstars} estimate the stellar mass expected in a single pixel as a function of surface brightness within the stacked halo from sample T. These estimates are based on stellar population mass-to-light ratios (in units of $M_\odot/L_\odot$) of $M/L_g=2.1$, $M/L_r=1.9$ and $M/L_i=1.7$, suitable for a 10 Gyr old, $Z=0.001$, $\tau=1$ Gyr stellar population with a \citet{2001MNRAS.322..231K} IMF throughout the mass range 0.01--120 $M_\odot$, as predicted by the P\'EGASE.2 code \citep{1999astro.ph.12179F}. Each pixel ($0.396$ arcsec) is moreover estimated to correspond to a linear distance of 120 pc for the T sample. In the region where the red halo colours are derived ($\mu_i\approx 28$--29 mag arcsec$^{-2}$), the population mass per pixel is only a few times $10^3\ M_\odot$. However, surface brightness fluctuations may be smeared out over the  SDSS seeing disk (FWHM 1.4\arcsec, i.e. an area covering $\approx$ 10 pixels), and this can boost the relevant mass per resolution element by a factor up to $\approx 10$ (indicated by the large arrow in Fig.~\ref{Mstars}). When comparing to the IMF sampling effects modelled by \citet{2004A&A...413..145C}, we moreover need to convert the current stellar mass (as derived from the adopted mass-to-light ratios) to the initial stellar mass. This can amount to an additional boost of up to 40\% (indicated by the small arrow). 

In conclusion, we may be probing a stellar population with an initial mass of a few times $10^4\ M_\odot$ per resolution element. The analysis by  \citet{2004A&A...413..145C} indicates that an initial stellar mass of $\gtsim 2\times 10^4\ M_\odot$ would be required (for an old (age $\geq 1$ Gyr) stellar population observed in the $gri$ wavelength range) to ensure that IMF sampling effects are not too serious. Hence, we are probably close to the surface brightness limit where IMF sampling effects are expected to become important. This means that one should in general avoid using median stacks when attempting to combine images of different low-redshift objects for the purpose of studying regions of very low surface brightness. Alternatively, many pixels within each image could be combined to raise the expected stellar mass per resolution element above the safety limit before stacking.

\subsection{A stellar halo with a bottom-heavy initial mass function}
\begin{figure*}
%  \vspace*{174pt}
	\includegraphics[width=8cm]{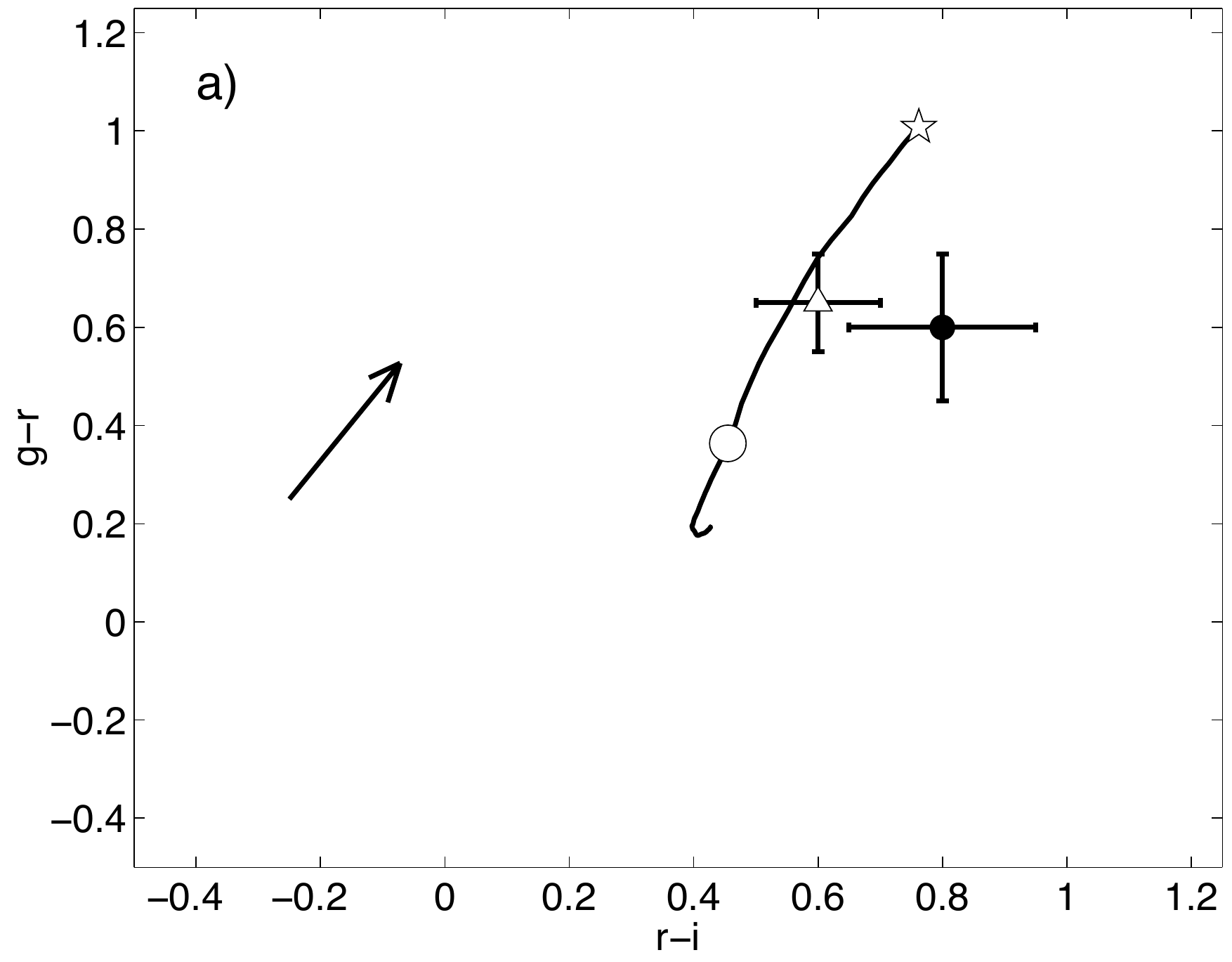}
	\includegraphics[width=8cm]{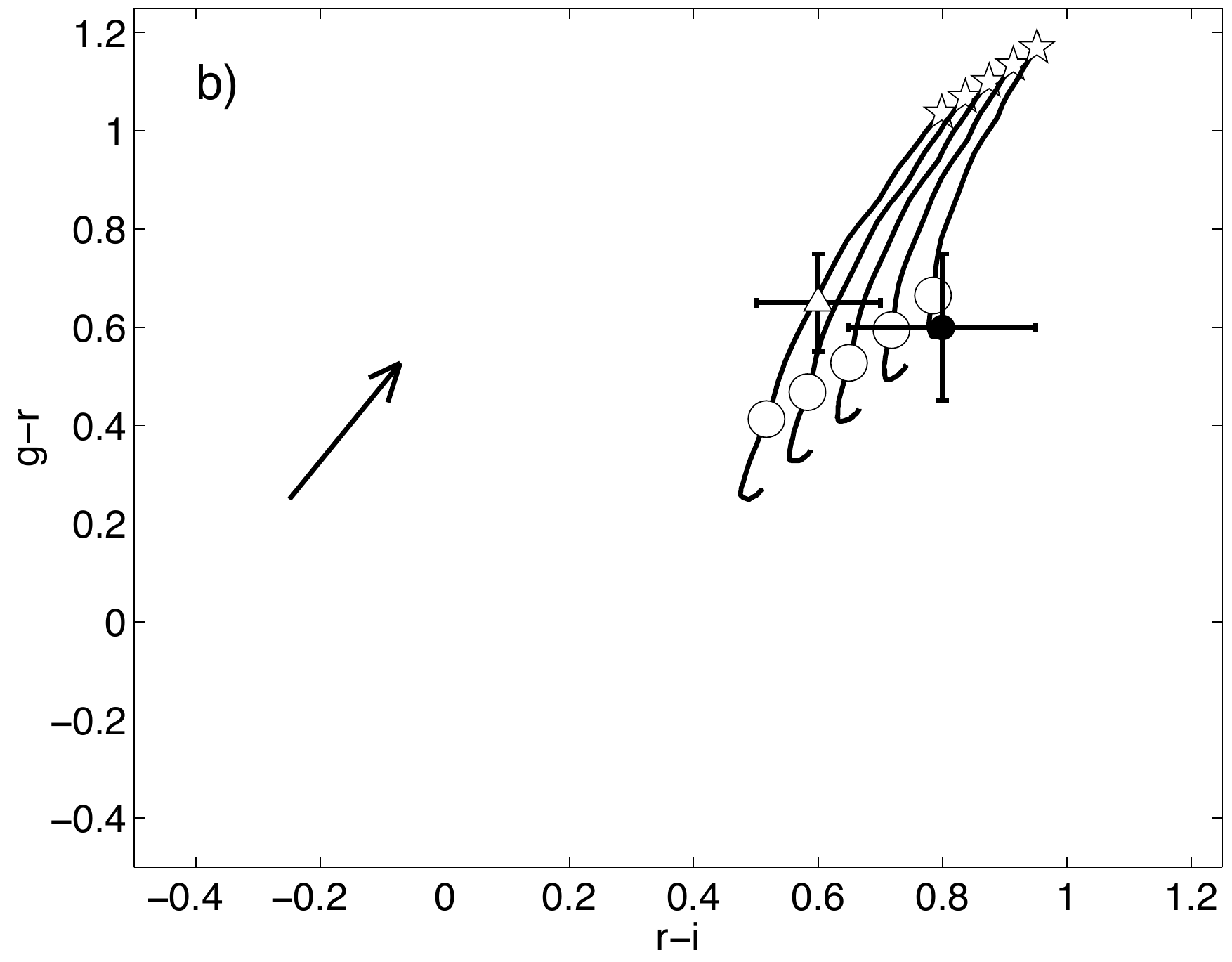}
\caption{The $r-i$ versus $g-r$ colours of the halo derived from sample T (filled circle with error bars), compared to the predicted spectral evolution of  stellar populations with various IMFs, based on the P\'EGASE.2 model \citep{1999astro.ph.12179F}. 
Also included are the colour derived by Z04 for stacked high surface brightness galaxies (white triangle with error bars). For all model tracks, a metallicity of $Z=0.008$ and an exponentially declining star formation rate ($\mathrm{SFR}(t)\propto \exp(-t/\tau)$ with $\tau=1$ Gyr) has been assumed. The markers along the evolutionary tracks indicate ages of 1 Gyr (circle) and 14 Gyr (star). The arrow on the left side of each plot represents the dust reddening vector in the case of a Milky Way extinction curve \citep{1992ApJ...395..130P} and a $g$-band extinction of $A(g)=1.0$ mag. {\bf a)} A stellar population with $\alpha=4.5$ (solid line). This model provides a reasonable fit to the halo colours derived by Z04, but not to the halo colours measured by us.  {\bf b)} Stellar populations with $\alpha=4.6$, 4.7, 4.8, 4.9 and 5.0 (solid lines, from left to right). Clearly, the $\alpha = 4.8$--5.0 IMFs (the three rightmost lines) can fit the halo colours from the T sample, but at the cost of having an unexpectedly young ($\leq 1$ Gyr) halo population.}

\label{IMFfig}
\end{figure*}
The stellar IMF has long been suspected to be universal, but recent observational studies based on unresolved stellar populations have given some support to the notion that it may vary both as a function of environment \citep{2008ApJ...675..163H,2009ApJ...695..765M}, and as a function of cosmic time \citep{2008MNRAS.385..147D, 2008ApJ...674...29V, 2008MNRAS.391..363W}. Attempts to constrain the IMF in the field population of the LMC using star counts \citep[e.g.][]{2006ApJ...641..838G} have also yielded results seemingly inconsistent with the standard IMF.

Using the P\'EGASE.2 model, \citet{2006ApJ...650..812Z} demonstrated that a bottom-heavy IMF with slope $\alpha=4.5$ ($\mathrm{d}N/\mathrm{d}M\propto M^{-\alpha}$) provided a reasonable fit to the red halos of both blue compact galaxies and the stacked high-surface brightness disks of Z04. This implies a stellar halo population with a huge number of low-mass stars and a very high mass-to-light ratio. 

Since a significant fraction of the cosmic baryons at low redshifts remain unaccounted for \citep[e.g.][]{2009and..book..419P}, one could envision that red halos with a bottom-heavy IMF may serve as reservoirs for some of them. However, \citet{2008ApJ...687..242Z} argue that scenarios of this type are inconsistent with direct star counts in the Milky Way halo, unless the low-mass stars are predominantly located in star clusters. A {\it smooth} halo population obeying this extreme IMF could remain viable if the surface brightness of the Z04 halo had been overestimated \citep[for instance due to scattered light;][]{2008MNRAS.388.1521D}, but the mass of this halo would then be too low to have any bearing on the missing baryons.

As shown in Fig.~\ref{Salpetercol}, the halo we detect in sample T is about 0.2 mag redder in $r-i$ than the Z04 halo. This places the halo in a region of the $r-i$ vs. $g-r$ diagram which is somewhat difficult to reconcile with current models based on a bottom-heavy IMF of the type proposed by \citet{2006ApJ...650..812Z}. The problem is illustrated in Fig.~\ref{IMFfig}, where we compare the halo colours observed by Z04 to the $\alpha=4.5$, $Z=0.008$ P\'EGASE.2 model favoured by \citet{2006ApJ...650..812Z}. This combination of IMF and metallicity is unable to explain the high $r-i$ colour of the halo from sample T at {\it any} age. The fit can be improved by adopting an even more extreme IMF ($\alpha>4.5$), but this also pushes the model track towards redder $g-r$ colours. This is shown in Fig.~\ref{IMFfig}b, where stellar populations with $Z=0.008$ and $\alpha = 4.6$, 4.7, 4.8, 4.9 and 5.0 (from left to right) are shown. Even though the model tracks of the $\alpha=4.8$--5.0 IMFs (the three rightmost evolutionary sequences) are formally within the error bars of the halo from sample T, this fit would imply a curiously young halo ($\ltsim 1$ Gyr old). Moreover, the required metallicity of $Z=0.008$ is unexpectedly high compared to the stellar halos of the Milky Way ($Z\ltsim 0.001$; e.g. \citealt{2007Natur.450.1020C}) and Andromeda ($Z\approx 0.005$ for the high surface brightness debris and $Z\approx 0.001$ for the smooth underlying halo; e.g. \citealt{2006ApJ...653..255C}, \citealt{2006ApJ...648..389K}, \citealt{2009MNRAS.396.1842R}). 

Just like in the case of Salpeter-like IMFs, this comparison between data and models neglects the non-zero redshift of the galaxies in the T sample (median redshift $z\approx 0.06$). Using the 
\citet{2001A&A...375..814Z} spectral synthesis code (which gives results similar to those of P\'EGASE.2 in the relevant $r-i$ vs. $g-r$ diagrams), we have assessed the likely $k$-corrections for bottom-heavy IMF populations with parameters identical to those used in Fig.~\ref{IMFfig}. Even though the $k$-corrections become somewhat larger than for stellar populations obeying Salpeter-like IMFs, they still remain smaller than the observational error bars on the halo colours. Hence, redshift corrections have no major impact on the outcome of our comparison. 

While {\it current} models for bottom-heavy IMFs seem hard-pressed when trying to fit the observed halo colours, the prescriptions for the evolution and spectra of low-mass stars used may still be too crude for the best-fit parameter values ($Z$, age and to some extent the IMF slope $\alpha$) to be taken at face value. These stars (with masses $M<0.8\ M_\odot$) give a negligible contribution to the integrated flux at optical wavelengths in the case of a normal IMF, but become important when bottom-heavy IMFs are considered. As demonstrated by \citet{2008MNRAS.389..585C}, the best models available for low-mass stars still produce temperatures (colours) that are considerably hotter (bluer) than observed for low-mass stars. While bottom-heavy IMFs are generically predicted to shift $r-i$ in the right direction to explain the red halo colours, attempting to attain an exact match to $g-r$ and $r-i$ may therefore be a futile exercise, given the limitations of current population synthesis codes. Hence, we conclude that a stellar population with an unusually high fraction of low-mass stars still represents a plausible explanation for colours of the halo, although this population probably contains no more than a tiny fraction of the missing baryons. A number of physical mechanisms that might explain why the IMF appears bottom-heavy in low-density environments are discussed by \citet{2004MNRAS.354..367E}.

\subsection{Extended red emission}
In principle, the red halo colours could be due to the dust luminescence phenomenon known as extended red emission (ERE). While the dust component responsible for the ERE has yet to be identified, ERE has been reported in a wide range of environments where far-ultraviolet radiation and dust are known to coexist, including reflection nebulae, HII regions, the halo region of the starburst galaxy M82 and the high-latitude diffuse interstellar medium and cirrus in the Milky Way \citep[for a review, see][]{2004ASPC..309..115W}. One may speculate that similar conditions could be met in the halos of LSB galaxies, provided that some dust is present in these regions and that far-ultraviolet radiation from the disk stars can escape in the polar direction (or, alternatively, that there are a few young stars sprinkled throughout the halo). The ERE manifests itself as a wide ($\sim$ 1000 \AA) emission feature in the wavelength range from 6100--9500 \AA{ }(i.e. inside either the SDSS $r$, $i$ or $z$ bands), with the peak wavelength varying from object to object. At the current time, we have no robust way of ruling out the possibility that the $i$-band excess detected in the halo of the stacked disks is due to ERE. However, there are a number of arguments that speak against this hypothesis. 

To begin with, the peak wavelength of the ERE is known to correlate with the density of the far-ultraviolet radiation field \citep{2002ApJ...565..304S}, and having the ERE peak in the SDSS $i$-band of sample T would (even after correcting for redshift effects) probably require a typical ionization parameter higher than that measured in the interstellar medium of the solar neighborhood. It seems reasonable that the typical far-ultraviolet radiation density should be lower than this at several kpc away from the disk of LSB galaxies. Studies of ERE from high-latitude clouds above the Milky Way \citep[e.g.][]{1998ApJ...494L..93S, 2008ApJ...679..497W} suggest an $R$-band ERE peak (at wavelengths intermediate between the SDSS $r$ and $i$ bands) for these objects -- which may be similar to the hypothetical ERE-emitting clouds above LSB disks -- although exceptions do exist. Most of the Galactic high-latitude clouds currently searched for ERE are moreover located within 300 pc of the Milky Way disk. Even if one assumes a radiation field as strong as that from the Milky Way, interstellar gas clouds at $\sim 10$ kpc (where the red halo colours are measured) would be expected to display ERE peaks at even shorter wavelengths. The ERE also tends to coexist with a very blue component of dust-scattered continuum radiation, whose relative contribution depends on the angle between the line of sight and the direction from which the ultraviolet photons are originating. Even if the ERE were to peak in the SDSS $i$-band, it is not clear that it would give rise to $g-r$ and $r-i$ colours similar to those observed. Differences in dust content between disk galaxies of high and low surface brightness may, however, drive the prediction in either direction. Increasing the optical depth of dust could reduce the far-ultraviolet flux at any given distance from the disk, but at the same time potentially increase the density of the ERE carriers. 

As pointed out by \citet{2007arXiv0708.0762Z}, explaining the $i$-band excess as due to ERE, would require different explanations for the red excesses reported in this band and at longer wavelengths. For example, the published surface photometry on NGC 5907 \citep{1994Natur.370..441S, 1998MNRAS.301..280J, 1997Natur.387..159R,1998A&A...334L...9L,1996A&A...312L...1L} suggests that that the red halo of NGC 5907 displays an excess flux -- compared to the expectations for a low-metallicity halo population with a standard IMF -- in both $I$, $J$ and $K$ bands. No ERE has yet been reported in this wavelength range. The red halos of blue compact galaxies have also been detected through an excess in the $K$-band \citep{2002A&A...390..891B, 2005mmgf.conf..355B,2006ApJ...650..812Z}. Occam's razor would argue for a single explanation for the red halo excesses of all objects and in all filters in the wavelength range from $I$ to $K$. On the other hand, this argument based on simplicity may be weakened if some of these detections turn out to be spurious. \citet{1999AJ....117.2757Z} has raised serious concerns about the reliability of the surface photometry data reported for the NGC 5907 halo (but see \citealt{2000AJ....119.1701Z} for a different view). Moreover, \citet{2009arXiv0902.4695Z} argue that, due to recent developments in the field of near-IR spectral synthesis modelling, the observed colours of many of the ``red halos'' of blue compact galaxies may now have other, more mundane explanations.

\subsection{Nebular emission}
\begin{figure}
\includegraphics[width=8cm]{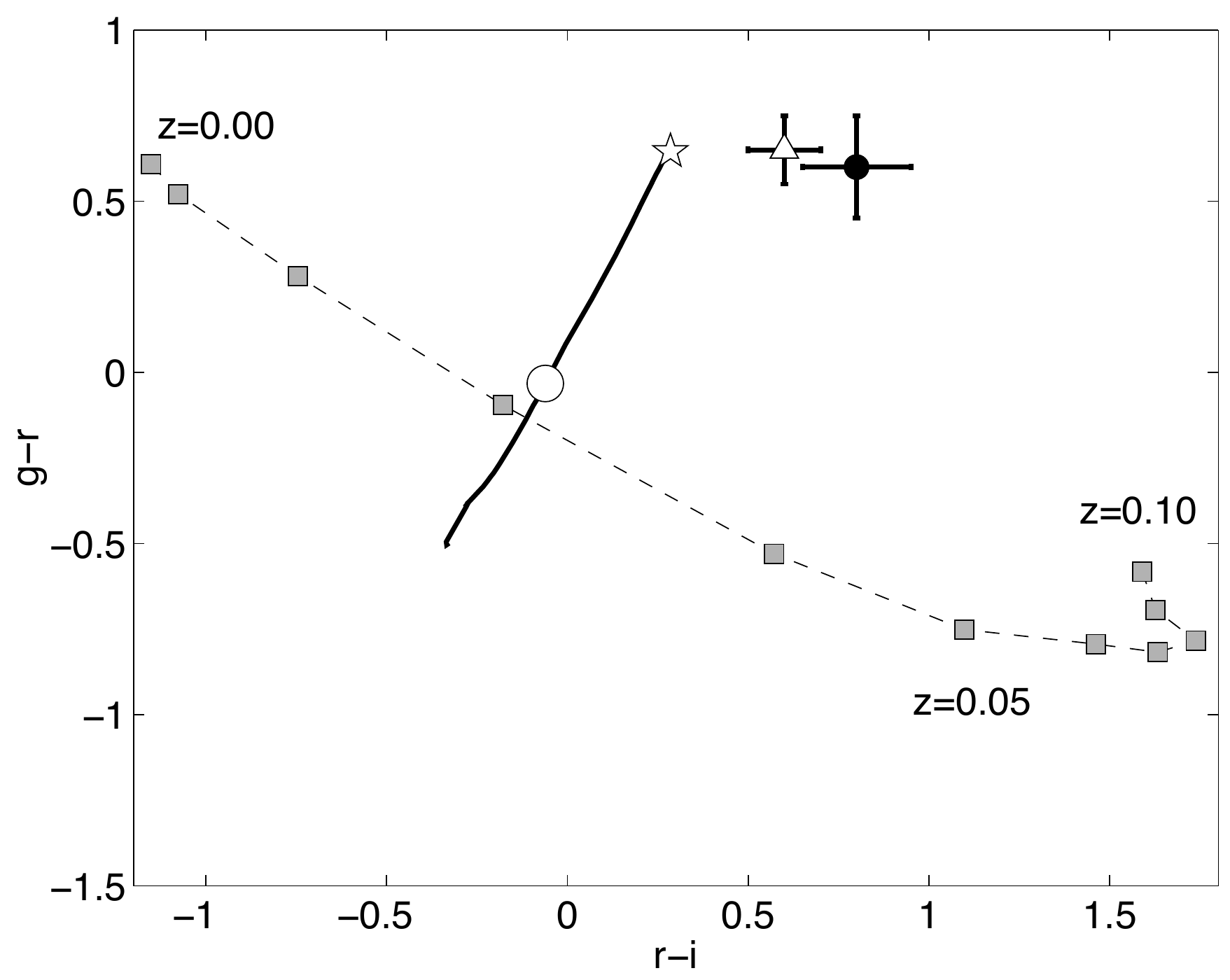}
\caption{The $r-i$ versus $g-r$ colours of the halo derived from the T sample (filled circle with error bars) and the Z04 halo (open triangle with error bars), compared to the predicted colours of a photoionized nebula (metallicity $Z=0.001$, hydrogen density $n(\mathrm{H})=10^{-3}$ cm$^{-3}$ and filling factor $f=0.1$) at redshifts from $z=0$ to $z=0.1$ (gray squares). The solid line indicates the spectral evolution of a halo-like stellar population ($Z=0.001$, a Salpeter IMF and an exponentially decaying star formation rate ($\mathrm{SFR}(t)\propto \exp(-t/\tau)$ with $\tau=1$ Gyr) as predicted by the \citet{2001A&A...375..814Z} model at $z=0.06$ (the median redshift of sample T). Markers along this line indicate ages of 1 Gyr (circle) and 14 Gyr (star). At redshifts $z\geq 0.04$, this model nebula can in principle generate the very red $r-i$ halo colours, but has difficulties in simultaneously explaining the $g-r$ data. A superposition of nebular emission and a normal stellar halo population also fails in this respect.}
\label{nebcol}
\end{figure}

Another possibility is that the halo colours could be contaminated by extra-planar gas photoionized by hot stars in the disk. This scenario was investigated by \citet{2006ApJ...650..812Z} and found to give $r-i$ colours that were much too blue to be reconciled with the colours derived by Z04. This would imply that, even though nebular emission could possibly account for some of the light detected at large projected distances from the disk, correcting for this contribution would just make the underlying halo redder and therefore even more difficult to explain. The models presented by \citet{2006ApJ...650..812Z} were, however, based on the approximation of zero redshift (i.e. no $k$-correction on the broadband fluxes). In reality, the redshift of the Z04 sample peaks at $z\approx 0.05$, whereas the median redshift of our sample T is $z\approx 0.06$. While these redshifts may seem small, they nonetheless turn out to have substantial effects on the predicted colours of photoionized gas.

Because of the high emission-line equivalent widths predicted for a purely nebular spectrum, $k$-corrections can become huge even for small redshifts. For the SDSS filters used here, the H$\alpha$ emission line shifts out of the $r$ filter and into $i$ at $z\approx 0.036$, which means that the the $r-i$ colour of photoionized gas turns very red at redshifts higher than this. This is demonstrated in Fig.~\ref{nebcol}, where we plot the redshift dependence (gray squares indicating the evolution from $z=0$ to $z=0.1$ in steps of $\Delta z=0.01$) of the colours predicted for an HII region with low density ($n(\mathrm{H})=10^{-3}$ cm$^{-3}$) and high filling factor ($f=0.1$), as expected for extraplanar diffuse ionized gas \cite[e.g.][]{2000A&A...359..433R}. This model nebula has been generated using the \citet{2001A&A...375..814Z} spectral synthesis code, assuming a low-metallicity ($Z_\mathrm{stars}=Z_\mathrm{gas}=0.001$), 10 Gyr old stellar population with a Salpeter-IMF and disk-like star formation history ($\mathrm{SFR}(t)\propto \exp(-t/\tau)$, $\tau=6$ Gyr) as the ionization source. The colours of the resulting nebula are seen to evolve dramatically as a function of redshift,  eventually (at $z\geq 0.05$) reaching $r-i$ colours {\it even redder} that those of the halo colours observed by Z04 (open triangle) and by us (filled circle). Hence, \citet{2006ApJ...650..812Z} may have been too hasty in dismissing nebular emission as the cause of the red halo colours.

As the H$\alpha$ emission line redshifts out of the $r$ filter, $g-r$ admittedly turns blue at the same time as $r-i$ turns red. Because of this, the nebular model presented in Fig.~\ref{nebcol} is unable to simulataneously reproduce both halo colours at any of the relevant redshifts. Mixing this nebular spectrum with that of a normal halo-like stellar population (solid line in Fig.~\ref{nebcol}) does not allow the observed $g-r$ and $r-i$ colours to be reproduced either. However, predicting the exact optical colours of extraplanar nebular gas is a very difficult endeavour, and we cannot rule out the possibility that a model more realistic than the one used in Fig.~\ref{nebcol} may be able to provide a better fit to the observations. 

The nebular model presented assumes a complete, ionization-bounded Str\"omgren sphere of constant density, whereas the halo observations are likely to probe only the outer regions of the ionized cloud surrounding the stacked disk. \citet{2006ApJ...650..812Z} attempted to model the spectra of individual sightlines through the outskirts of HII regions, and predicted a large scatter in the predicted optical colours. The presence of shocks, turbulent mixing layers, density gradients and a multiphased medium would make the problem even more complex. One way to test whether nebular emission really is responsible would be to stack a sample of galaxies all located at $z<0.036$. If a halo with an anomalously red $r-i$ colour is detected in such a sample, nebular emission can be ruled out as the primary reason. Unfortunately, our T sample is not sufficiently large for this exercise. Attempts to remove all $z\geq 0.036$ objects from the sample results in a halo colour profile with too large error bars to reliably claim the detection of a red excess.

While the nebular models used here and in \citet{2006ApJ...650..812Z} have turned out to have a very strong redshift dependence in $r-i$, this is not the case for colours like $V-K$. Therefore, the \citet{2006ApJ...650..812Z} argument -- that nebular emission would be far too blue to explain the anomalously red $V-K$ colours reported around other types of galaxies -- still holds. Postulating photoionization as the primary explanation for the flux excess detected in the $i$-band (Z04 and the present paper) would require a combination of two different mechanisms to explain red halos in general. Hence, the same issues raised against extended red emission apply to nebular emission as well.

\subsection{The extinction of extragalactic background light}
One of the obvious obstacles when attempting to do surface photometry at surface brightness levels as faint as $\mu_i\approx 29$ mag arcsec$^{-2}$ is the challenge of subtracting the sky with sufficient accuracy. The night sky has a typical surface brightness of $\mu_i\approx 20.3$ mag arcsec$^{-2}$ at the site of the SDSS telescope (Apache Point), which means that the signal we are measuring is about 3000 times fainter than the sky. To measure a surface brightness with an error of $\pm 0.1$ mag at such isophotal levels requires that the sky can be subtracted with an accuracy of about one part in 30000. While we have argued in section 3 that the night sky can indeed be measured with the required accuracy by stacking $\sim 1500$ SDSS images, there may be a subtle systematic problem involved in the actual subtraction which we cannot fully correct for. 
\begin{figure*}
\includegraphics[width=8cm]{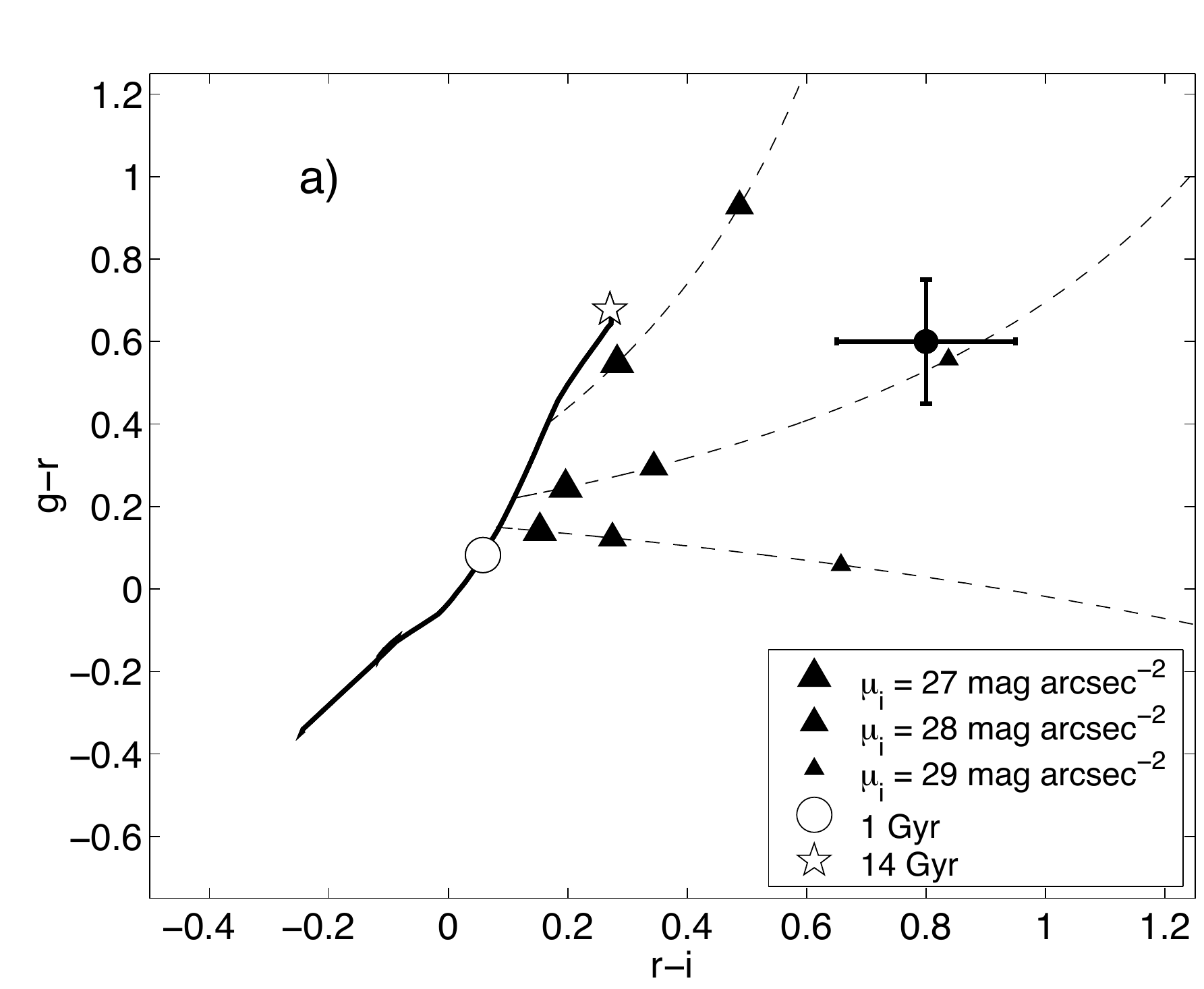}
\includegraphics[width=8cm]{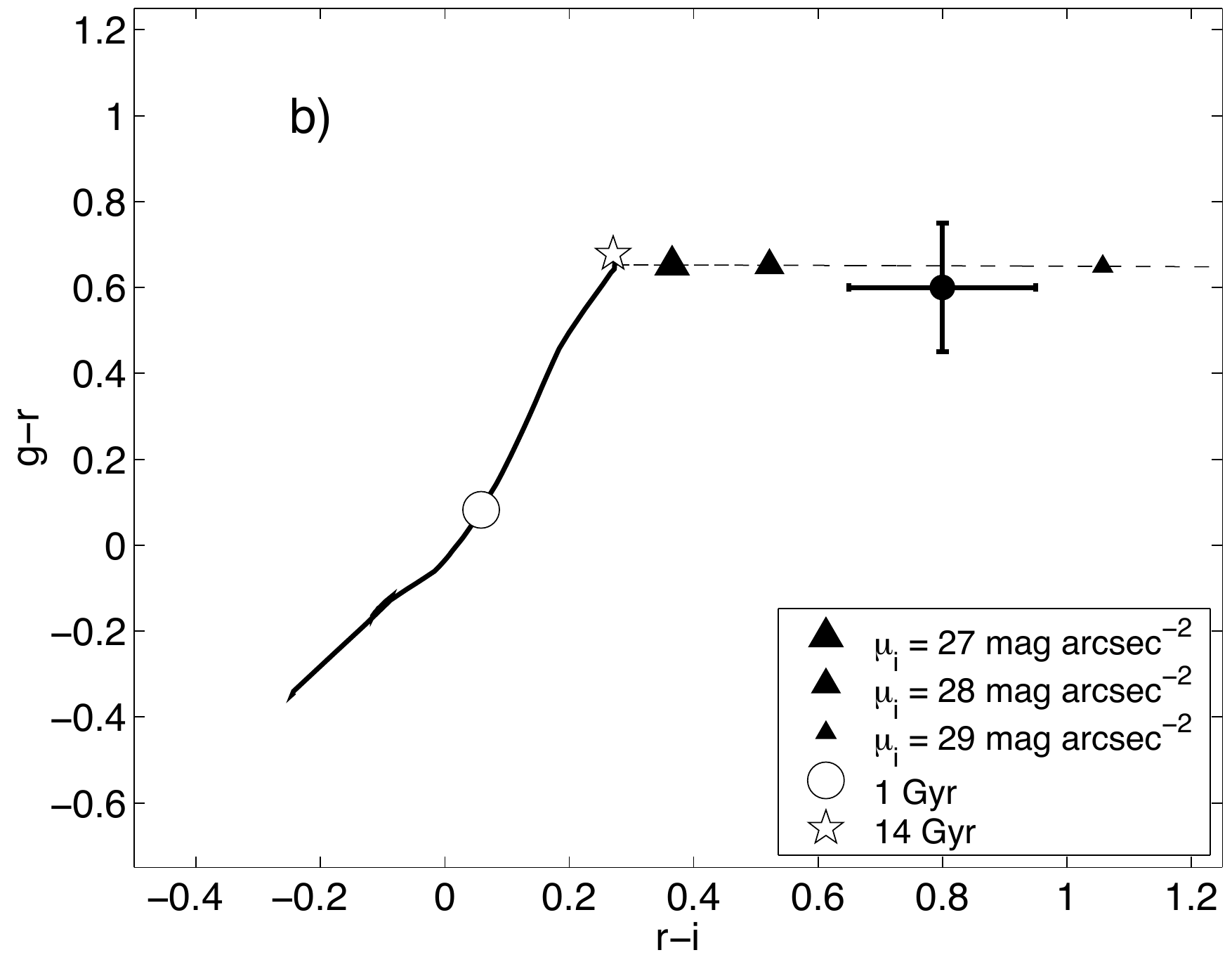}
\caption{The effect of EBL extinction in a diagram of $r-i$ vs. $g-r$. The solid line indicates the evolution of a metal-poor ($Z=0.001$) stellar population with a Salpeter initial mass function and an exponentially decaying star formation rate ($\mathrm{SFR}\propto\exp{-t/\tau}$ with $\tau=1$ Gyr), based on \citet{2008A&A...482..883M} isochrones. White markers represent ages of 1 Gyr (circle) and 14 Gyr (star). The filled circle represents the colours of the halo detected in the T sample (at $\mu_i\approx 28$--29 mag arcsec$^{-2}$).  Dashed lines indicate how a spatially constant EBL extinction would shift the observed colours away from the model predictions under the assumption of a Milky Way extinction curve. The differently-sized triangles along these tracks correspond to surface brightness levels of $\mu_i=27$, 28 and 29 mag arcsec$^{-2}$ (largest to smallest triangles).
{\bf a)} Results assuming an EBL extinction of $A(g)=0.15$ mag and fiducial diffuse EBL levels \citet{2009MNRAS.397.2057Z}.  At $\mu_i=29$ mag arcsec$^{-2}$, EBL extinction can indeed bring the colours of a normal stellar population into agreement with those observed. This particular solution does, however, require a fairly young ($\approx 3$ Gyr) stellar halo population. {\bf b)} Results assuming an EBL extinction of $A(g)=0.03$ mag and diffuse EBL levels slightly offset from the fiducial ones (see main text for details). In this case, EBL extinction effects can reconcile a 10 Gyr old stellar halo with the colours observed at $\mu_i\approx 28$--29 mag arcsec$^{-2}$.}
\label{EBLfig}
\end{figure*}

This problem \citep[first recognized by][]{2009MNRAS.397.2057Z} stems from to the fact that while most of the sky flux originates from regions between the telescope and the galaxies that we target, a small fraction - the extragalactic background light (EBL) - comes from behind. The EBL at optical wavelengths is believed to be the product of direct and reprocessed starlight emitted over the entire star formation history of the Universe, and hence stems from objects at vastly different redshifts. Provided that the existing EBL measurements are correct, most of this light appears to be diffuse (i.e. unresolved with all existing instruments). Unlike the other components of the night sky flux, the EBL can be subject to extinction by dust present in the halos of our target galaxies, thereby invalidating an implicit assumption adopted in all current surface photometry measurements, namely that the sky flux and the flux from the target objects are unrelated. The extinction of the EBL may lead to a slight depression of the overall sky flux across the galaxies (and their halos) in our sample. Since our sky subtraction method is based on estimates of the sky flux well away from the target galaxies -- where the EBL suffers less extinction than in the regions where the halo colours are measured -- a systematic oversubtraction of the sky may result. As argued by \citet{2009MNRAS.397.2057Z}, even very small amounts of extinction can in principle lead to spurious features in radial surface brightness profiles and colour maps of extended objects. 

\citet{2009MNRAS.397.2057Z} showed that even though the halo colours derived by Z04 could in principle be attributed to EBL extinction effects, this requires both an unresolved EBL level at the high end of what is allowed by current error bars, and an EBL extinction of $A(g)=0.2$ mag in the region where the halo colours are measured ($\approx 8$ kpc above the disk in the Z04 sample). Our halo colours are, however, derived at $\mu_i\approx 28$--29 mag arcsec$^{-2}$, which is about 1--2 mag fainter than that of the region where Z04 note the onset of anomalous colours ($\mu_i\approx 26.7$ mag arcsec$^{-2}$). This means that a similar colour shift can be produced for a somewhat lower EBL level and/or a lower halo opacity.

This is demonstrated in Fig.~\ref{EBLfig}, where we have adopted the formalism from \citet{2009MNRAS.397.2057Z} to predict the colour shifts in a diagram of $r-i$ vs. $g-r$ resulting from a failure to account for EBL extinction. The solid line represents the predicted evolution of an old, metal-poor ($Z=0.001$) stellar halo population modelled using \citet{2008A&A...482..883M} isochrones with a Salpeter initial mass function. The dashed lines in Fig.~\ref{EBLfig}a indicate how the observed colours would shift due to EBL extinction if we assume an extinction of $A(g)=0.15$ mag in the halo. The shift becomes increasingly severe at fainter surface brightness levels (triangles along the dashed lines indicating the colours at the $\mu_i=27$, 28 and 29 mag arcsec$^{-2}$ isophotes). As discussed by Zackrisson et al., the direction of the shift depends on the intrinsic colours of the halo, which is why three different dashed lines are included. Indeed, one of the dashed lines passes very close to the observed colours at $\mu_i\approx 29$ mag arcsec$^{-2}$. No artificial boost of the observed EBL level (like that used by Zackrisson et al. to fit the Z04 data) is required. This particular line originates from a fairly young ($\approx 3$ Gyr) halo population, but similar fits for older populations can be achieved by assuming slightly different colours for the EBL (well within the observational errorbars). This is shown in Fig.~\ref{EBLfig}b, where we have adopted $\mu_g=26.1$ mag arcsec$^{-2}$, $\mu_r=25.1$ mag arcsec$^{-2}$ and $\mu_i=25.4$ mag arcsec$^{-2}$ for the diffuse EBL (0.2 mag fainter in $gi$ and 0.3 mag brighter in $r$ than the fiducial values). In this case, the observed colours are reproduced using a 10 Gyr old halo population and a halo extinction of just $A(g)=0.03$ mag.

The anomalous colours of the halo in sample T turn up at $\approx 80$--110 pixels from the plane of the disk (see Fig.~\ref{col_halo}), which converts into $\approx 10$--13 kpc (assuming 120 pc per pixel). It is difficult to judge whether an extinction of $A(g)=0.03$--0.15 mag is reasonable in this region, since no studies of the opacity profiles of halos have specifically targeted low surface brightness galaxies. However, \citet{1994AJ....108.1619Z} and \citet{2009arXiv0902.4240M} have measured optical extinctions in the $\approx 0.03$--0.15 mag in the halos of other types of galaxies (at even greater projected distances than what we probe here). Dust is also well known to exist several kpc above the plane of edge-on discs in general \citep[e.g.][]{2000A&AS..145...83A}. 

The dust column density at the site of our halo observations can be estimated from
\begin{equation}
S_\mathrm{dust} = \frac{0.4 A}{K_\mathrm{ext}\log(e)}
\end{equation}
where $A$ is the extinction in magnitudes, $K_\mathrm{ext}$ is the dust extinction coefficient and $S_\mathrm{dust}$ is the dust column density. The dust extinction coefficients in the V band for the Milky Way, LMC and SMC are given by \citet{2001ApJ...548..296W}. Since the effective wavelength of the $g$ band is close to that of Johnson B, we will assume $E(g-V) = E(B-V)\sim A(V)/3$. Thus $A(g)=4 A(V)/3$. We expect the metallicity in the halo to be low and therefore the most appropriate value to use here is that of SMC, $K_\mathrm{ext}(V)$ = $1.54\times 10^4$ cm$^2$g$^{-1}$ and thus $K_\mathrm{ext}(g)$ = $2.05\times 10^4$ cm$^2$g$^{-1}$. $A(g)=0.03$--0.15 mag then corresponds to dust column densities of 1.4$\times 10^{-6}$ -- 6.7$\times 10^{-6}$ g cm$^{-2}$. Assuming a pixel size of 120 pc, these numbers correspond to a a range in dust mass surface density of $\Sigma_\mathrm{dust} =  90$--460 $M_\odot\ \mathrm{pixel}^{-1}$. We may compare these values with the stellar mass density per pixel that we derived in Section 4.2. In the $g$ band, at a surface brightness of $\mu \sim 29$ mag arcsec$^{-2}$, we obtain $\Sigma_\mathrm{stars}\approx 3.6 \times 10^3\ M_\odot \ \mathrm{pixel}^{-1}$. Thus the dust-to-stellar mass density ratio is $M_\mathrm{dust}/ M_\mathrm{stars}\approx 2$--13\%. How does this compare to the global dust properties of LSB galaxies? \citet{2007ApJ...663..908R} recently presented dust masses and total stellar luminosities for a few LSB galaxies. From these data, we estimate a typical dust-to-stellar mass ratio of $M_\mathrm{dust}/ M_\mathrm{stars} \approx 1$\% by assuming a stellar mass-to-light ratio of $M/L \approx 1$ \citep{2005A&A...435...29Z}. 

This indicates, that to accomodate a halo extinction of $A(g)=0.03$--0.15 mag, the opacity would have to decline slower than the stellar surface brightness as one moves in the polar direction away from the disk. A dust distribution more extended than that of stars has already been confirmed within high surface brightness disks \citep[e.g.][]{2005A&A...444..109H,2009ApJ...703.1569M}. Considering that halo opacity profiles \citep{1994AJ....108.1619Z,2009arXiv0902.4240M} appear to fall off much slower than halo surface brightness profiles (Z04 and this paper) would suggest a similar trend for the dust-to-stellar mass ratios in halos, although the selection criteria of these different studies do not exactly match.

If the red halo colours are due to EBL extinction effects, one may naively expect the thin disk to display a similar colour anomaly at comparable surface brightness levels. However, the colour profiles along the thin disk (Fig.~\ref{col_thin}) in sample T show no obvious sign of this, and neither do the disk profiles presented by Z04. \citet{2009MNRAS.397.2057Z} discuss two possible explanations. Firstly, the colour shifts are very sensitive to the intrinsic colours of the affected stellar population (as seen in Fig.~\ref{EBLfig}a), and a disk population is not expected to have colours identical to that of a stellar halo. Moreover, the extinction curve may be grayer in the disk \citep[e.g.][]{2005AJ....129.1396H}. Both of these effects may contribute to diminish or alter the offset vector of any colour shifts related to EBL extinction. Hence, the lack of a red excess in the disk does not rule out EBL extinction as the cause of the red excess in the halo. As argued by \citet{2009MNRAS.397.2057Z}, EBL extinction may also be able to explain the anomalous halo colours detected around other types of galaxies and at longer wavelengths.

\section{Summary and conclusions}

We present the results of an investigation of the statistical photometric properties of a homogeneous set of 1510 LSB disk galaxies drawn from the SDSS DR5. There have been claims that the faint outskirts of normal disk galaxies and blue compact galaxies tend to show colours that are redder than predicted by stellar evolutionary models - the so called 'red halo' phenomenon. We have therefore focused on deriving the characteristic colours of the faintest regions in the halos of our LSB galaxies. For this purpose we use a statistical approach to reach the faintest levels. Thus we have rescaled, aligned and stacked the $g,r$ and $i$ images, thereby increasing the S/N of the combined image. In the finally stacked images we typically reach a surface brightness of $\mu \sim$  31 mag arcsec$^{-2}$. 

We derive surface brightness and colour profiles based on median filtering and averaging of the images in the stack. When we reach $\mu \sim$ 28--29 mag arcsec$^{-2}$, a significant red excess appears in $r-i$.  We find no support for the notion that this effect can be attributed to scattered light.  After a careful assessment of PSF effects, the combined colours, $g-r = 0.60\pm0.15$ and $r-i = 0.80\pm0.15$, are incompatible with colours of a normal stellar population. The fact that no similar colour anomaly is seen at comparable surface brightness levels along the disk rules out a sky subtraction residual as the source of the extreme halo colours. We discuss possible alternatives to explain the red colours. These include a bottom heavy IMF, nebular emission, extended red emission and dust extinction of extragalactic background light. We find two of these scenarios to be broadly consistent with our results and other claims of a red excess in the optical/near-IR. The first is extinction of extragalactic background light, which would cause a colour-dependent oversubtraction of the sky background in the region where dust is present. In this case, a halo opacity of A(g) =  0.03--0.15 mag at a surface brightness of $\mu_g \sim$ 29 mag arcsec$^{-2}$ would be needed to explain the observations. The second alternative is an abnormal IMF, dominated by low-mass stars. 

\label{sec:conclus}

\section*{Acknowledgments}

N. Bergvall acknowledges support from the Swedish Research Council.
E. Zackrisson acknowledges research grants from the Academy of Finland, the Royal Swedish Academy of Sciences and the Royal Physiographical Society of Lund. The authors would like to thank Luca Casagrande for advice concerning the reliability of current models for low-mass stars, and Zhongmu Li for input on the effects of binary evolution. We also want to thank the referee, Ramin Skibba, whose many detailed comments helped to improve the quality of the paper. Funding for the Sloan Digital Sky Survey (SDSS) has been provided by the Alfred P. Sloan Foundation, the Participating Institutions, the National Aeronautics and Space Administration, the National Science Foundation, the U.S. Department of Energy, the Japanese Monbukagakusho, and the Max Planck Society. The SDSS Web site is http://www.sdss.org/.

 The SDSS is managed by the Astrophysical Research Consortium (ARC) for the Participating Institutions. The Participating Institutions are The University of Chicago, Fermilab, the Institute for Advanced Study, the Japan Participation Group, The Johns Hopkins University, Los Alamos National Laboratory, the Max-Planck-Institute for Astronomy (MPIA), the Max-Planck-Institute for Astrophysics (MPA), New Mexico State University, University of Pittsburgh, Princeton University, the United States Naval Observatory, and the University of Washington.

\bibliographystyle{mn2e}
\bibliography{lsbhalo}

\bsp

\label{lastpage}

\end{document}